\documentclass[11pt]{article}
\usepackage{amsmath,amsthm,amssymb,cite,enumerate}
\usepackage{multirow}
\usepackage[a4paper,left=0.5in,right=1in]{geometry}
\usepackage[usenames]{color}
\usepackage{graphicx}
\usepackage{float}
\usepackage{epsfig}
\usepackage{geometry}
\usepackage{array} 
\usepackage{mathrsfs}
\usepackage{dcolumn}
\usepackage{bm}
\usepackage{authblk} 
\usepackage{soul} 
\usepackage{t1enc}	
\usepackage{comment}
\usepackage{xcolor}
\usepackage{longtable}
\begin{document}

\def \d {{\rm d}}

\def \bm {\mbox{\boldmath{$m$}}}
\def \bmt {\mbox{\boldmath{$\tilde m$}}}

\def \bF {\mbox{\boldmath{$F$}}}
\def \bA {\mbox{\boldmath{$A$}}}
\def \cF {\mbox{\boldmath{$\cal F$}}}
\def \bH {\mbox{\boldmath{$H$}}}
\def \bC {\mbox{\boldmath{$C$}}}
\def \bSS {\mbox{\boldmath{$S$}}}
\def \bS {\mbox{\boldmath{${\cal S}$}}}
\def \bV {\mbox{\boldmath{$V$}}}
\def \bff {\mbox{\boldmath{$f$}}}
\def \bT {\mbox{\boldmath{$T$}}}
\def \bk {\mbox{\boldmath{$k$}}}
\def \bl {\mbox{\boldmath{$\ell$}}}
\def \bn {\mbox{\boldmath{$n$}}}
\def \bbm {\mbox{\boldmath{$m$}}}
\def \tbbm {\mbox{\boldmath{$\bar m$}}}
\def \bet {\mbox{\boldmath{$\eta$}}}
\def \H {{\cal H}}
\def \bmu {\mbox{\boldmath{$\mu$}}}
\def \bSig {\mbox{\boldmath{$\Sigma$}}}
\def \T {\bigtriangleup}
\newcommand{\msub}[2]{m^{(#1)}_{#2}}
\newcommand{\msup}[2]{m_{(#1)}^{#2}}

\newcommand{\be}{\begin{equation}}
\newcommand{\ee}{\end{equation}}

\newcommand{\beqn}{\begin{eqnarray}}
\newcommand{\eeqn}{\end{eqnarray}}
\newcommand{\AdS}{anti--de~Sitter }
\newcommand{\AAdS}{\mbox{(anti--)}de~Sitter }
\newcommand{\AAN}{\mbox{(anti--)}Nariai }
\newcommand{\AS}{Aichelburg-Sexl }
\newcommand{\pa}{\partial}
\newcommand{\pp}{{\it pp\,}-}
\newcommand{\ba}{\begin{array}}
\newcommand{\ea}{\end{array}}

\newcommand*\bR{\ensuremath{\boldsymbol{R}}}

\newcommand*\BF{\ensuremath{\boldsymbol{F}}}
\newcommand*\BR{\ensuremath{\boldsymbol{R}}}
\newcommand*\BS{\ensuremath{\boldsymbol{S}}}
\newcommand*\BC{\ensuremath{\boldsymbol{C}}}
\newcommand*\bg{\ensuremath{\boldsymbol{g}}}
\newcommand*\bE{\ensuremath{\boldsymbol{E}}}

\newcommand*\bh{\ensuremath{\boldsymbol{h}}}
\newcommand*\bZ{\ensuremath{\boldsymbol{Z}}}

\def \bo {\mbox{\boldmath{$\omega$}}}
\def \bot {\mbox{\boldmath{$\tilde\omega$}}}
\def \bE {\mbox{\boldmath{$e$}}}
\def \bEt {\mbox{\boldmath{$\tilde e$}}}
\def \bG {\mbox{\boldmath{$\Gamma$}}}
\def \bGt {\mbox{\boldmath{$\tilde\Gamma$}}}
\def \bTt {\mbox{\boldmath{$\tilde\Theta$}}}

\newcommand{\M}[3] {{\stackrel{#1}{M}}_{{#2}{#3}}}
\newcommand{\m}[3] {{\stackrel{\hspace{.3cm}#1}{m}}_{\!{#2}{#3}}\,}

\newcommand{\tr}{\textcolor{red}}
\newcommand{\tb}{\textcolor{blue}}
\newcommand{\tg}{\textcolor{green}}

\newcommand{\thorn}{\mathop{\hbox{\rm \th}}\nolimits}

\def\a{\alpha}
\def\g{\gamma}
\def\de{\delta}

\def\E{{\cal E}}
\def\B{{\cal B}}
\def\R{{\cal R}}
\def\F{{\cal F}}
\def\L{{\cal L}}

\def\e{e}
\def\bb{b}

\newtheorem{theorem}{Theorem}[section] 
\newtheorem{cor}[theorem]{Corollary} 
\newtheorem{lemma}[theorem]{Lemma} 
\newtheorem{proposition}[theorem]{Proposition}
\newtheorem{definition}[theorem]{Definition}
\newtheorem{remark}[theorem]{Remark}

\title{Algebraic and optical properties of generalized Kerr-Schild spacetimes in arbitrary dimensions}

\author[1,2]{Aravindhan Srinivasan\thanks{srinivasan(at)math(dot)cas(dot)cz}}

\affil[1]{Institute of Mathematics, Czech Academy of Sciences, \newline \v Zitn\' a 25, 115 67 Prague 1, Czech Republic}

\affil[2]{Institute of Theoretical Physics, Faculty of Mathematics and Physics, \newline
 Charles University, V Hole\v{s}ovi\v{c}k\'{a}ch 2, 180 00 Prague 8, Czech Republic}

\maketitle
\abstract{ We study the class of generalized Kerr-Schild (GKS) spacetimes in dimensions $n\geq 3$ and analyze their geometric and algebraic properties in a completely theory-independent setting. First, considering the case of a general null vector $\mathbf{k}$ defined by the GKS metric, we obtain the conditions under which it is geodesic. Assuming $\mathbf{k}$ to be geodesic for the remainder of the paper, we examine the alignment properties of the curvature tensors, namely the Ricci and Weyl tensors. We show that the algebraic types of the curvatures of the full (GKS) geometry are constrained by those of the respective background curvatures, thereby listing all kinematically allowed combinations of the algebraic types for the background and the full geometry. A notable aspect of these results is that, unlike the case of Kerr-Schild (KS) spacetimes, the Weyl types of the GKS spacetimes need not be type $II$ or more special. Then, focusing on the case of an expanding $\mathbf k$, we derive the conditions for it to satisfy the optical constraint, extending the previous results of KS spacetimes. We illustrate the general results using the example of (A)dS-Taub-NUT spacetimes in $n=4$, where we also comment on their KS double copy from a GKS perspective. Finally, as an application of our general results, we obtain the full family of GKS spacetimes with a geodesic, expanding, twistfree, and shearfree $\mathbf k$, satisfying the vacuum Einstein equations, and identify it with a subset of the higher-dimensional vacuum Robinson-Trautman solutions. In passing, we also determine the subcase of these solutions that manifests the KS double copy.

}

\vspace{.2cm}
\noindent

%
\tableofcontents
\section{Introduction}
\label{intro}

\subsection{Background}\label{background_intro}

The Kerr-Schild (KS) class, which encompasses several important spacetimes, has been a significant topic of study for quite some time. The KS class is defined by the following metric form \cite{Kerr:1965wfc}
\begin{align}
    \mathbf{g}&=\mathbf{\Bar{g}} -2H \mathbf{k}\otimes \mathbf{k},\label{GKS}
\end{align}
where the background $\mathbf{\Bar{g}}$ is the flat spacetime $\bet$, $H$ is a scalar function, and $\mathbf{k}$ is a $1$-form that is null with respect to the background and hence also with respect to the full metric $ \mathbf{g}$. Notable examples of four-dimensional KS spacetimes include Kerr, Kerr-Newman, Vaidya, and pp-waves \cite{Kerr:1963ud,Kerr_Newman,Vaidya:1951zz,Peres_PPwaves}. Moreover, the KS ansatz was instrumental in obtaining the Myers-Perry solutions \cite{Myers:1986un}, which generalize Kerr black holes to higher dimensions. Additionally, the KS class has also attracted increased attention recently in the form of the KS double copy prescription \cite{double_copy_Connel_2014}, which relates exact vacuum solutions of GR with Maxwell or Yang-Mills solutions in flat spacetime. 

There have also been studies that focus on the geometric and algebraic properties of the KS class in both four and higher dimensions, which are particularly useful in the context of the classification of exact solutions \cite{Pet54,Pirani,Debever1959,PENROSE1960171,Coley:2004jv,Milson:2004jx}. In four-dimensional vacuum spacetimes, the KS vector $\mathbf{k}$ generates a geodesic, shearfree null congruence, coinciding with the repeated principal null direction (PND) of the Weyl tensor \cite{Stephani:2003tm}. This is consistent with the four-dimensional Goldberg-Sachs theorem \cite{Gold_Sachs_62,newman1962approach}, which states that \textit{a null congruence in a non-conformally flat Einstein spacetime is geodesic and shearfree if and only if it forms a repeated PND of the corresponding Weyl tensor.} Although the Goldberg-Sachs theorem does not straightforwardly generalize to higher dimensions, there has been partial success in developing some notions of generalizations. Notably, a result for the ``geodesic part'' of the theorem was proved in \cite{Durkee:2009nm}. Similarly, several results concerning the ``shearfree part'' have been obtained for specific dimensions \cite{Ortaggio:2012hc, Tintera:2019hil}, particular algebraic types \cite{VPravda_2004,type_III_N}, or under certain simplifying assumptions on the Weyl aligned null direction (WAND) \cite{Gold_Sachs_HD}. In this context, \cite{HD_KS} explored several geometric features of higher-dimensional vacuum KS spacetimes and showed that the KS congruence is a geodesic multiple WAND (mWAND),\footnote{The higher-dimensional generalization of the repeated PND is the mWAND \cite{Coley:2004jv,HD_alg_review}.} analogous to its four-dimensional counterpart. Moreover, for the case of an expanding KS vector (defined in equation \eqref{optical_scalars}, cf. also Section \ref{expanding ks}), the optical matrix of $\mathbf{k}$ (defined in Section \ref{Prelims}) satisfies what is known as the optical constraint (cf. Section \ref{Optical constraint}), which can be regarded as a `weaker' version of the Goldberg-Sachs theorem in higher dimensions, restricted to the case of KS spacetimes \cite{HD_KS}. The results of \cite{HD_KS} were later extended to KS spacetimes with maximally symmetric backgrounds in \cite{Malek:2010mh}.\footnote{Hereafter, the term KS spacetimes (or class) is to be understood as including its extensions with maximally symmetric backgrounds.}

In this paper, as a further extension of \cite{HD_KS} and \cite{Malek:2010mh}, we study the class of generalized Kerr-Schild (GKS) spacetimes defined by ansatz \eqref{GKS} and satisfying the following conditions
\begin{enumerate}[I.]
\item \label{GKS_assum_1} $\mathbf{\Bar{g}}$ is a general background which, in principle, can be any Lorentzian metric in $n$ spacetime dimensions.
\item \label{GKS_assum_2} There exist parameters $\mu_\alpha \hspace{1mm} (\alpha=1,2, \dots, \mathcal{J})$, which in general can be local or non-constant, such that the background $\mathbf{\Bar g}$ is independent of them and $(H\mathbf{k}\otimes \mathbf{k})|_{\mu_\alpha=0}=0$, implying that $\mathbf{\Bar{g}}= \mathbf{g}|_{\mu_\alpha=0}$. If not for this condition, any metric $\mathbf{g}$ can be cast into the GKS form \eqref{GKS} by defining the background tautologically as $\mathbf{\Bar{g}}=\mathbf{g}+2H \mathbf{k}\otimes \mathbf{k}$ for arbitrary $H$ and $\mathbf{k}$. 
\item \label{GKS_assum_3}By means of the rescaling freedom $H\rightarrow \Omega^2 H$, $\mathbf{k}\rightarrow \Omega^{-1}\mathbf{k}$, the $\mu_{\alpha}$-dependence of $(H\mathbf{k}\otimes \mathbf{k})$ can be completely transferred to the scalar function $H$, so that $\mathbf{k}$ is independent of the parameters $\mu_\alpha$, i.e., $\mathbf{\Bar{k}}\equiv\mathbf{k}|_{\mu_\alpha=0}=\mathbf{k}$, and hence $H|_{\mu_\alpha}=0$. We will implicitly assume that $H$ and $\mathbf{k}$ in \eqref{GKS} have already been rescaled to such a form.   
\end{enumerate}
Note that condition \ref{GKS_assum_3} does not fully fix the rescaling freedom, as one can still rescale $H$ and $\mathbf{k}$ by $\mu_\alpha$-independent functions. Conditions \ref{GKS_assum_2} and \ref{GKS_assum_3} imply that the background geometry and the quantities defined on it are immune to changes in $H$ caused by changes in the parameters $\mu_\alpha$. Throughout the paper, this property will be referred to as $H$-independence of the background.

Let us illustrate the conditions \ref{GKS_assum_1}--\ref{GKS_assum_3} using the Kerr-Newman metric \cite{Kerr_Newman}, which can be cast into the GKS form with the following three interpretations:
\begin{enumerate}[(i)]
\item The standard KS form of Kerr-Newman, with a flat background $\bet$, given by \cite{Debney:1969zz}
\begin{align}
& \Bar{\mathbf{g}}=\bet=-\d u^2+2\d r(\d u+a\sin^2\theta\d\phi)+(r^2+a^2\cos^2\theta)\d\theta^2+(r^2+a^2)\sin^2\theta\d\phi^2 , \label{flatbg} \\
& \bk=\d u+a\sin^2\theta\d\phi , \label{KS_vector_kerr}\\
& 2H=2H_{\mbox{\tiny KN}}\equiv-\frac{2Mr-Q^2}{r^2+a^2\cos^2\theta} ,
\label{H_KN}
\end{align}
where $M$ and $Q$ are respectively the mass and electric charge parameters. Clearly, the background $\bet$ is independent of the parameters $M$ and $Q$, and setting these parameters to zero leads to the vanishing of $H_{\mbox{\tiny KN}}$. One can therefore identify these as the parameters described in condition \ref{GKS_assum_2}. The null $1$-form $\mathbf{k}$ is independent of the charge and mass, in agreement with condition \ref{GKS_assum_3}. 
\item GKS form with an uncharged Kerr background, which can be obtained by absorbing the $M$ dependent term of \eqref{H_KN} into $\Bar{\mathbf{g}}$ as
\begin{align}
& \Bar{\mathbf{g}}=\mathbf{g}_{\mbox{\tiny Kerr}}=\bet -\frac{2Mr}{r^2+a^2\cos^2\theta} \mathbf{k}\otimes \mathbf{k}, \quad 2H=\frac{Q^2}{r^2+a^2\cos^2\theta}.
\end{align}
In this line of interpretation, we identify $Q$ as the only parameter that fits with the notions of conditions \ref{GKS_assum_2} and \ref{GKS_assum_3}. 
\item Finally, one can interpret Kerr-Newman to be in a GKS form with the background being the massless limit of the full Kerr-Newman solution.
\begin{align}
& \Bar{\mathbf{g}}=\mathbf{g}_{\mbox{\tiny KN}}|_{M=0}=\bet +\frac{Q^2}{r^2+a^2\cos^2\theta} \mathbf{k}\otimes \mathbf{k}, \\
\label{Kerr_bg}
& 2H=-\frac{2Mr}{r^2+a^2\cos^2\theta}.
\end{align}
In this case, $M$ is the only parameter complying with conditions \ref{GKS_assum_2} and \ref{GKS_assum_3}. 
\end{enumerate}
For the Kerr-Newman case, the parameters $``\mu_\alpha"$ identified in $H$ are constants. However, as mentioned earlier, one could have cases where the parameters are local. For example, the Vaidya metric \cite{Vaidya:1951zz} belongs to the standard KS class (and hence also GKS \cite{vaidya_73}) with a flat background and is given by
\begin{align}
& \Bar{\mathbf{g}}=\bet=-\d u^2+2\d r \d u+r^2(\d\theta^2+\sin^2\theta\d\phi^2 ),\nonumber  \\
& \bk=\d u, \quad  2H=-\frac{2M(u)r}{r^2},
\end{align}
where now the parameter in $H$ is the time-dependent mass function $M(u)$.

The GKS ansatz, first introduced in \cite{Thompson} (and almost simultaneously in \cite{EDELEN}), has been the subject of many studies \cite{Grses1975LorentzCT,xanthopoulos1978, Taub:1981evj, XANTHOPOULOS_83, bilge, Dereli:1986cm,martin1986petrov ,nahmad_GKS_88, Senovilla_88, Stephani:2003tm,Barrientos:2024uuq}, with most of them focusing on the four-dimensional case. The GKS metric ansatz \eqref{GKS} can also be interpreted as a map between the two metrics $\mathbf{\Bar{g}}$ and $\mathbf{g}$, known as the GKS transformation \cite{Stephani:2003tm, bilge, Thompson}, whose group theoretic aspects in arbitrary spacetime dimensions were explored in \cite{Coll:2000rm}. Similar to the KS class, the GKS class also has some significance in the context of the KS double copy when the proposal is extended as a map between GKS solutions and Maxwell (or Yang-Mills) solutions in curved spacetimes \cite{Bahjat_double_copy_curved_bg_1,double_copy-curved_3,Gurses:2018ckx,doublecopy_curved_4,Prabhu_double_copy_curved}.

Although there have been extensive studies on the geometric and algebraic properties of four-dimensional GKS spacetimes (see \cite{Stephani:2003tm} and references therein), the study of higher-dimensional GKS spacetimes has primarily focused on generating specific exact solutions \cite{Dereli:1986cm,Barrientos:2024uuq}, with the notable exception of the group theoretic studies \cite{Coll:2000rm} mentioned earlier. Therefore, this work aims to derive some general geometric and algebraic properties of GKS spacetimes in arbitrary spacetime dimensions $n\geq3$, analogous to those in $n=4$  \cite{Stephani:2003tm},  thereby forming a GKS extension of the corresponding KS results obtained in \cite{HD_KS,Malek:2010mh}. Even though all the examples of GKS spacetimes discussed in this paper will be those that solve some gravitational field equations, for the purposes of obtaining general geometric results, we will not assume anything about the dynamics; hence, the results will hold true for any theory. 
\subsection{Summary of results}\label{summary_intro}

After setting up the notation in Section \ref{Prelims}, in Section \ref{GKS_vs_Bg_connection} we derive some basic relations for the Ricci rotation coefficients and frame covariant derivatives for the full geometry, $\mathbf{g}$, in terms of the corresponding quantities on the background, $\mathbf{\Bar g}$, and the function $H$ (of the GKS form \eqref{GKS}). In Section \ref{Geode_k}, we derive the conditions under which $\mathbf{k}$ is geodesic. Assuming the geodesicity of $\mathbf{k}$ thereafter, we obtain the following key results
\begin{itemize}
    \item In Section \ref{GKS_with_geo_k}, analyzing the alignment properties of the Ricci and Weyl tensors, we obtain several results on aligned null directions (ANDs) of these tensors for the background and the full geometry. In particular, we show that the curvature tensors of the full geometry cannot be algebraically more special than their background counterparts. Moreover, unlike KS spacetimes \cite{HD_KS,Malek:2010mh}, GKS spacetimes admit Weyl types less special than type $II$. Based on our deductions, we formulate Tables \ref{table_Ricci_types} and \ref{table_Weyl_types}, listing respectively all the kinematically allowed Ricci and Weyl types of the full geometry for all possible corresponding algebraic types of the background.\footnote{The results obtained for the Weyl tensor in Section \ref{GKS_with_geo_k} straightforwardly translate to the Riemann tensor. However, to keep the discussion as brief as possible, we have omitted the alignment properties of the Riemann tensor. }
    \item In Section \ref{expanding ks}, assuming an expanding $\mathbf{k}$ (cf. Section \ref{Prelims}), we derive the conditions under which it satisfies the optical constraint (cf. \eqref{opt_constraint}) and, analogous to \cite{HD_KS}, draw parallels with the four-dimensional Goldberg-Sachs theorem. We discuss the example of four-dimensional (A)dS-Taub-NUT spacetimes (cf. Section \ref{NUT_4d}) to illustrate the results on the optical constraint as well as those from Section \ref{GKS_with_geo_k} on algebraic properties. Further, for the (A)dS-Taub-NUT example, we make some comments on the KS double copy in curved backgrounds from a GKS viewpoint (cf. Section \ref{double_copy}).
    \item In Section \ref{RT_GKS_section}, we specialize to the case of expanding, twistfree, shearfree $\mathbf{k}$ (cf. Section \ref{Prelims} for the definitions), and as an application of our general results (discussed in the previous sections), we derive the full family of such spacetimes in $n>4$ that satisfy the vacuum Einstein equations \eqref{vacuum_Einstein}. The solutions are given by two branches, described respectively by equations \eqref{RT_GKS_met_1}, \eqref{guu_RT_M_nonzero} and by equations \eqref{RT_GKS_met_1}, \eqref{guu_fn_for_RT_case_2}, and can be identified with a subset of the higher-dimensional vacuum Robinson-Trautman spacetimes \cite{Podolsky_Ort_RT,Ortaggio_RT}.
 We also observe that the solutions described by \eqref{RT_GKS_met_1}, \eqref{guu_RT_M_nonzero} admit a notion of the KS double copy in curved backgrounds, analogous to the (A)dS-Taub-NUT example.
\end{itemize}
In addition to the aforementioned examples, we briefly discuss the GKS structure of the five-dimensional minimal supergravity (SUGRA) solution of \cite{CCLP}, in the context of Weyl types and the optical constraint (cf. Sections \ref{subsection_remarks_alge_types}, \ref{GKS_in_GR_subsection}). In Appendix \ref{Examples_apeendix}, we present further examples of GKS spacetimes with expanding $\mathbf{k}$, including an example beyond general relativity (GR),\footnote{For GKS spacetimes, it is well known that the mixed Ricci tensor $R^{a}_{\hspace{1mm}b}$, and hence the Einstein equation, is linear in $H$ \cite{Grses1975LorentzCT,xanthopoulos1978,XANTHOPOULOS_83,Taub:1981evj, Dereli:1986cm, Stephani:2003tm} (cf. also Appendix \ref{Curvature}). Interestingly, it was shown in \cite{Gurses:2016moi} that, for a subclass of vacuum Kerr-Schild-Kundt spacetimes (recently shown to be equivalent to Kundt metrics of Weyl and traceless Ricci type $N$ in \cite{Ortaggio_Jose_Jakub}), the equations of motion in the most general theory of gravity (built out of curvatures and their derivatives) reduce to equations linear in $H$ (cf. also \cite{Gullu:2011sj, Gurses:2012db, Gurses:2013jua} and \cite{Hervik:2013cla, Kuchynka}).} and use them as additional illustration of our general results. The remaining appendices contain several results that supplement the main text.

\subsection{Notation}
\label{Prelims}

Let us set up the notation that will be used throughout the paper. We will work in the mostly positive signature $(-,+,\dots, +)$. We will use the following indices with ranges as indicated
\begin{align}
    &a,b,\dots=0,1,\dots,n-1,\quad
    i,j,\dots= 2,3,\dots, n-1.
\end{align}
Consistent with the notation $\mathbf{\Bar{g}}$ for the background metric, we will denote all the quantities defined on the background spacetime with bars. The inverse of \eqref{GKS} is given by

\begin{align}
    g^{ab}= \Bar{g}^{ab} + 2H k^a k^b,\label{inverse_GKS}
\end{align}
where $\Bar{g}^{ab}$ is the inverse of the background metric and $k^{a}=\Bar{g}^{ab}k_b=g^{ab}k_b$.\footnote{For quantities such as the KS vector, which are invariant under GKS transformations, the raising and lowering of the tensor indices can be consistently defined with respect to both metrics.
} The lowering and raising of indices $a, b, \dots$ for the unbarred quantities are defined using the GKS metric (and its inverse), whereas those of the barred quantities are defined using the background metric (and its inverse). The indices $i, j, \dots$, on the other hand, are raised and lowered with the spatial part of the metric, $g_{ij}$, and its inverse, $g^{ij}$, which, in the spatial orthonormal basis defined below, become $\delta_{ij}$. A null direction is defined as an equivalence class of null vectors that are related to each other by means of scaling functions. It will be useful to note that the null directions defined by two null vectors, $\mathbf{l}_1$ and $\mathbf{l}_2$, are distinct if and only if (iff) $\mathbf{l}_1 \cdot \mathbf{l}_2 \neq 0$. From now on, by abuse of notation, we will denote both a null vector and the null direction it defines by the same symbol. Moreover, when we say $\mathbf{l}_1 \neq \mathbf{l}_2$, it should be implicitly understood that the null vectors $\mathbf{l}_1$ and $\mathbf{l}_2$ define two distinct null directions. \\

We will work in a null-frame \cite{Coley:2004jv,VPravda_2004} adapted to \eqref{GKS}, denoted by 
\begin{align}
   \{\mathbf{m}_{(0)} = \mathbf{k}, \mathbf{m}_{(1)} = \mathbf{n}, \mathbf{m}_{(i)}\}, \label{frame_GKS}
\end{align}
 where $\mathbf{n}$ is a null vector and $\mathbf{m}_{(i)}$ are $(n-2)$ spacelike vectors satisfying
\begin{align}
    &n^{a}k_a=1, \quad m_{(i)}^a k_{a}=0=m_{(i)}^a n_{a} , \quad
     m_{(i)}^a m^{(j)}_a= \delta_{ij}.
\end{align}
 We will use the same symbol $\mathbf{k}$ to denote both the vector field $\mathbf{k} = k^a \partial_a$ and the covector field $\mathbf{k} = k_a dx^a$. The distinction will not matter in most cases, and whenever it does, it will be clear from the context. To align with the terminology used in  \cite{HD_KS,Malek:2010mh,Ort_Srini, Ort_Pra_Pra_KS_double}, we will refer to $\mathbf{k}$ of the GKS metric \eqref{GKS} as the KS vector. Frame-projected covariant derivatives defined with respect to the GKS metric will be denoted by
 \begin{align}
     D\equiv m^a_{(0)}\nabla_a = k^a \nabla_a, \quad \Delta \equiv m^a_{(1)}\nabla_a =  n^a \nabla_a, \quad \delta_i \equiv  m^{a}_{(i)}\nabla_a. \label{Frame_derivatives_GKS}
 \end{align}
The Ricci rotation coefficients for the GKS metric are defined as 
\begin{align}
    L_{ab}= m^{c}_{(a)}m^{d}_{(b)} k_{c;d}, \quad  N_{ab}= m^{c}_{(a)}m^{d}_{(b)} n_{c;d}, \quad  \overset{i}{M}_{ab}= m^{c}_{(a)}m^{d}_{(b)} m^{(i)}_{c;d}, \label{GKS_Ricci_rot}
\end{align}
and they satisfy the relations 
\begin{align}
  L_{0a}=N_{1a}=N_{0a}+L_{1a}=\M{i}{0}{a} + L_{ia} = \M{i}{1}{a}+N_{ia}=\M{i}{j}{a}+\M{j}{i}{a}=0.  \label{Ricci_rot_identities}
\end{align}
The Ricci rotation coefficient $L_{ij}$, known as the optical matrix, can be used to define the following quantities \cite{VPravda_2004, Ricci_in_HD, HD_KS}
\begin{align}
&S_{ij}\equiv L_{(ij)} = \sigma_{ij} + \theta \delta_{ij}, \quad A_{ij}\equiv L_{[ij]}, \label{optical_decompose}\\
& \theta \equiv \frac{1}{(n-2)} S_{ii}, \quad \sigma^2 \equiv \sigma_{ij}\sigma_{ij}, \quad \omega^2 \equiv A_{ij}A_{ij}. \label{optical_scalars}
\end{align}
The scalars $\theta$, $\sigma^2$, and $\omega^2$ are respectively called the expansion, shear, and twist and collectively form the optical scalars.
\section{Basic features of GKS spacetimes for general $\mathbf{k}$}
\label{gen_props}
\subsection{GKS vs. the background connection}
\label{GKS_vs_Bg_connection}
The GKS metric $\mathbf{g}$ can be expressed in the null frame \eqref{frame_GKS} as
\begin{align}
    \mathbf{g}= \mathbf{k}\otimes \mathbf{n}+  \mathbf{n}\otimes \mathbf{k} + \delta_{ij}\mathbf{m}^{i}\mathbf{m}^{j}. \label{GKS_in_frame}
\end{align}
From \eqref{GKS_in_frame} and \eqref{GKS}, one has the following expression for the background metric
\begin{align}
     \mathbf{\Bar{g}}= \mathbf{k}\otimes \mathbf{\Bar{n}}+  \mathbf{\Bar{n}}\otimes \mathbf{k} + \delta_{ij}\mathbf{m}^{i}\mathbf{m}^{j},\label{bakcground_in_frame}
\end{align}
where 
\begin{align}
    \Bar{n}_a= n_a + H k_a \iff \Bar{n}^a= n^a - H k^a. \label{n_bg}
\end{align}
Therefore, one can define a null frame for the background, analogous to \eqref{frame_GKS},\footnote{The definition \eqref{n_bg} removes the $H$-dependence from $\mathbf{n}$, resulting in an $H$-independent null vector $\mathbf{\Bar{n}}$, consistent with the fact that $\mathbf{\Bar{g}}$ is $H$-independent.} as
\begin{align}
   \{\mathbf{\Bar{m}}_{(0)} = \mathbf{k}, \mathbf{\Bar{m}}_{(1)}= \mathbf{\Bar{n}}, \mathbf{\Bar{m}}_{(i)}=\mathbf{m}_{(i)}\}.\label{frame_background}
\end{align}
This allows us to define the frame-projected covariant derivatives and the Ricci rotation coefficients for the background spacetime as
\begin{align}
     &\Bar{D}\equiv \Bar{m}^a_{(0)}\overline{\nabla}_a = k^a \overline{\nabla}_a, \quad \Bar{\Delta}\equiv \Bar{m}^a_{(1)}\overline{\nabla}_a = \Bar {n}^a \overline{\nabla}_a, \quad \Bar{\delta}_i \equiv \Bar{m}^{a}_{(i)}\overline{\nabla}_a= m^{a}_{(i)}\overline{\nabla}_a, \label{Frame_derivatives_bg}\\
     &  \Bar{L}_{ab}= \Bar{m}^{c}_{(a)}\Bar{m}^{d}_{(b)} k_{c\Bar{;}d}, \quad  \Bar{N}_{ab}= \Bar{m}^{c}_{(a)}\Bar{m}^{d}_{(b)} \Bar{n}_{c\Bar{;}d}, \quad  \overset{i}{\Bar{M}}_{ab}= \Bar{m}^{c}_{(a)}\Bar{m}^{d}_{(b)} \Bar{m}^{(i)}_{c\Bar{;}d}, \label{Ricci_rot_bg}
 \end{align}
where $\Bar{;}$ denotes the covariant derivative with respect to $\mathbf{\Bar{g}}$ and the background Ricci rotation coefficients satisfy identities analogous to \eqref{Ricci_rot_identities}. Following \eqref{Frame_derivatives_GKS}, \eqref{GKS_Ricci_rot}, \eqref{Frame_derivatives_bg}, and \eqref{Ricci_rot_bg}, it will be useful to set up the definitions for the corresponding frame components of a general tensorial quantity $\mathbf{T}_{a_1 \dots a_n}$ defined on the full geometry and its background counterpart $\mathbf{\Bar{T}}_{a_1 \dots a_n}\equiv \mathbf{T}_{a_1 \dots a_n}|_{H=0}$. The definitions are given by
\begin{align}
   &T_{b_1\dots b_n} \equiv  \mathbf{T}_{a_1 \dots a_n} \mathbf{m}^{a_1}_{(b_1)}\dots \mathbf{m}^{a_n}_{(b_n)}, \quad
   \Bar T_{b_1\dots b_n} \equiv  \mathbf{\Bar T}_{a_1 \dots a_n} \mathbf{\Bar m}^{a_1}_{(b_1)}\dots \mathbf{\Bar m}^{a_n}_{(b_n)}.
\end{align}
From the definition of the background frame \eqref{frame_background}, it follows that the Ricci rotation coefficients of the full geometry are related to those of the background as \cite{Ort_Srini}
\begin{align}
  & L_{i0}=\Bar{L}_{i0} , \quad L_{10}=\Bar{L}_{10} , \quad L_{ij}=\Bar{L}_{ij} , \quad \overset{i}{M}_{j0}=\overset{i}{\Bar{M}}_{j0}, \quad \overset{i}{M}_{jk}=\overset{i}{\Bar{M}}_{jk},\label{Li0_GKS} \\
  & N_{i0}=\Bar N_{i0} , \quad L_{i1}=\Bar L_{i1} , \quad L_{1i}=\Bar L_{1i}-H \Bar L_{i0} , \quad N_{ij}=\Bar N_{ij}+H\Bar L_{ji} , \label{Nij_GKS} \\
  &\overset{i}{M}_{j1}=\overset{i}{\Bar M}_{j1}+H\big(\overset{i}{\Bar M}_{j0}+2\Bar L_{[ij]}\big) , \quad  L_{11}=\Bar L_{11}-H\Bar L_{10}-\Bar D H , \label{L11_GKS} \\
  & N_{i1}=\Bar N_{i1}+H \big(\Bar N_{i0}+2\Bar L_{1i}-H \Bar L_{i0}-\Bar L_{i1}\big)+ \Bar \delta_i H .  \label{Ni1_GKS}
\end{align}
Likewise, the frame-projected covariant derivatives satisfy the relations
\begin{align}
Df &= \Bar{D}f , \quad \delta_i f = \Bar{\delta}_i f , \quad \Delta f = \Bar{\Delta}f + H \Bar{D}f, \label{Cov_der_bg_vs_GKS}
\end{align}
for any scalar function $f$. The relations \eqref{Li0_GKS}--\eqref{Ni1_GKS}, together with equations \eqref{Cov_der_bg_vs_GKS}, encode the information relating the covariant derivatives of the full geometry to those of the background. From \eqref{Li0_GKS} and \eqref{Nij_GKS}, we also see that the optical matrix (and hence the optical scalars) and the parallel transport of the frame vectors \eqref{frame_GKS} are unaffected by a GKS transformation \cite{Stephani:2003tm}.
\subsection{Geodesicity of $\mathbf{k}$}
\label{Geode_k}
Using \eqref{GKS_Ricci_rot}, we have   $k^b k_{a;b}= L_{10}k_{a} + L_{i0} m^{(i)}_{a}$, 
and from \eqref{Ricci_rot_bg}, a similar relation holds true with respect to the background. Hence, by definition, $\mathbf{k}$ is geodesic in the full (background) geomtery iff $L_{i0} \hspace{1mm} (\Bar{L}_{i0})$ vanishes.
From the first of \eqref{Li0_GKS}, we have $L_{i0} = \Bar{L}_{i0}$. Therefore, $\mathbf{k}$ is geodesic in the background geometry $\mathbf{\Bar g}$ iff it is geodesic in the full geometry $\mathbf{g}$. As with $\mathbf{k}$, we will hereafter omit bars for quantities that are the same in both the background and the full geometry.

The relations \eqref{Li0_GKS}--\eqref{Ni1_GKS} and \eqref{Cov_der_bg_vs_GKS}, in combination with the Ricci identities $(11a)-(11p)$ of \cite{Ricci_in_HD}, allow us to relate the frame components of the Riemann tensor of the full spacetime with that of the background. The final results will be presented directly in Appendix \ref{Curvature}. Here, we outline the steps for relating the component $R_{0i0j}$ with $\Bar{R}_{0i0j}$, which will be used to deduce some results on the geodesicity of $\mathbf{k}$. Consider equation $(11g)$ from \cite{Ricci_in_HD} calculated for the full geometry \eqref{GKS} in its null frame \eqref{frame_GKS}
\begin{align}
    R_{0i0j}  =&- DL_{ij}+\delta_{j}L_{i0}+L_{10}L_{ij} - L_{i0}(2L_{1j}+ N_{j0}) - L_{i1}L_{j0}+ 2L_{k[0|}\overset{k}{M}_{i|j]}\nonumber\\
    &- L_{ik}(L_{kj}+ \overset{k}{M}_{j0}).\label{Ricci_(11)g_1}
\end{align}
Using \eqref{Nij_GKS}, we rewrite \eqref{Ricci_(11)g_1} as
\begin{align}
    R_{0i0j}  =&- DL_{ij}+\delta_{j}L_{i0}+L_{10}L_{ij} - L_{i0}( 2\Bar L_{1j}- 2H L_{i0}+ N_{j0}) - L_{i1}L_{j0}+ 2L_{k[0|}\overset{k}{M}_{i|j]}\nonumber\\
    &- L_{ik}(L_{kj}+ \overset{k}{M}_{j0}). \label{Ricci_11g_2}
\end{align}
Further using the expression analogous to \eqref{Ricci_(11)g_1} for the background geometry, we find
\begin{align}
       R_{0i0j} = \Bar R_{0i0j} + 2 H L_{i0}L_{j0}. \label{Riemann0i0j}
\end{align}
Contracting \eqref{Riemann0i0j} with $\delta_{ij}$, we have
\begin{align}
    R_{00}= \Bar R_{00} + 2H L_{i0}L_{i0}.\label{Ricci_00}
\end{align}
From \eqref{Riemann0i0j} and \eqref{Ricci_00}, we can state the following propositions
\begin{proposition}\label{prop_geod}
    Let $\mathbf{k}$ be the KS vector of a GKS spacetime. Then the following are equivalent
    \begin{enumerate}
        \item $\mathbf{k}$ is geodesic.
        \item $  (R_{ab}-\Bar R_{ab})k^a k^b=0$.
        \item $  k_{[e}  R_{a]bc[d}k_{f]}k^b k^c = k_{[e} \Bar R_{a]bc[d}k_{f]}k^b k^c$.\footnote{Note that the condition given in statement $3$ of Proposition \ref{prop_geod} is a basis independent way of expressing $R_{0i0j}=\Bar{R}_{0i0j}$, akin to the Bel-Debever criteria \cite{Milson:2004jx,Ort_Bel_Deb}. }
    \end{enumerate}
\end{proposition}
From Proposition \ref{prop_geod}, we can make the following remark
\begin{remark}\label{geod_remark2}  
     When the background $\mathbf{\Bar{g}}$ is a spacetime of constant curvature, it follows that $\Bar{R}_{00}=0=\Bar R_{0i0j}$. Consequently, Proposition \ref{prop_geod} reduces to the results on the geodesicity of $\mathbf{k}$ in KS spacetimes \cite{HD_KS,Malek:2010mh}.
\end{remark}
 \begin{proposition}
 \label{Riemann_AND_full}
   If $\mathbf{k}$ is a Riemann AND of the full geometry, it follows that $\mathbf{k}$ is geodesic and also a Riemann AND of the background.
 \end{proposition}
 \noindent \textbf{Proof.} When $\mathbf{k}$ is a Riemann AND of the full geometry, then from \eqref{Riemann0i0j} we have
 \begin{align}
   0 = \Bar R_{0i0j} + 2 H L_{i0}L_{j0}. \label{R_0i0j_for_proof} 
\end{align}
By definition, the background quantities are independent of $H$, so for \eqref{R_0i0j_for_proof} to be consistent for all values of parameters ``$\mu_\alpha$'' in $H$, the $H$-dependent and $H$-independent terms must vanish independently. This leads to $\Bar R_{0i0j}=0=L_{i0}$, thus concluding the proof.\\

By applying the same line of arguments as in Proposition \ref{Riemann_AND_full} to equation \eqref{Ricci_00}, we obtain the following result
 \begin{proposition}
 \label{Ricci_AND_full}
    If $\mathbf{k}$ is a Ricci AND of the full geometry, it follows that $\mathbf{k}$ is geodesic and also a Ricci AND of the background.
 \end{proposition}
\begin{remark}
    Both Propositions \ref{Riemann_AND_full} and \ref{Ricci_AND_full} form sufficient conditions for $\mathbf{k}$ to be geodesic. However, let us note that Proposition \ref{Riemann_AND_full} is stronger, as $\mathbf{k}$ being a Riemann AND of the full geometry automatically implies that it is also a Ricci AND of the full (as well as the background) geometry. In that case, $\mathbf{k}$ also becomes a WAND of both the background and the full geometry.
\end{remark}
   When the full geometry satisfies the Einstein equations
   \begin{align}
    R_{ab} -\frac{1}{2}g_{ab}R+ \Lambda g_{ab}= \kappa T_{ab}, \label{Einstein_eqns}
\end{align}
  using arguments similar to those of Proposition \ref{Riemann_AND_full}, one finds that the background geometry must also satisfy the Einstein equations, with its energy-momentum tensor given by $\Bar{T}_{ab}=T_{ab}|_{H=0}$. Therefore, using Proposition \ref{prop_geod}, we can make the following remark about the geodesicity of $\mathbf{k}$
   \begin{remark}\label{geod_remark3} 
    If the GKS spacetime satisfies the Einstein equations \eqref{Einstein_eqns}, then $\mathbf{k}$ is geodesic iff $(T_{00} - \Bar{T}_{00}) = 0$. One can retrieve the results associated with KS spacetimes by setting $\Bar{T}_{00}=0$; explicitly, $\mathbf{k}$ is geodesic iff $T_{00} = 0$ \cite{HD_KS,Malek:2010mh}. 
\end{remark} 
We also note from Proposition \ref{prop_geod} or \ref{Ricci_AND_full} that $\mathbf{k}$ of a GKS spacetime is geodesic when it satisfies the vacuum Einstein equations\footnote{As can be seen from \eqref{vacuum_Einstein}, we actually mean $\Lambda$-vacuum; however, for brevity, we will keep referring to it as vacuum throughout the paper.} 
    \begin{align}
R_{ab} = \frac{2\Lambda}{n-2}g_{ab}, \label{vacuum_Einstein}
\end{align}
or when the matter fields are aligned with $\mathbf{k}$, for example, $F_{ab}k^b \sim k_a$. This observation is similar to those made for KS spacetimes in \cite{HD_KS,Malek:2010mh}.
\section{GKS spacetimes with geodesic $\mathbf{k}$}
\label{GKS_with_geo_k}
In the rest of the paper, we will assume the simplifying condition that the KS vector $\mathbf{k}$ is geodesic ($L_{i0}=0$) and, without loss of generality, that the geodesic $\mathbf{k}$ is affinely parametrized (i.e., $L_{10}=0$), with $r$ being an affine parameter. This applies to all examples discussed in this paper. This section will be devoted to discussing the algebraic types of the curvature tensors of GKS spacetimes.\footnote{See \cite{Kuchynka} for a similar discussion on GKS-transformed Einstein-Kundt metrics.} 

\subsection{Alignment properties of the Ricci tensor}\label{Ricci_types}
The frame components of the Ricci tensor are listed in Appendix \ref{Curvature}. Let us use them to deduce some results concerning the Ricci types of the GKS spacetimes. From the relations given in \eqref{R00_GKS}, we can state the following two results
\begin{proposition}\label{prop_Ricci_AND_k}
    The KS direction $\mathbf{k}$ is a single AND of $\Bar R_{ab}$ iff it is a single AND of $ R_{ab}$.
\end{proposition}
\begin{proposition}\label{prop_Ricci_double_AND_k}
      The KS direction $\mathbf{k}$ is a multiple AND of $\Bar R_{ab}$ iff it is a multiple AND of $ R_{ab}$.
\end{proposition}
\begin{remark}
    In Proposition \ref{prop_Ricci_double_AND_k}, it should be noted that the multiplicity of $\mathbf{k}$ with respect to $\Bar R_{ab}$ and $ R_{ab}$ need not be the same. Proposition \ref{prop_k_multiple} discusses the constraint on the multiplicities of $\mathbf{k}$ with respect to the two geometries.
\end{remark}
 In passing we note that, according to Proposition 4.10 of \cite{Hervik:2012jn}, the only possible Ricci types for any spacetime are $G$, $I_i$, $II$, $II_i$, $D$, $III$, $III_i$, $N$, and $O$ (Ricci-flat) \cite{HD_alg_review}; i.e., Ricci type $I$ is forbidden.
\begin{proposition}\label{prop_l_AND}
Let $\mathbf{l} \neq \mathbf{k}$ be an AND of $R_{ab}$. Then, the null vector $\mathbf{\Bar{l}}$, defined by $\Bar{l}^a = l^a - H k^a$, forms an AND of $\Bar{R}_{ab}$.
\end{proposition}
\noindent \textbf{Proof.} Since $\mathbf{l} \neq \mathbf{k}$, after appropriately normalizing $\mathbf{l}$, we can have $\mathbf{l} \cdot \mathbf{k} = 1$. Therefore, let us choose the second null vector $\mathbf{n}$ of the null frame \eqref{frame_GKS} to be $\mathbf{l}$, which means choosing $\mathbf{\Bar{n}}$ to be $\mathbf{\Bar{l}}$ as per \eqref{n_bg}. By assumption, $\mathbf{l}$ is an AND of $R_{ab}$. Thus, using \eqref{R11_GKS_1}, we have
\begin{align}
    0=&\Bar R_{11}+H^{2}\Bar R_{00}   +\delta_{i}\delta_{i}H +(4L_{1i}-2L_{i1}+\overset{i}{M}_{kk})\delta_{i}H+\Bar N_{ii}DH-S_{ii}\Bar \Delta H \nonumber \\
 & +2H\Big(\delta_{i}L_{1i}- \Bar \Delta S_{ii}+4L_{1i}L_{[1i]}-L_{ki} \Bar N_{ki}+L_{1k}\overset{k}{M}_{ii}-2S_{ik}\overset{k}{\Bar M}_{i1}\Big).\label{R_11_for_proof}
\end{align}
As in Propositions \ref{Riemann_AND_full} and \ref{Ricci_AND_full}, for \eqref{R_11_for_proof} to be consistent for all values of parameters ``$\mu_\alpha$'' in $H$, the $H$-dependent and $H$-independent terms must vanish separately. In particular, we have $\Bar R_{11}=0$, which proves the claim of the proposition.
\begin{remark}
    Let us note that the converse of the above proposition is not true in general, i.e., given a Ricci AND $\mathbf{\Bar l}$ of the background, distinct from $\mathbf{k}$, one cannot conclude just from a kinematical analysis that $\mathbf{l}$ must be an AND of $R_{ab}$.
\end{remark}
\begin{proposition}\label{prop_k_multiple}
   Let $\mathbf{k}$ be an AND of $R_{ab}$ and hence also of $\Bar{R}_{ab}$, with respective multiplicities $\tau$ and $\Bar{\tau}$. Then, $\Bar{\tau} \geq \tau$.
\end{proposition}
 \noindent \textbf{Proof.} The multiplicities of Ricci ANDs in general take values in the set $\{1,\dots,4\}$ \cite{HD_alg_review}. We will look at different values of $\tau$ case by case and show that it cannot exceed $\Bar{\tau}$.
  \begin{itemize}
      \item \textbf{Case 1}: $\tau$=1\\
      \noindent  The case is defined by $R_{00}=0\neq R_{0i}$. From Proposition \ref{prop_Ricci_AND_k}, we see that $\mathbf{k}$ must be a single AND of $\Bar{R}_{ab}$, i.e., $\Bar \tau=1$. Therefore, in this case, $\tau=1=\Bar \tau$. 
      \item \textbf{Case 2} : $\tau$=2 \\
      \noindent  In this case $R_{00}=0 =R_{0i}$ and at least some of the boost weight (b.w.) $0$ components of $R_{ab}$ must be non-vanishing.\footnote{See \cite{HD_alg_review} for the definition of boost weight.} From Proposition \ref{prop_Ricci_double_AND_k}, we see that $\mathbf{k}$ must at least be a double AND of $\Bar{R}_{ab}$. Moreover, since $\Bar{R}_{ab}$ is $H$-independent, one could in principle have any (or all) of the remaining background Ricci components to be vanishing and still retain all of the remaining $R_{ab}$ components \eqref{Rij_GKS}-\eqref{R11_GKS_1} as non-zero due to the $H$-dependent terms present in each of these. Therefore, from a kinematical perspective, we can only conclude that $\Bar{R}_{ab}$ can range from type $II$ up to type $O$. Hence, $\Bar{\tau} \geq 2 = \tau$.
 \item \textbf{Case 3}: $\tau=3$\\
 All the non-negative b.w. components of $R_{ab}$ vanish in this case. In particular, the relations \eqref{Rij_GKS} and \eqref{R01_GKS} for b.w. $0$ components give us the conditions
\begin{align}
0 =&\Bar R_{ij} - 2H \Bar R_{0i0j} + 2HL_{ik}L_{jk} - 2S_{ij}\left[DH + (n-2)\theta H\right],\label{Rij_bg_proof}\\
0 =& \Bar R_{01} - H \Bar R_{00} - (D^2H + (n-2)\theta DH + 2H\omega^2).\label{R01_bg_proof}
\end{align}

As in the case of Propositions \ref{Riemann_AND_full}, \ref{Ricci_AND_full}, and \ref{prop_l_AND}, relations \eqref{Rij_bg_proof} and \eqref{R01_bg_proof} can be consistent for all values of the parameters ``$\mu_\alpha$'' in $H$ only if $\Bar{R}_{ij} = 0 = \Bar{R}_{01}$. This, in combination with the results from previous cases, ensures that all the non-negative b.w. components of $\Bar{R}_{ab}$ vanish. In addition, as in case $2$, some or all of the remaining components of $\Bar R_{ab}$ could also vanish. We therefore have $\Bar \tau \geq 3 = \tau$.

 \item \textbf{Case 4} : $\tau =4$\\
Except for $R_{11}$, all the other components of $R_{ab}$ vanish. Therefore, upon direct application of the conclusion of case $3$, all the non-negative b.w. components of $\Bar{R}_{ab}$ must necessarily be zero. Moreover, relation \eqref{R1i_GKS} with $R_{1i}=0$ leads to the vanishing of its $H$-independent part, i.e., $\Bar R_{1i}=0$.  Hence, we have $\Bar{\tau} = 4 = \tau$.
  \end{itemize}
\begin{proposition}\label{prop_multiple_l_AND}
    Let $\mathbf{l}\neq k$ be an AND of $R_{ab}$ and hence $\mathbf{\Bar{l}}$ be that of $\Bar{R}_{ab}$, with respective multiplicities $\chi$ and $\bar \chi$. Then, $\Bar \chi \geq \chi$.
\end{proposition}
 \noindent \textbf{Proof.} As in Proposition \ref{prop_l_AND}, we choose the null vector $\mathbf{n}$ of the null frame \eqref{frame_GKS} to be $\mathbf{l}$, and hence $\mathbf{\Bar{n}} = \mathbf{\Bar{l}}$. The proof then follows exactly the same line of arguments as that of Propositions \ref{prop_l_AND} and \ref{prop_k_multiple}, i.e., the vanishing of the LHS of any of the equations \eqref{R00_GKS}–\eqref{R11_GKS_1} implies that the $H$-dependent and $H$-independent terms on their RHS vanish separately. This, in particular, means that if any component of $R_{ab}$ in the frame given by \eqref{frame_GKS} vanishes, then the corresponding component of $\Bar{R}_{ab}$ defined with respect to the background frame \eqref{frame_background} must also vanish. However, the converse is not true in general. Hence, $\Bar \chi \geq \chi$.
\subsection{Alignment properties of the Weyl tensor}\label{Weyl_types}
Using equations \eqref{C0i0j}-\eqref{C1i1j}, which relate the frame components of the Weyl tensor of the full geometry to those of the background, we can obtain results on Weyl types that are analogous to those for Ricci types. Moreover, since the proofs of these results are exactly similar to those presented in the previous subsection, we will not repeat the details here. However, as the Weyl tensor vanishes identically for $n=3$, the results of this subsection (and Table \ref{table_Weyl_types}) apply only to $n \geq 4$. From \eqref{C0i0j}, we can state the following two propositions

\begin{proposition}\label{prop_WAND_k}
    The KS direction $\mathbf{k}$ is a single WAND of the background iff it is a single WAND of the full geometry.
\end{proposition}
\begin{proposition}\label{prop_double_WAND_k}
      The KS direction $\mathbf{k}$ is an mWAND of the background iff it is an mWAND of the full geometry.
\end{proposition}
\begin{remark}
   As in Proposition \ref{prop_Ricci_double_AND_k}, the multiplicity of the mWAND $\mathbf{k}$ can be different for the two geometries, subject to the constraint given by Proposition \ref{prop_k_multiplicity_WAND}. 
\end{remark}
Using equation \eqref{C1i1j}, we obtain the following Weyl analogue of Proposition \ref{prop_l_AND}
\begin{proposition}\label{prop_l_WAND}
Let $\mathbf{l} \neq \mathbf{k}$ be a WAND of the full geometry. Then, the null vector $\mathbf{\Bar{l}}$, defined by $\Bar{l}^a = l^a - H k^a$, forms a WAND of the background.
\end{proposition}
\begin{remark}
    As in the case of the Ricci tensor, the converse of the above proposition does not hold in general. Hence, given a WAND $\mathbf{\Bar l}$ of the background geometry, distinct from $\mathbf{k}$, one cannot conclude that $\mathbf{l}$ forms a WAND of the full geometry.
\end{remark}
\noindent Finally, similar to Propositions \ref{prop_k_multiple} and \ref{prop_multiple_l_AND}, we have the following two results concerning the multiplicities of the WANDs
\begin{proposition}\label{prop_k_multiplicity_WAND}
   Let $\mathbf{k}$ be a WAND of the full geometry, and hence also of the background, with respective multiplicities $\tau$ and $\Bar{\tau}$. Then, $\Bar{\tau} \geq \tau$.
\end{proposition} 
\begin{proposition}\label{prop_multiple_l_WAND}
    Let $\mathbf{l}\neq k$ be a WAND of the full geometry and hence $\mathbf{\Bar{l}}$ be that of the background, with respective multiplicities $\chi$ and $\bar \chi$. Then, $\Bar \chi \geq \chi$.
    \end{proposition}

\subsection{All kinematically allowed Ricci and Weyl types }\label{subsection_remarks_alge_types}
From the results of Section \ref{Ricci_types}, we formulate Table \ref{table_Ricci_types}, which describes the kinematically allowed combinations of Ricci types (and the associated ANDs) for the background and the full geometry. Similarly, based on the results of Section \ref{Weyl_types}, we form an analogous list for the Weyl types in Table  \ref{table_Weyl_types}.\footnote{Similar to the Ricci and Weyl tensors, one can produce analogous results for the Riemann tensor.} Since our results are independent of dynamics, any GKS spacetime in any theory must conform to one of the combinations in each table, provided $\mathbf{k}$ is geodesic. Nevertheless, it is to be noted that the possible cases for the ANDs\footnote{Since a large part of this discussion applies to both Ricci ANDs and WANDs, unless necessary, we will simply refer to them commonly as ANDs.} are mentioned in a broader sense, and one can further refine or elaborate upon them. For example, based on the classification scheme of \cite{Coley:2004jv,Milson:2004jx,HD_alg_review}, type $I_i$ tensors are defined by having at least two single ANDs while having no multiple ANDs. Therefore, while discussing the type $I_i$ case in the tables, we list only two single ANDs, with the possibility of more than two single ANDs implicitly included. Likewise, type $II_i$ is defined by exactly one double AND and at least one single AND, and even in this case, we discuss only the bare minimum as required by the definition, while keeping open the possibility of more than one single AND. The same applies to the case of $III_i$, where one may, in principle, also have multiple single ANDs.

For type $O$ backgrounds, by virtue of Proposition \ref{prop_Ricci_double_AND_k} for the case of the Ricci tensor and by Proposition \ref{prop_double_WAND_k} for the Weyl tensor, $\mathbf{k}$ is at least a double AND. In accordance with this, we list the possible types of the full geometry as $II$, $II_i$, $D$, $III$, $III_i$, $N$, and $O$. Among these, when the full geometry is of type $D$ we mention only the double AND $\mathbf{k}$, with the implicit understanding that it has (at least) one more double AND by definition. Similarly, in many instances (in both the tables) where the full geometry is of type $I_i$, $II_i$, or $III_i$, with $\mathbf{k}$ being respectively a single, double, or triple AND, the focus is on the AND $\mathbf{k}$, with the existence of (other) single AND(s) implicitly understood.

Another point to be noted in relation to the tables is that we have used the results (and remarks) from Appendix \ref{gen_result_type_D_tensors} (cf. also the references therein) for the type $D$ cases. In particular, from Proposition \ref{Prop_rank_2_type_D} and Remark \ref{remark_single_RAND}, we see that in $n>3$, type $D$ Ricci tensors may also admit single ANDs, in addition to the double ANDs. A similar observation holds for the Weyl tensor in $n>4$ (cf. Remark \ref{remark_single_WAND_typeD_4d} and references therein). Based on these, we explicitly list the possibility of single ANDs for the type $D$ cases wherever necessary, while in other places such possibilities are implicit.

\newpage
\begin{table}[H]
    \centering
   \tiny
   \resizebox{0.8\textwidth}{!}{
    \begin{tabular}{|>{\centering\arraybackslash}p{3cm}|>{\centering\arraybackslash}p{5.5cm}|>{\centering\arraybackslash}p{3cm}|>{\centering\arraybackslash}p{5.5cm}|} \hline 
       \vspace{0.2mm}  Background Ricci type&  \vspace{0.2mm} Possible cases for the AND(s) of $\Bar{R}_{ab}$& \vspace{0.2mm} Ricci type of the full geometry&  \vspace{0.2mm}  Possible cases for the AND(s) of $R_{ab}$\\ \hline \hline

        \vspace{0.2mm}  $G$& \vspace{0.2mm} no ANDs & \vspace{0.2mm} $G$ & \vspace{0.2mm}no ANDs   \\ \hline\hline 
           \multirow{3}{*}{ $I_i$} &\vspace{0.2mm}  \multirow{2}{*}{$\mathbf{k}\neq$ AND; $\mathbf{\Bar l}_1, \mathbf{\Bar l}_2=$ single ANDs} &\vspace{0.2mm} $G$ & \vspace{0.2mm} no ANDs   \\  \cline{3-4}
            &\vspace{0.2mm}  & \vspace{0.2mm} $I_i$ & \vspace{0.2mm} $\mathbf{ l}_1, \mathbf{l}_2=$ single ANDs\\
  \cline{2-4}
  &\vspace{0.2mm} $\mathbf{k}, \mathbf{\Bar l}=$ single ANDs& \vspace{0.2mm} $I_i$ & \vspace{0.2mm} $\mathbf{k},\mathbf{l}=$ single ANDs   \\ \hline \hline
       \multirow{3}{*}{ $II$} & \vspace{0.2mm} \multirow{3}{*}{ $ \mathbf{k}\neq$ AND; $\mathbf{\Bar l}=$ double AND; no other ANDs} &\vspace{0.2mm} $G$ & \vspace{0.2mm} no ANDs   \\ 
  \cline{3-4}
    & \vspace{0.2mm}&\vspace{0.2mm}  $II$ & \vspace{0.2mm}  $\mathbf{ l}=$ double AND; no other ANDs\\ \cline{2-4}
  & \vspace{0.2mm} $\mathbf{k}=$ double AND; no other ANDs &\vspace{0.2mm}  $II$ & \vspace{0.2mm} $\mathbf{k}=$ double AND; no other ANDs  \\ \hline \hline
   \multirow{8}{*}{ $II_i$ } &  \vspace{0.2mm} \multirow{4}{*}{ $\mathbf{k}\neq$ AND; $\mathbf{\Bar l}_1=$ double AND; $\mathbf{\Bar l}_2=$ single AND }  &  \vspace{0.2mm}$G$ &  \vspace{0.2mm}no ANDs   \\ 
  \cline{3-4}
  &  & \vspace{0.2mm} $I_i$ &  \vspace{0.2mm} $\mathbf{l}_2,\mathbf{l}_3=$ single ANDs; \mbox{$\mathbf{\Bar l}_3=$ single AND of background;} \mbox{$\mathbf{k}, \mathbf{l}_1\neq$ AND}   \\ \cline{3-4}
  &  & \vspace{0.2mm}$II$ &  \vspace{0.2mm} $\mathbf{l}_1=$ double AND; no other ANDs  
      \\ \cline{3-4}
      &  & \vspace{0.2mm}$I_i/II_i$ &  \vspace{0.2mm} $\mathbf{l}_1=$ single/double AND; $\mathbf{l}_2=$single AND;  \mbox{ $\mathbf{k}\neq$ AND}  \\ \cline{2-4}
  &  \vspace{0.2mm}\multirow{2}{*}{$ \mathbf{k}=$ single AND; $\mathbf{\Bar l}=$ double AND } & \vspace{0.2mm}$I_i$ &  \vspace{0.2mm} $\mathbf{k}=$ single AND; no double AND   \\ \cline{3-4}
   &  & \vspace{0.2mm}$II_i$ & \vspace{0.2mm} $\mathbf{k}=$ single AND; $\mathbf{ l}=$ double AND  \\ \cline{2-4}
   &\vspace{0.2mm} \multirow{2}{*}{ $\mathbf{k}=$ double AND; $\mathbf{\Bar l}=$ single AND }  & \vspace{0.2mm}$II$ & \vspace{0.2mm} $\mathbf{k}=$ double AND; no other ANDs   \\ \cline{3-4}
    & & \vspace{0.2mm}$II_i$ & \vspace{0.2mm} $\mathbf{k}=$ double AND; $\mathbf{ l}=$ single AND  \\ \hline \hline
     \multirow{17}{*}{ $D$ } &  \vspace{0.2mm}  \multirow{9}{*}{ $\mathbf{k}\neq$ AND; $\mathbf{\Bar l}_1, \mathbf{\Bar l}_2=$ double ANDs } &  \vspace{0.2mm}$G$ &  \vspace{0.2mm}no ANDs  \\ 
  \cline{3-4}
   &  \vspace{0.2mm}  &  \vspace{0.2mm}\multirow{4}{*}{ $I_i$ } &  \vspace{0.2mm} $\mathbf{ l}_1, \mathbf{ l}_2=$ single ANDs; $\mathbf{k} \neq$ AND \\ 
  \cline{4-4}
  &  \vspace{0.2mm}  &  \vspace{0.2mm}  &  \vspace{0.2mm}  $\mathbf{ l}_1, \mathbf{ l}_3=$ single ANDs; \mbox{$\mathbf{ \Bar l}_3=$ single AND of background;} $\mathbf{k},  \mathbf{ l}_2  \neq$ AND  \\  \cline{4-4}
  &  \vspace{0.2mm}  &  \vspace{0.2mm}  &  \vspace{0.2mm} $\mathbf{ l}_3, \mathbf{ l}_4=$ single ANDs; \mbox{$\mathbf{ \Bar l}_3, \mathbf{ \Bar l}_4=$ single ANDs of background;} \mbox{$\mathbf{k}, \mathbf{ l}_1,  \mathbf{ l}_2  \neq$ AND}  \\  
    \cline{3-4}
    &  \vspace{0.2mm} &  \vspace{0.2mm}  $II$ &  \vspace{0.2mm} $\mathbf{l}_1=$ double AND; no other ANDs \\ 
  \cline{3-4}
   &  \vspace{0.2mm}  &  \vspace{0.2mm}  \multirow{2}{*}{ $II_i$ }&  \vspace{0.2mm} $\mathbf{l}_1=$ double AND; $\mathbf{l}_2=$ single AND; $\mathbf{k}\neq$ AND \\ 
  \cline{4-4}
   &  \vspace{0.2mm}  &  \vspace{0.2mm} &  \vspace{0.2mm} $\mathbf{l}_1=$ double AND; $\mathbf{l}_3=$ single AND;  \mbox{$\mathbf{\Bar l}_3=$ single AND of background;} $\mathbf{k}, \mathbf{l}_2\neq$ AND \\  \cline{3-4}
   &  \vspace{0.2mm}  &  \vspace{0.2mm} $D$ &  \vspace{0.2mm} $\mathbf{l}_1, \mathbf{l}_2=$ double ANDs; $\mathbf{k}\neq$ AND \\ 
  \cline{2-4}
  & \multirow{3}{*}{ \vspace{0.2mm}$\mathbf{k}=$ single AND; $\mathbf{\Bar l}_1,\mathbf{\Bar l}_2=$ double ANDs }  & \vspace{0.2mm}$I_i$ & \vspace{0.2mm} $\mathbf{k}=$ single AND; no double AND  \\ \cline{3-4}
  &  & \vspace{0.2mm}$II_i$ & \vspace{0.2mm} $\mathbf{k}=$ single AND;  $\mathbf{ l}_1=$ the only double AND   \\ \cline{3-4} 
      &  & \vspace{0.2mm}$D$ & \vspace{0.2mm} $\mathbf{k}=$ single AND;  $\mathbf{ l}_1,\mathbf{ l}_2=$ double ANDs   \\ \cline{2-4} 
     & \vspace{0.2mm} \multirow{5}{*}{ $\mathbf{k},\mathbf{\Bar l}=$ double ANDs } & \vspace{0.2mm}$II$ & \vspace{0.2mm} $\mathbf{k}=$ double AND; no other ANDs   \\ \cline{3-4}
     &  & \vspace{0.2mm}\multirow{3}{*}{ $II_i$ } & \vspace{0.2mm} $\mathbf{k}=$ the only double AND; $\mathbf{ l}=$ single AND    \\ \cline{4-4} 
      &  &  & \vspace{0.2mm} $\mathbf{k}=$ the only double AND; $\mathbf{ l}'=$ single AND;  $\mathbf{\Bar l}'=$ single AND of background; $\mathbf{l}\neq$ AND   \\  
      \cline{3-4}
    &  & \vspace{0.2mm}$D$ & \vspace{0.2mm}  $\mathbf{k},\mathbf{ l}=$ double ANDs
    \\ \hline \hline
     \multirow{3}{*}{ $III$ } &  \vspace{0.2mm}\multirow{2}{*}{$ \mathbf{k}\neq$ AND; $\mathbf{\Bar l}=$ triple AND } &  \vspace{0.2mm}$G$ &  \vspace{0.2mm}no ANDs  \\ 
  \cline{3-4}
    & \vspace{0.2mm} & \vspace{0.2mm} $II/III$ &  \vspace{0.2mm} $\mathbf{l}=$ double/triple AND; \mbox{no other ANDs} \\  \cline{2-4}
  & \vspace{0.2mm}$ \mathbf{k}=$ triple AND; no other ANDs  & \vspace{0.2mm}$II/III$ &  \vspace{0.2mm} $\mathbf{k}=$ double/triple AND; no other ANDs      \\ \hline \hline
    \multirow{8}{*}{ $III_i$ } &  \vspace{0.2mm}  \multirow{4}{*}{ $ \mathbf{k}\neq$ AND; $\mathbf{\Bar l}_1=$ triple AND; $\mathbf{\Bar l}_2=$ single AND } &  \vspace{0.2mm}$G$ &  \vspace{0.2mm}no ANDs   \\ 
  \cline{3-4}
  &   & \vspace{0.2mm} $I_i$ &  \vspace{0.2mm} $\mathbf{l}_2,\mathbf{l}_3=$ single ANDs; \mbox{$\mathbf{\Bar l}_3=$single AND of background;} $\mathbf{k}, \mathbf{l}_1\neq$ AND    \\ \cline{3-4} 
   &   & \vspace{0.2mm} $II/III$ &  \vspace{0.2mm} $\mathbf{l}_1=$ double/triple AND; no other ANDs   \\ \cline{3-4} 
    &   & \vspace{0.2mm} $I_i/II_i/III_i$ &  \vspace{0.2mm} $\mathbf{l}_1=$ single/double/triple AND; \mbox{$\mathbf{l}_2=$ single AND;} $\mathbf{k}\neq$ AND   \\ \cline{2-4} 
  &  \multirow{2}{*}{ \vspace{0.2mm}$ \mathbf{k}=$ single AND; $\mathbf{\Bar l}=$ triple AND }  & \vspace{0.2mm}$I_i$ &  \vspace{0.2mm} $\mathbf{k}=$ single AND; no multiple ANDs   \\ \cline{3-4} 
  &  & \vspace{0.2mm}$II_i/III_i$ &  \vspace{0.2mm} $\mathbf{k}=$ single AND;  $\mathbf{ l}=$ double/triple AND   \\ \cline{2-4}
   &  \multirow{2}{*}{  \vspace{0.2mm}$ \mathbf{k}=$ triple AND; $\mathbf{\Bar l}=$ single AND } & \vspace{0.2mm}$II/III$ &  \vspace{0.2mm} $\mathbf{k}=$ double/triple AND; no other ANDs   \\ \cline{3-4} 
    &   & \vspace{0.2mm}$II_i/III_i$ &  \vspace{0.2mm} $\mathbf{k}=$ double/triple AND;  $\mathbf{ l}=$ single AND   \\  \hline \hline
    \multirow{3}{*}{ $N$ } &  \vspace{0.2mm}\multirow{3}{*}{ $ \mathbf{k}\neq$ AND; $\mathbf{\Bar l}=$ quadruple AND } &  \vspace{0.2mm}$G$ &  \vspace{0.2mm}no ANDs   \\ 
  \cline{3-4}
    &   & \vspace{0.2mm}$II/III/N$ &  \vspace{0.2mm} $\mathbf{l}=$ double/triple/quadruple AND; \mbox{no  other ANDs}   \\ 
  \cline{2-4}
  &   \vspace{0.2mm}$ \mathbf{k}=$ quadruple AND  & \vspace{0.2mm}$II/III/N$ &  \vspace{0.2mm} $ \mathbf{k}=$ double/triple/quadruple AND; \mbox{no  other ANDs}  \\ 
  \hline \hline
   \multirow{4}{*}{ $O$ } &  \multirow{4}{*}{   \vspace{0.2mm} }  &  \vspace{0.2mm}$II, II_i, D$ &  \vspace{0.2mm}$ \mathbf{k}=$ double AND   \\ 
  \cline{3-4}
& &  \vspace{0.2mm}$III, III_i$ &  \vspace{0.2mm}$ \mathbf{k}=$ triple AND \\ 
  \cline{3-4}
 & &  \vspace{0.2mm}$N$ &  \vspace{0.2mm}$ \mathbf{k}=$ quadruple AND  \\ 
  \cline{3-4}
  & &  \vspace{0.2mm}$O$ &  \vspace{0.2mm}    \\  \hline
    \end{tabular}}
     \caption{\footnotesize All kinematically allowed combinations of Ricci types for the background and the full geometry. A null direction distinct from $\mathbf{k}$, say $\mathbf{l}$, and its barred counterpart are defined to be related analogously to \eqref{n_bg}. Note that the cases where type $D$ Ricci tensor admits a single AND are forbidden in $n=3$ (cf. Proposition \ref{Prop_rank_2_type_D} and Remark \ref{remark_single_RAND}).}\label{table_Ricci_types}
\end{table}

\newpage
\begin{table}[H]
    \centering
   \tiny
   \resizebox{0.8\textwidth}{!}{
    \begin{tabular}{|>{\centering\arraybackslash}p{2.6cm}|>{\centering\arraybackslash}p{6.2cm}|>{\centering\arraybackslash}p{2.5cm}|>{\centering\arraybackslash}p{7.5cm}|} \hline 
       \vspace{0.2mm}  Background Weyl type&  \vspace{0.2mm} Possible cases of the background WAND(s)& \vspace{0.2mm} Weyl type of the full geometry&  \vspace{0.2mm}  Possible cases of WAND(s) for the full geometry\\ \hline \hline

         \vspace{0.2mm} $G$& \vspace{0.2mm} no WANDs  & \vspace{0.2mm} $G$ & \vspace{0.2mm}no WANDs   \\ \hline\hline 
           \multirow{3}{*}{ $I$} & \vspace{0.2mm} \multirow{2}{*}{ $\mathbf{k}\neq$ WAND; $\mathbf{\Bar l}=$ single WAND; no other WANDs} &\vspace{0.2mm} $G$ & \vspace{0.2mm} no WANDs  \\ \cline{3-4}
           &  &\vspace{0.2mm} $I$ & \vspace{0.2mm} $\mathbf{ l}=$ single WAND; no other WANDs  \\ \cline{2-4}
  &\vspace{0.2mm} $\mathbf{k}=$single WAND; no other WANDs & \vspace{0.2mm} $I$ & \vspace{0.2mm}  $\mathbf{k}=$ single WAND; no other WANDs   \\ \hline \hline
   \multirow{5}{*}{ $I_i$} &\vspace{0.2mm} \multirow{3}{*}{ $\mathbf{k}\neq$ WAND; $\mathbf{\Bar l}_1, \mathbf{\Bar l}_2=$ single WANDs} &\vspace{0.2mm} $G$ & \vspace{0.2mm} no WANDs  \\ 
  \cline{3-4}
 & &\vspace{0.2mm} $I$ & \vspace{0.2mm}  $\mathbf{ l}_1=$ single WAND ; no other WANDs \\ 
  \cline{3-4}
   & &\vspace{0.2mm} $I_i$ & \vspace{0.2mm}  $\mathbf{ l}_1,\mathbf{ l}_2=$ single WANDs; $\mathbf{k}\neq$ WAND   \\ 
  \cline{2-4}
  &\vspace{0.2mm}  \multirow{2}{*}{ $\mathbf{k}, \mathbf{\Bar l}=$ single WANDs} & \vspace{0.2mm} $I$ & \vspace{0.2mm} $\mathbf{k}=$ single WAND; no other WANDs\\ \cline{3-4} 
   &\vspace{0.2mm}  & \vspace{0.2mm} $I_i$ & \vspace{0.2mm} $\mathbf{k},\mathbf{l}=$ single WANDs \\ \hline \hline
       \multirow{3}{*}{ $II$} &\vspace{0.2mm}  \multirow{2}{*}{ $ \mathbf{k}\neq$ WAND; $\mathbf{\Bar l}=$ double WAND; no other WANDs} &\vspace{0.2mm} $G$ & \vspace{0.2mm} no WANDs   \\ 
  \cline{3-4}
   &   &\vspace{0.2mm} $I/II$ & \vspace{0.2mm} $\mathbf{ l}=$ single/double WAND; no other WANDs  \\   \cline{2-4}
  & \vspace{0.2mm} $\mathbf{k}=$ double WAND; no other WANDs &\vspace{0.2mm}  $II$ & \vspace{0.2mm} $\mathbf{k}=$ double WAND; no other WANDs  \\ \hline \hline
   \multirow{9}{*}{ $II_i$ } & \vspace{0.2mm} \multirow{5}{*}{ $ \mathbf{k}\neq$ WAND; $\mathbf{\Bar l}_1=$ double WAND; $\mathbf{\Bar l}_2=$ single WAND }  &  \vspace{0.2mm}$G$ &  \vspace{0.2mm}no WANDs   \\ 
  \cline{3-4}
   & \vspace{0.2mm}  &  \vspace{0.2mm}$I$ &  \vspace{0.2mm} $\mathbf{ l}_2=$ single WAND; no other WANDs   \\ 
  \cline{3-4}
  & \vspace{0.2mm}  &  \vspace{0.2mm} $I_i$ &  \vspace{0.2mm} $\mathbf{ l}_2,\mathbf{ l}_3=$ single WANDs; $\mathbf{ \Bar l}_3=$ single WAND of background; \mbox{$\mathbf{k}, \mathbf{l}_1\neq$ WAND}  \\ 
  \cline{3-4}
   & \vspace{0.2mm}  &  \vspace{0.2mm}$I/II$ &  \vspace{0.2mm} $\mathbf{ l}_1=$ single/double WAND; no other WANDs   \\ 
  \cline{3-4}
  & \vspace{0.2mm}  &  \vspace{0.2mm}$I_i/II_i$ &  \vspace{0.2mm} $\mathbf{l}_1=$ single/double WAND; $\mathbf{ l}_2=$ single WAND; $ \mathbf{k}\neq$ WAND \\ 
  \cline{2-4}
  & \multirow{2}{*}{ \vspace{0.2mm}$ \mathbf{k}=$ single WAND; $\mathbf{\Bar l}=$ double WAND } & \vspace{0.2mm}$I,I_i$ &  \vspace{0.2mm} $\mathbf{k}=$ single WAND; no double WAND   \\ \cline{3-4}
   &  & \vspace{0.2mm}$II_i$ & \vspace{0.2mm} $\mathbf{k}=$ single WAND; $\mathbf{ l}=$ double WAND \\ \cline{2-4}
   & \multirow{2}{*}{ \vspace{0.2mm}$\mathbf{k}=$ double WAND; $\mathbf{\Bar l}=$ single WAND }  & \vspace{0.2mm}$II$ & \vspace{0.2mm} $\mathbf{k}=$ double WAND; no other WANDs   \\ \cline{3-4}
    & & \vspace{0.2mm}$II_i$ & \vspace{0.2mm} $\mathbf{k}=$ double WAND; $\mathbf{ l}=$ single WAND  \\ \hline \hline
     \multirow{18}{*}{ $D$ } &  \vspace{0.2mm} \multirow{11}{*}{ $\mathbf{k}\neq$ WAND; $\mathbf{\Bar l}_1, \mathbf{\Bar l}_2=$ double WANDs } &  \vspace{0.2mm}$G$ &  \vspace{0.2mm}no WANDs  \\ 
  \cline{3-4}
  &  \vspace{0.2mm} &  \vspace{0.2mm} \multirow{3}{*}{ $I$ }  &  \vspace{0.2mm} $\mathbf{ l}_1=$ single WAND; no other WANDs \\ 
  \cline{4-4}
    &  \vspace{0.2mm}  &  \vspace{0.2mm} &  \vspace{0.2mm}  $\mathbf{ l}_1, \mathbf{ l}_2\neq $ WAND; $\mathbf{l}_3=$ single WAND; \mbox{$ \mathbf{\Bar l }_3 =$ single WAND of background;} \mbox{no other WANDs} \\ 
  \cline{3-4}
    &  \vspace{0.2mm}  &  \vspace{0.2mm} \multirow{5}{*}{ $I_i$ } &  \vspace{0.2mm} $\mathbf{ l}_1, \mathbf{ l}_2=$ single WANDs; $\mathbf{k} \neq$ WAND \\ 
  \cline{4-4}
  &  \vspace{0.2mm}  &  \vspace{0.2mm}  &  \vspace{0.2mm}  $\mathbf{ l}_1, \mathbf{ l}_3=$ single WANDs; $\mathbf{ \Bar l}_3=$ single WAND of background; \mbox{$\mathbf{k},  \mathbf{ l}_2  \neq$ WAND}   \\ 
 \cline{4-4}
  &  \vspace{0.2mm}  &  \vspace{0.2mm}  &  \vspace{0.2mm}  $\mathbf{ l}_3, \mathbf{ l}_4=$ single WANDs; $\mathbf{ \Bar l}_3, \mathbf{ \Bar l}_4=$ single WANDs of background; $\mathbf{k}, \mathbf{ l}_1,  \mathbf{ l}_2  \neq$ WAND  \\   
 \cline{3-4}
    &  \vspace{0.2mm} &  \vspace{0.2mm}  $II$ &  \vspace{0.2mm} $\mathbf{l}_1=$ double WAND; no other WANDs \\ 
  \cline{3-4}
   &  \vspace{0.2mm}  &  \vspace{0.2mm}  \multirow{2}{*}{ $II_i$ }&  \vspace{0.2mm}  $\mathbf{l}_1=$ double WAND; $\mathbf{l}_2=$ single WAND; $\mathbf{k}\neq$ WAND \\ 
  \cline{4-4}
   &  \vspace{0.2mm}  &  \vspace{0.2mm} &  \vspace{0.2mm} $\mathbf{l}_1=$ double WAND; $\mathbf{l}_3=$ single WAND;  \mbox{$\mathbf{\Bar l}_3=$ single WAND of background;} \mbox{$\mathbf{k}, \mathbf{l}_2\neq$ WAND}   \\  \cline{3-4}
   &  \vspace{0.2mm}  &  \vspace{0.2mm} $D$ &  \vspace{0.2mm} $\mathbf{l}_1, \mathbf{l}_2=$ double WANDs; $\mathbf{k}\neq$ WAND \\ 
  \cline{2-4}
  & \multirow{3}{*}{ \vspace{0.2mm}$\mathbf{k}=$ single WAND; $\mathbf{\Bar l}_1,\mathbf{\Bar l}_2=$ double WANDs }  & \vspace{0.2mm}$I,I_i$ & \vspace{0.2mm} $\mathbf{k}=$ single WAND; no double WAND  \\ \cline{3-4}
  &  & \vspace{0.2mm}$II_i$ & \vspace{0.2mm} $\mathbf{k}=$ single WAND;  $\mathbf{ l}_1=$ the only double WAND  \\ \cline{3-4} 
      &  & \vspace{0.2mm}$D$ & \vspace{0.2mm} $\mathbf{k}=$ single WAND;  $\mathbf{ l}_1,\mathbf{ l}_2=$ double WANDs  \\ \cline{2-4} 
     &  \multirow{2}{*}{\vspace{0.2mm} $\mathbf{k},\mathbf{\Bar l}=$ double WANDs } & \vspace{0.2mm}$II,II_i$ & \vspace{0.2mm} $\mathbf{k}=$ the only double WAND  \\ \cline{3-4}
    &  & \vspace{0.2mm}$D$ & \vspace{0.2mm}  $\mathbf{k},\mathbf{ l}=$ double WANDs  \\ \hline \hline
     \multirow{3}{*}{ $III$ } &  \vspace{0.2mm}\multirow{3}{*}{ $ \mathbf{k}\neq$ WAND; $\mathbf{\Bar l}=$ triple WAND; no other WANDs } &  \vspace{0.2mm}$G$ &  \vspace{0.2mm}no WANDs  \\ \cline{3-4}
  &   & \vspace{0.2mm} $I/II/III$ &  \vspace{0.2mm}$\mathbf{l}=$ single/double/triple WAND; no other WANDs  \\ 
  \cline{2-4}
  & \multirow{2}{*}{\vspace{0.2mm}$ \mathbf{k}=$ triple WAND; no other WANDs }  & \vspace{0.2mm}$II/III$ &  \vspace{0.2mm} $\mathbf{k}=$ double/triple WAND; no other WANDs   \\ \hline \hline
    \multirow{11}{*}{ $III_i$ } &  \vspace{0.2mm} \multirow{7}{*}{ $\mathbf{k}\neq$ WAND; $\mathbf{\Bar l}_1=$ triple WAND; $\mathbf{\Bar l}_2=$ single WAND } &  \vspace{0.2mm}$G$ &  \vspace{0.2mm}no WANDs   \\ 
  \cline{3-4}
   &  \vspace{0.2mm}   &  \vspace{0.2mm}   \multirow{2}{*}{ $I$ } &  \vspace{0.2mm} $\mathbf{l}_1=$ single WAND; no other WANDs \\ 
  \cline{4-4}
   &  \vspace{0.2mm} &  \vspace{0.2mm} &  \vspace{0.2mm}  $\mathbf{l}_2=$ single WAND; no other WANDs\\ 
  \cline{3-4}
   &  \vspace{0.2mm} &  \vspace{0.2mm} \multirow{2}{*}{ $I_i$ } &  \vspace{0.2mm} $\mathbf{l}_1, \mathbf{l}_2=$ single WANDs; $\mathbf{k}\neq$ WAND  \\ 
  \cline{4-4}
   &  \vspace{0.2mm} &  \vspace{0.2mm} &  \vspace{0.2mm} $\mathbf{l}_2, \mathbf{l}_3=$ single WANDs; $\mathbf{\Bar l}_3=$ single WAND of background;  \mbox{$\mathbf{k},\mathbf{l}_1 \neq$ WAND} \\ 
  \cline{3-4}
   &  \vspace{0.2mm} &  \vspace{0.2mm} $II/III$&  \vspace{0.2mm} $\mathbf{l}_1=$ double/triple WAND; no other WANDs \\ 
  \cline{3-4}
    &  \vspace{0.2mm} &  \vspace{0.2mm} $II_i/III_i$&  \vspace{0.2mm}  $\mathbf{l}_1=$ double/triple WAND;  $\mathbf{l}_2=$ single WAND; $\mathbf{k} \neq$ WAND \\ 
  \cline{2-4}
  &  \multirow{2}{*}{ \vspace{0.2mm}$ \mathbf{k}=$ single WAND; $\mathbf{\Bar l}=$ triple WAND }  & \vspace{0.2mm}$I,I_i$ &  \vspace{0.2mm} $\mathbf{k}=$ single WAND  \\ \cline{3-4} 
  &  & \vspace{0.2mm}$II_i/III_i$ &  \vspace{0.2mm} $\mathbf{k}=$ single WAND;  $\mathbf{ l}=$ double/triple WAND  
   \\ \cline{2-4} 
   &  \multirow{2}{*}{  \vspace{0.2mm}$ \mathbf{k}=$ triple WAND; $\mathbf{\Bar l}=$ single WAND } & \vspace{0.2mm}$II/III$ &  \vspace{0.2mm} $\mathbf{k}=$ double/triple WAND; no other WANDs   \\ \cline{3-4} 
    &   & \vspace{0.2mm}$II_i/III_i$ &  \vspace{0.2mm} $\mathbf{k}=$ double/triple WAND;  $\mathbf{ l}=$ single WAND  
    \\ \hline \hline
    \multirow{3}{*}{ $N$ } &  \vspace{0.2mm} \multirow{2}{*}{ $\mathbf{k}\neq$ WAND; $\mathbf{\Bar l}=$ quadruple WAND }  &  \vspace{0.2mm}$G$ &  \vspace{0.2mm}no WANDs   \\ 
  \cline{3-4}
 &  \vspace{0.2mm}  &  \vspace{0.2mm}$I/II/III/N$ &  \vspace{0.2mm} $\mathbf{ l}=$ single/double/triple/quadruple WAND; no other WANDs \\ 
  \cline{2-4}
  &  \vspace{0.2mm}$ \mathbf{k}=$ quadruple WAND  &  \vspace{0.2mm}$II/III/N$ &  \vspace{0.2mm}$ \mathbf{k}=$ double/triple/quadruple WAND; no  other WANDs  \\ 
  \hline \hline
   \multirow{4}{*}{ $O$ } &  \multirow{4}{*}{   \vspace{0.2mm} }  &  \vspace{0.2mm}$II, II_i, D$ &  \vspace{0.2mm}$ \mathbf{k}=$ double WAND  \\ 
  \cline{3-4}
& &  \vspace{0.2mm}$III, III_i$ &  \vspace{0.2mm}$ \mathbf{k}=$ triple WAND \\ 
  \cline{3-4}
 & &  \vspace{0.2mm}$N$ &  \vspace{0.2mm}$ \mathbf{k}=$ quadruple WAND  \\ 
  \cline{3-4}
  & &  \vspace{0.2mm}$O$ &  \vspace{0.2mm}    \\  \hline
    \end{tabular}}
     
     \caption{\footnotesize All kinematically allowed combinations of Weyl types for the background and the full geometry. Similar to Table \ref{table_Ricci_types}, barred null directions are defined by equation \eqref{n_bg}. Note that type $G$, as well as type $D$ Weyl tensor with a single WAND, are forbidden in $n=4$ (cf. [14,24] and Remark D.4). }\label{table_Weyl_types}
\end{table}
\newpage

Having provided a `guide' to interpret the tables, let us highlight some interesting properties of GKS spacetimes from an algebraic viewpoint and draw comparisons with the KS class. From the tables, as well as the discussions of Sections \ref{Ricci_types}, \ref{Weyl_types}, we see that the algebraic type of the full geometry is constrained by that of the background. In particular, the Ricci and Weyl tensors of the full geometry cannot be more special than the respective tensors of the background. Further, we note that the KS class, which has a maximally symmetric background, belongs to the category of GKS spacetimes that have a conformally flat (Weyl type $O$) background. Hence, the full geometry in this case is at least of Weyl type $II$, in agreement with the results of \cite{HD_KS,Malek:2010mh}. However, as shown in Table \ref{table_Weyl_types}, for a non-conformally flat background, the full geometry can take on Weyl types less special than type $II$, including the general type $G$. Below, we briefly discuss an example: the five-dimensional minimal SUGRA solution of Chong-Cveti{\v c}-L{\" u}-Pope (CCLP) \cite{CCLP}, where the spacetime is in particular of Weyl type $I_i$.

The minimal SUGRA solution of CCLP \cite{CCLP}, which describes a charged rotating black hole in five dimensional Einstein-Maxwell-Chern-Simons theory, admits the following form \cite{Aliev_CCLP} 
\begin{align}
   & \mathbf{g}= \mathbf{g}_0 -2 f \mathbf{k}\otimes \mathbf{k} - V(\mathbf{k}\otimes \mathbf{m} + \mathbf{m}\otimes \mathbf{k}), \label{CCLP_metrc_xKS}\\
   & f=-\frac{M}{\rho^2}+ \frac{Q^2}{2\rho^4}, \quad V= -\frac{Q}{\rho^2}, \quad  \rho^2= r^2 + a^2 \cos ^2 \theta + b^2 \sin ^2 \theta, \label{soln_fns_CCLP}
\end{align}
where $\mathbf{g_{0}}$ represents a maximally symmetric metric, $\mathbf{k}$ is a null (co)-vector field and $\mathbf{m}$ is a spacelike (co)-vector field. Further, $M$, $Q$ are respectively the constant mass and charge parameters, and $a$, $b$ denote the two constant rotation parameters. While the precise forms of $\mathbf{g_{0}}$ and the vector fields can be found in \cite{Aliev_CCLP}, we note that the $1$-form potential of the Maxwell-Chern-Simons field is given by $\mathbf{A}=-\frac{\sqrt{3}Q}{2\rho^2} \mathbf{k}$ \cite{CCLP,Aliev_CCLP}.

The metric ansatz \eqref{CCLP_metrc_xKS}, which has a splitting with respect to both null and spacelike vector fields, is referred to as the extended Kerr-Schild (xKS) form \cite{Ett_Kastor_xKS,Malek_xKS}. Noting that the dependence on parameters $Q$ and $M$ enters the metric only through the functions $f$ and $V$ defined in \eqref{soln_fns_CCLP} , one can massage the xKS form of the metric to recast it into the GKS form \eqref{GKS} as\footnote{A different GKS splitting of the CCLP solution was already explored in \cite{Srini_Kubiz_Hassaine}, in the context of the `extremal Kerr-Schild' form, where they define the background to be the extremal limit of the full (black hole) metric. In contrast, we consider the $M=0$ limit to be the background. See Section \ref{discussion} for a brief discussion on different possible background choices.}
\begin{align}
  &  \mathbf{\Bar g}= \mathbf{g}_0 - \frac{Q^2}{ \rho^4} - V(\mathbf{k}\otimes \mathbf{m} + \mathbf{m}\otimes \mathbf{k}),\label{bg_GKS_CCLP}\\
  & H= -\frac{M }{\rho^2},\label{fn_H_CCLP_GKS}
\end{align}
with $\mathbf{k}$ identified as the KS vector.

In \cite{Malek_xKS}, where the CCLP solutions were studied in the context of xKS metrics, it was shown that $\mathbf{k}$ defines an expanding geodesic null congruence, with the coordinate $r$ identified as its affine parameter. Moreover, it was shown there that the spacetime is of Weyl type $I_i$, with $\mathbf{k}$ being a WAND.\\

To conclude the discussion of this section, we now turn to some remarks on the Ricci types. For the KS class, as the background is maximally symmetric and hence an Einstein spacetime, any null direction is a double Ricci AND of the background (cf. Remark \ref{rem_Eins}) and so is $\mathbf{k}$. Hence, Propositions \ref{prop_Ricci_double_AND_k} and \ref{prop_k_multiple} ensure that $\mathbf{k}$ is a double Ricci AND of also the full geometry. In contrast, for GKS spacetimes with a non-Einstein background, $\mathbf{k}$ need not be a Ricci AND, as seen from the long list of other options available in Table \ref{table_Ricci_types}. An example of this is provided by the Schwarzschild-Melvin black holes \cite{Ernst:1976mzr}, discussed in Appendix \ref{matter-matter}, which form a GKS spacetime with geodesic $\mathbf{k}$ and the background being the Bonnor-Melvin magnetic universe \cite{WBBonnor_1954,Melvin:1963qx,Melvin}. In this case, the background and the full geometry are of Ricci type $D$ (cf. \cite{Stephani:2003tm,Griffiths_Podolský_2009} and references therein; also \cite{Ort_Melvin}), with $\mathbf{k}$ not being a Ricci AND. Therefore, from Proposition \ref{prop_l_AND}, the two double Ricci ANDs of the full geometry must be related to their background counterparts in a manner analogous to that in \eqref{n_bg}. Interestingly, the Schwarzschild-Melvin black holes also form an example of GKS spacetimes of Weyl type less special than $II$. In particular, the black holes are of Weyl type $I_i$ \cite{Bose_Ernst_algebra_type} (cf. also \cite{Ort_Melvin}) while the Bonnor-Melvin background is of Weyl type $D$ \cite{Wild_Melvin_type_D}. Since this is an example in $n=4$, from Table \ref{table_Weyl_types}, it can further be concluded that $\mathbf{k}$ is also not a WAND.

\section{Expanding $\mathbf{k}\quad (\theta \neq 0)$ } \label{expanding ks}
In this section, we will discuss some results concerning GKS spacetimes with an expanding $\mathbf{k}$, i.e., $\theta \neq 0$, and illustrate the results using some examples. The case of non-vanishing expansion is relevant in the context of black holes and other spacetimes with horizons (for example, Taub-NUT, cf. \cite{Griffiths_Podolský_2009} and references therein). Moreover, as we shall see in Section \ref{Optical constraint}, an expanding $\mathbf{k}$ allows for a weak generalization of the shearfree part of the Goldberg-Sachs theorem, applicable only to GKS spacetimes, thereby extending the previous analogous results of KS spacetimes \cite{HD_KS, Malek:2010mh}.\footnote{It was shown in \cite{Ortaggio_Jose_Jakub} (cf. also references therein; and also \cite{HD_KS,Malek:2010mh}) that Einstein-KS spacetimes with a non-expanding $\mathbf{k}$ are equivalent to type-$N$ Kundt spacetimes. Therefore, in these spacetimes, the optical matrix of the KS vector is identically zero and hence shearfree. Moreover, the KS vector forms the corresponding mWAND. These aspects, however, do not straightforwardly generalize to GKS spacetimes with a non-expanding $\mathbf{k}$ and will therefore require a dedicated study elsewhere. }

\subsection{Optical constraint and comments on the Goldberg-Sachs theorem}\label{Optical constraint}
The Goldberg-Sachs theorem \cite{Stephani:2003tm} is an important result in four dimensions, as it significantly simplifies the integration of the Newman-Penrose equations for algebraically special spacetimes \cite{newman1962approach}, thereby aiding in the derivation of exact solutions to the vacuum Einstein equations. Although, as mentioned in Section \ref{background_intro}, the theorem does not have an obvious higher-dimensional generalization, reasonable generalizations have been formulated for several special cases \cite{Ortaggio:2012hc, Gold_Sachs_HD, Tintera:2019hil, VPravda_2004,type_III_N}.\footnote{See also \cite{Taghavi-Chabert:2010ngx,Taghavi-Chabert:2011kmm} for a different approach to higher-dimensional generalization of Goldberg-Sachs theorem.} The outcomes of such generalized results are often in the form of constraints on the optical matrix for the WAND and, analogous to the four-dimensional case, typically help simplify the integration of the Einstein equations. For example, the results of \cite{Ortaggio:2012hc} led to the complete integration and classification of five-dimensional Einstein spacetimes with an expanding mWAND \cite{BernardideFreitas:2015vke, Wylleman:2015avh, BernardideFreitas:2015lyp} (cf. also \cite{Gomez-Lobo:2011xwu, Reall:2012_5d_integr}), and likewise, there has been partial progress in six dimensions \cite{6D_MP_uniqueness}. Along similar lines, the complete class of Einstein spacetimes, in even dimensions $n\geq 6$, characterized by shearfree-twisting null geodesic congruences was obtained in \cite{Taghavi-Chabert:2020hzw}.\footnote{In obtaining this class of spacetimes, an assumption on the Weyl tensor (which also constrains the geodesic twisting-shearfree null congruence) is made. See \cite{Taghavi-Chabert:2020hzw} for details.}  

In the context of constraining the optical matrix and its potential use in classifying exact solutions, it was shown in \cite{HD_KS,Malek:2010mh} that the (expanding) KS vector $\mathbf{k}$ of an Einstein-KS spacetime satisfy the following relation, known as the optical constraint \cite{HD_KS,Malek:2010mh,type_III_N,Gold_Sachs_HD}
\begin{align}
     L_{ik}L_{jk}= \frac{L_{lk}L_{lk}}{(n-2)\theta} S_{ij}. \label{opt_constraint}
\end{align}
Furthermore, it was shown that, as a consequence of the optical constraint, the optical matrix can be brought into a canonical (block-diagonal) form (see equations \eqref{L_ij_canonical}-\eqref{diag_block_Lij_canonical}), which helps in the partial integration of the function $H$ associated with the KS metric \cite{HD_KS,Malek:2010mh}. Moreover, the optical constraint \eqref{opt_constraint}, in combination with the results of \cite{Taghavi-Chabert:2020hzw}, has proven useful in classifying all higher-dimensional expanding-twisting KS solutions in the Einstein-Maxwell theory \cite{Ort_Srini}.\footnote{In \cite{Ort_Srini}, the electrovacuum KS solutions are obtained under the additional assumption that the vector potential of the Maxwell field is proportional to the KS vector, i.e., $\mathbf{A} \propto \mathbf{k}$.} In what follows, we shall derive theory-independent conditions under which the optical constraint and the canonical form of $L_{ij}$ generalize to GKS spacetimes.

\begin{proposition}\label{prop_opt_contr}
  Let $\mathbf{k}$ be an AND of the Riemann tensor of the background and hence also the full spacetime. Then, $\mathbf{k}$ satisfies the optical constraint \eqref{opt_constraint} iff $ (R_{ij} -\Bar R_{ij})\propto   S_{ij}$.
\end{proposition}
  \noindent\textbf{Proof.} By assumption, $\mathbf{k}$ is an AND of the Riemann tensor. Therefore, equation \eqref{Rij_GKS} reduces to 
  \begin{align}
  & R_{ij} = \Bar R_{ij} + 2HL_{ik}L_{jk} - 2S_{ij}\left[DH + (n-2)\theta H\right]. \label{Rij_opt_1}
  \end{align} 
  First, suppose that $\mathbf{k}$ satisfies the optical constraint \eqref{opt_constraint}. Then, equation \eqref{Rij_opt_1} can be written as 
  \begin{align}
  & R_{ij} - \Bar R_{ij} = \mathcal{B}S_{ij},
  \end{align} 
  with the proportionality function $\mathcal{B} = 2 \left[H \frac{L_{lk}L_{lk}}{(n-2)\theta} - DH - (n-2)\theta H\right]$. 
  
  Conversely, suppose that $R_{ij} - \Bar R_{ij} = \mathcal{B}S_{ij}$ for some proportionality function $\mathcal{B}$. Then, one can rewrite \eqref{Rij_opt_1} as
  \begin{align} 
  & \mathcal{B}S_{ij} = 2HL_{ik}L_{jk} - 2S_{ij}\left[DH + (n-2)\theta H\right]. \label{Rij_opt_2} 
  \end{align}
  Contracting equation \eqref{Rij_opt_2} with $\delta_{ij}$, one obtains
  \begin{align}
 \mathcal{B}=2\left[H \frac{L_{lk}L_{lk}}{(n-2)\theta} -DH- (n-2)\theta H\right].\label{Rij_B_opt_const}
\end{align}
Substituting expression \eqref{Rij_B_opt_const} back into equation \eqref{Rij_opt_2} gives
\begin{align}
    H(L_{ik}L_{jk}-   \frac{L_{lk}L_{lk}}{(n-2)\theta} S_{ij})=0.
\end{align}
Since $H\neq 0$ for the notion of the GKS form to make sense, the above relation leads to the optical constraint \eqref{opt_constraint}.\\

When the KS vector $\mathbf{k}$ of a GKS spacetime is a Riemann AND satisfying the optical constraint \eqref{opt_constraint}, as a direct extension of the results of \cite{HD_KS}, one can cast the optical matrix into the following (canonical) block diagonal form \cite{Malek:2010mh,type_III_N,Gold_Sachs_HD,HD_alg_review}\footnote{The block diagonalization is achieved by making use of the optical constraint \eqref{opt_constraint} in combination with the Ricci identity $(11g)$ from \cite{Ricci_in_HD}, while choosing the spatial vectors $\mathbf{m}_{i}$ to be parallel transported along $\mathbf{k}$, see \cite{HD_KS} for details.}
\beqn
L_{ij}=\left(\begin {array}{cccc} \fbox{${\cal L}_{(1)}$} & & &  \\
& \ddots & & \\ 
& & \fbox{${\cal L}_{(q)}$} & \label{L_general} \\
& & & \fbox{$\begin {array}{ccc} & & \\ \ \ & \tilde{\cal L} \ \ & \\ & & \end {array}$}
\end {array}
\right) , \label{L_ij_canonical}
\eeqn
where the first $q$ blocks are $2\times 2$ matrices and the last block $\tilde{\cal L}$ is an $(n-2-2q)\times(n-2-2q)$-dimensional diagonal matrix. The expressions for these blocks are given by
\beqn
& & {\cal L}_{(\nu)}=\left(\begin {array}{cc} s_{(2\nu)} & A_{2\nu,2\nu+1} \nonumber \\
-A_{2\nu,2\nu+1} & s_{(2\nu)} 
\end {array}
\right) \qquad (\nu=1,\ldots, q) , \\
& & s_{(2\nu)}=\frac{r}{r^2+(a^0_{(2\nu)})^2} , \qquad A_{2\nu,2\nu+1}=\frac{a^0_{(2\nu)}}{r^2+(a^0_{(2\nu)})^2} , \label{s_A} \\
& &  \tilde{\cal L}=\frac{1}{r}\mbox{diag}(\underbrace{1,\ldots,1}_{(m-2q)},\underbrace{0,\ldots,0}_{(n-2-m)}) , \label{diag_block_Lij_canonical}
\label{diagonal}
\eeqn
where $0\le 2q\le m\le n-2$, $r$ is an affine parameter of the geodesic $\mathbf{k}$ so that $k^a \partial_a= \partial_r$, and $a^0_{(2\nu)}$ are $r$-independent spin functions. The rank of $L_{ij}$ is denoted by the integer $m$; hence, $L_{ij}$ is non-degenerate when $m=n-2$.

On using the above expression for $L_{ij}$, one finds that the optical scalars \eqref{optical_scalars} take the form
\beqn
 & & (n-2)\theta=2\sum_{\nu=1}^q\frac{r}{r^2+(a^0_{(2\nu)})^2}+\frac{m-2q}{r} , \label{exp} \\
 & & \omega^2=2\sum_{\nu=1}^q\left(\frac{a^0_{(2\nu)}}{r^2+(a^0_{(2\nu)})^2}\right)^2 , \label{twist} \\
 & & \sigma^2=2\sum_{\nu=1}^q\left(\frac{r}{r^2+(a^0_{(2\nu)})^2}-\theta\right)^2+(m-2q)\left(\frac{1}{r}-\theta\right)^2+(n-2-m)\theta^2 . \label{shear} 
\eeqn
Let us note that $\theta \neq 0 \Leftrightarrow m \neq 0$, and likewise $\omega \neq 0 \Leftrightarrow q \neq 0$. For non-vanishing expansion, one has the following two alternatives for the vanishing of shear
\begin{enumerate}[(i)]
    \item For $q = 0$: $\sigma=0 \Leftrightarrow m=n-2 \Leftrightarrow \theta = \frac{1}{r}$.\\
    This corresponds to the subclass of Robinson-Trautman geometries \cite{Stephani:2003tm, Griffiths_Podolský_2009, Podolsky_Ort_RT, Podolsky_Svarc_RT } that intersect with the GKS class, when $\mathbf{k}$ is a geodesic Riemann AND (cf. Section \ref{RT_GKS_section}). 
    \item For $q\neq 0$: $\sigma=0 \Leftrightarrow m=2q= n-2, (a^0_{(2)})^2=(a^0_{(4)})^2=\ldots=(a^0_{(2q)})^2$.\\
   This means that when $\omega \neq 0$, the shear can vanish only in even spacetime dimensions, when all the spin functions $(a^0_{(2\nu)})^2$ coincide. An example of this is given by (A)dS-Taub-NUT spacetimes in even dimensions $n\geq 4$ \cite{Taub:1950ez,Newman:1963yy,DNPage_1987,Lor_Dieter,Taylor:1998fd,Awad:2000gg}, with the spin functions being the (equal) NUT parameters. We will further discuss the four-dimensional (A)dS-Taub-NUT in Section \ref{NUT_4d}.
\end{enumerate}

Proposition \ref{prop_opt_contr} relies on the prior assumption that $\mathbf{k}$ is a Riemann AND. Below, we present a complementary result where $\mathbf{k}$ being a Riemann AND is not assumed a priori but rather is part of the equivalence condition claimed in the proposition. As we shall see, this form of the result on optical constraint not only mimics the Goldberg-Sachs theorem for GKS spacetimes but is also more suitable for the discussion in the next subsection.
\begin{proposition}
\label{prop_opt_contr_2}
    Let $R_{ij}-\Bar{R}_{ij}=0$. Then, $\mathbf{k}$ is a Riemann AND iff it is a Ricci AND satisfying the optical constraint \eqref{opt_constraint}.
\end{proposition}
 \noindent \textbf{Proof.} Let us first assume that $\mathbf{k}$ is a Riemann AND. Then, it follows that $\mathbf{k}$ is also a Ricci AND. Moreover, since $R_{ij}-\Bar{R}_{ij}\propto S_{ij}$ holds trivially with the proportionality factor being zero, one can directly apply Proposition \ref{prop_opt_contr} to conclude that $\mathbf{k}$ satisfies the optical constraint.
 
Conversely, suppose that $\mathbf{k}$ is a Ricci AND satisfying optical constraint. Upon setting $R_{ij}-\Bar{R}_{ij}=0$ in \eqref{Rij_GKS} and subsequently using the optical constraint, one obtains 
\begin{align}
     H\Bar R_{0i0j} +S_{ij}\Big(   DH+ (n-2)\theta H-H\frac{L_{lk}L_{lk}}{(n-2)\theta} \Big)=0, \label{Riem_AND_opt_cont}
\end{align}
Taking the trace of the above equation with respect to $\delta_{ij}$ and then using the Ricci AND condition $R_{00}=\Bar{R}_{00}=0$, we have
\begin{align}
   \Big(   DH+ (n-2)\theta H-H\frac{L_{lk}L_{lk}}{(n-2)\theta} \Big)=0. \label{vacuum_Rij_eqn}
\end{align}
Combining \eqref{Riem_AND_opt_cont} and \eqref{vacuum_Rij_eqn}, along with the assumption that the GKS transformation is non-trivial, i.e., $H\neq0$, we find that $  R_{0i0j}= \Bar R_{0i0j}=0$, thus concluding the converse.
\begin{remark}
    \label{remark_optic_contr_2}
    It immediately follows from the above proposition that when $R_{00}=0=R_{ij}-\Bar{R}_{ij}=0$, $\mathbf{k}$ satisfies the optical constraint iff it is a WAND.
\end{remark}
Some comments related to the canonical form of the optical matrix and the Goldberg-Sachs theorem are in order, with most of these being minor extensions of the observations made in \cite{HD_KS}. Firstly, it is straightforward to check that an optical matrix with the canonical structure \eqref{L_ij_canonical}-\eqref{diag_block_Lij_canonical} automatically satisfies the optical constraint. Secondly, the Goldberg-Sachs theorem for four dimensional Einstein spacetimes (not specifically GKS) can be stated equivalently as follows (cf. Section $5.4$ of \cite{HD_KS} and references therein): \textit{A null congruence $\mathbf{l}$ of a non-conformally flat Einstein spacetime is a repeated PND iff $\mathbf{l}$ is geodesic and its optical matrix can be cast into the form
\beqn
  &&  L_{ij}=\left(\begin {array}{cc} s & A  \\
-A & s 
\end {array} \right), \label{can_form_4d_opt}
\eeqn
for some functions $s$ and $A$.} Finally, for the GKS case, the canonical form of the optical matrix \eqref{L_ij_canonical}-\eqref{diag_block_Lij_canonical} associated with the KS vector can be regarded as being composed of several shearfree $2\times2$ blocks, each resembling the form \eqref{can_form_4d_opt} specified by the four-dimensional Goldberg-Sachs theorem. However, such an interpretation works up to the exception of $m$ or (and) $n$ being odd, as these cases result in an unpaired $1$ or $0$ entry in the $ \mathcal{\Tilde L}$ block \eqref{diag_block_Lij_canonical}. Therefore, barring exceptions, one can sum up the above comments to interpret Proposition \ref{prop_opt_contr_2} as a weak generalization of the shearfree part\footnote{Note that geodesicity of $\mathbf{k}$ is not part of the result of Proposition \ref{prop_opt_contr_2}; rather, it is a prior assumption (which was made starting from Section \ref{GKS_with_geo_k}). Thanks to Propositions \ref{Riemann_AND_full} and \ref{Ricci_AND_full}, one can drop the (prior) geodesicity assumption and incorporate it into the result of the proposition by refining the statement as follows: \textit{Let $R_{ij}-\Bar{R}_{ij}=0$. Then, $\mathbf{k}$ is a Riemann AND of the full geometry iff it is a Ricci AND of the full geometry satisfying the optical constraint.} } of the Goldberg-Sachs theorem, which holds only in GKS spacetimes with respect to the KS vector $\mathbf{k}$. Although this is analogous to such an interpretation in KS spacetimes \cite{HD_KS}, unlike the standard Goldberg-Sachs theorem or the KS case \cite{HD_KS}, Proposition \ref{prop_opt_contr_2} involves a Riemann AND, which need not be of multiplicity greater than one.

\subsubsection{GKS spacetimes in the Einstein theory with $T_{ab}=\Bar{T}_{ab}$}\label{GKS_in_GR_subsection}

When the GKS spacetime and its background are solutions to the Einstein equations \eqref{Einstein_eqns}, with their on-shell energy-momentum tensors coinciding, i.e., $T_{ab}=\Bar{T}_{ab}$, it follows that $R_{ab}-\Bar{R}_{ab}=0$.\footnote{The GKS transformation term $H\mathbf{k}\otimes \mathbf{k}$ can also contribute to the cosmological constant term of the Einstein equations. Consequently, the background and the full geometry can have different cosmological constants. However, if the GKS splitting does not depend solely on the cosmological constant, the background can always be redefined to absorb the difference in cosmological constants. For the current discussion, we assume this to be the case; hence, the background and the full geometry share the same cosmological constant. Spacetimes where this assumption does not hold (see, for instance, equation \eqref{H_fn_for_RT_case_2}) will require a case-by-case treatment.} Hence, Proposition \ref{prop_opt_contr_2} and the subsequent comments apply in this case. Assuming $\mathbf{k}$ to be a Riemann AND, we can therefore conclude that $\mathbf{k}$ is a Ricci AND satisfying the optical constraint. Consequently, this also results in $H$ satisfying \eqref{vacuum_Rij_eqn}. Using the canonical structure of the optical matrix \eqref{L_ij_canonical}-\eqref{diag_block_Lij_canonical} and the consequent expressions for the optical scalars \eqref{exp}-\eqref{shear}, equation \eqref{vacuum_Rij_eqn} can be solved to fix the $r$-dependence of $H$ as \cite{HD_KS}
\begin{align}
    H= \frac{H_0}{r^{m-2q-1}}\prod_{\nu=1}^{q} \frac{1}{r^2+ (a^0_{(2\nu)})^2 }, \label{r-dpend_H_Einstein}
\end{align}
where $H_0$ is an $r$-independent integration function. Further, the equation for $H$ corresponding to the $R_{01}$ component of the Einstein equation, which reads
\begin{align}
    D^2H+ (n-2)\theta DH+2H\omega^2=0,\label{D^2H_eqn_vacuum_GKS}
\end{align}
is automatically satisfied by the solution in \eqref{r-dpend_H_Einstein}.

The assumptions on the energy-momentum tensor are, in particular, satisfied when the full geometry (and hence the background) satisfies the vacuum Einstein equations \eqref{vacuum_Einstein}. In the next subsection, we discuss in detail the example of (vacuum) Taub-NUT spacetimes in $n=4$, where we illustrate the application of Proposition \ref{prop_opt_contr_2} and observe agreement with the aforementioned results on the function $H$. Here, we briefly comment on a non-vacuum example: the five-dimensional charged-rotating black hole solution of CCLP \cite{CCLP} mentioned in Section \ref{subsection_remarks_alge_types}.

 In the GKS form of the CCLP solution, the charge associated with the Maxwell-Chern-Simons field, $Q$, appears only in the background, and the function $H$ depends solely on the mass.\footnote{One may consider interpretation (iii) of the GKS form of Kerr-Newman discussed in Section \ref{background_intro} also as a non-vacuum example for the current discussion. However, it is somewhat trivial, as absorbing the charge into the background in the case of Kerr-Newman is not necessary for its (G)KS representation, unlike the CCLP example where it is inevitable (see equation \eqref{bg_GKS_CCLP}).} Therefore, it satisfies $R_{ab}-\Bar{R}_{ab}=0$. In particular, since $R_{00}-\Bar R_{00}=0$, the KS vector must be geodesic, and as already noted in Section \ref{subsection_remarks_alge_types}, the results of \cite{Malek_xKS} indeed show that this is the case. Moreover, it turns out that $T_{00}=0$ (cf. \cite{Malek_xKS}). Combining this with the fact that $\mathbf{k}$ is a WAND \cite{Malek_xKS} with a non-vanishing expansion (see again Section \ref{subsection_remarks_alge_types}), we have from Proposition \ref{prop_opt_contr_2} (or Remark \ref{remark_optic_contr_2}) that $\mathbf{k}$ satisfies the optical constraint \eqref{opt_constraint}, which was explicitly demonstrated in \cite{Malek_xKS}. Finally, from \eqref{fn_H_CCLP_GKS}, we see that the function $H$ also agrees with the form given by \eqref{r-dpend_H_Einstein}, with $m=3$, $q=1$, and $(a^0_{2})^2$ identified with the $r$-independent function $(a^2 \cos ^2 \theta + b^2 \sin ^2 \theta)$ (see equations \eqref{soln_fns_CCLP}, \eqref{fn_H_CCLP_GKS}), and $H_0 = -M$.
\subsection{An illustrative example: (A)dS-Taub-NUT spacetimes in $n=4$}
\label{NUT_4d}
The four-dimensional (A)dS-Taub-NUT metric, which forms a solution to the vacuum Einstein equations \eqref{vacuum_Einstein}, can be written as \cite{Taub:1950ez,Newman:1963yy,Griffiths_Podolský_2009}\footnote{The optical and algebraic properties of (A)dS-Taub-NUT are well known and can be obtained from the references cited in this subsection (and the references therein). The purpose of the illustration is to highlight the GKS connection of these aspects, which we identify in retrospect. The same applies to all the examples discussed in this paper.}
\begin{align} 
&\mathbf{g}=-2\mathcal{H} \left( dt +  \frac{l}{P} (xdy-ydx)\right)^2 + \frac{dr^2}{2\mathcal{H}} + (r^2 + l^2) \frac{(dx^2+dy^2)}{P^2}, \label{NUT_1}\\
& 2\mathcal{H}=\frac{1}{r^2+l^2}\Big(K(r^2-l^2)-2Mr-\lambda(r^4+6l^2r^2-3l^4) \Big), \quad \lambda= 3\Lambda, \quad P= 1+ \frac{K}{4}(x^2+y^2), \label{cal_H_NUT}\end{align} 
where $M$ and $l$ respectively are mass and NUT parameters. Furthermore, $K$ is a constant and hence the two-dimensional base space $\mathbf{h}\equiv \frac{dx^2+dy^2}{P^2}$ is a space of constant curvature, with its Ricci scalar given by $\mathcal{R}=2K$. One can perform the coordinate transformation $du= dt + \frac{dr}{2\mathcal{H}}$ to recast \eqref{NUT_1} into the form
\begin{align}
    \mathbf{g}=2dr \left( du +  \frac{l}{P} (xdy-ydx)\right)+ (r^2 + l^2)\mathbf{h} -2\mathcal{H}\left( du +  \frac{l}{P} (xdy-ydx)\right)^2. \label{GKS_NUT}
\end{align}
It can be easily checked that the $1$-form
\begin{align} \mathbf{k}\equiv\left( du + \frac{l}{P} (xdy-ydx)\right) \label{KS_vector_NUT}\end{align}
forms an affinely parametrized null geodesic, with $r$ being an affine parameter. Upon identifying it with the KS vector, we can immediately recognize the GKS structure in metric \eqref{GKS_NUT}, with the background $\mathbf{\Bar{g}}$ and scalar function $H$ given by\footnote{Note that in accordance with the discussion in Section \ref{GKS_in_GR_subsection}, we have absorbed the cosmological constant term into the background.} 
\begin{align}
   & \mathbf{\Bar g}=dr\otimes\mathbf{k}+ \mathbf{k}\otimes dr + (r^2+l^2)\mathbf{h}-2f\mathbf{k}\otimes\mathbf{k}, \label{NUt_GKS_bg}\\
    & H=-\frac{Mr}{r^2+l^2}, \quad f= \mathcal{H}-H.\label{H_fn_NUT}
\end{align}

The background and the full geometry satisfy the vacuum Einstein equations with the same cosmological constant. Hence, Proposition \ref{prop_opt_contr_2} and the observations of Section \ref{GKS_in_GR_subsection} apply in this case. To explore this further, let us introduce the following null (co)-frame
\begin{align} 
 \mathbf{m}^{(0)}=n_a dx^{a}\equiv dr-\mathcal{H}\mathbf{k}, \quad \mathbf{m}^{(1)}=\mathbf{k}, \quad \mathbf{m}^{(2)} = \frac{l}{P}\left(dy+ \frac{r}{l}dx\right), \quad \mathbf{m}^{(3)} = \frac{l}{P}\left(dx- \frac{r}{l}dy  \right). \label{coframe_NUT}
\end{align} 
It is then straighforward to show that $\overset{i}{M}_{j0}=0$, i.e., the spatial frame vectors $\mathbf{m}_{(i)}$ are parallel transported along $\mathbf{k}$. Moreover, the optical matrix is given by
\begin{align}
    S_{ij}= \theta\delta_{ij}=\frac{r}{r^2+l^2}\delta_{ij}, \quad A_{ij}=\frac{l}{r^2+l^2}\varepsilon_{ij}, \label{NUT_opt_matrix}
\end{align}
where $\varepsilon_{ij}$ is the two dimensional Levi-Civita symbol, defined as $\varepsilon_{12}=1$. Clearly, the optical matrix of $\mathbf{k}$ is in the canonical form discussed in Section \ref{Optical constraint}, with 
\begin{align}
  m=2=2q, \quad a^0_{(2)}=l ,\label{m,q,a0_NUT}  
\end{align}
and hence automatically satisfies the optical constraint \cite{Pravda_type_D,type_III_N}. Therefore, applying Proposition  \ref{prop_opt_contr_2} and Remark \ref{remark_optic_contr_2}, we see that $\mathbf{k}$ must be a WAND (or equivalently a Riemann AND). In fact, in the above choice of null frame, one finds that the only non-vanishing Weyl components are the following
\begin{align}
 &C_{01ij}= 2W_1\varepsilon_{ij}, \quad C_{0i1j}=W_2 \delta_{ij}+ W_1 \varepsilon_{ij}, \quad C_{0101}=2W_2=-C_{1212} , \label{Weyl_NUT_1}\\
 & W_1=-l\frac{\Big((K-4\lambda  l^2)(r^3-3rl^2)-M(3r^2-l^2)  \Big)}{(r^2+l^2)^3}, \quad W_2=\frac{\Big( (K-4\lambda l^2)(l^4-3r^2l^2) +M (3rl^2-r^3) \Big)}{(r^2+l^2)^3}. \label{Weyl_NUT_2}
\end{align}
Thus, $\mathbf{k}$ is indeed a (double) WAND \cite{Stephani:2003tm,PLEBANSKI197698,Griffiths_Podolský_2009}. Furthermore, we observe that the function $H$ given in \eqref{H_fn_NUT} matches the form \eqref{r-dpend_H_Einstein}, with $m$, $q$, and $a_{(2)}^0$ specified by \eqref{m,q,a0_NUT} and $H_0 = -M$. Therefore, the example is consistent with the discussion of Section \ref{GKS_in_GR_subsection}.

We now turn to a comparison of the algebraic types of the background and the full geometry. From \eqref{Weyl_NUT_1} and \eqref{Weyl_NUT_2}, we see that the full geometry is of Weyl type $D$, with $\mathbf{n}$ being the other double WAND \cite{Stephani:2003tm, Griffiths_Podolský_2009, PLEBANSKI197698}. The background, which is the $M=0$ limit of the full geometry, is also of Weyl type $D$ in general, with $\mathbf{k}$ and $\mathbf{\Bar n} = \mathbf{n} + H\mathbf{k} = dr - f\mathbf{k}$ being its two double WANDs. Moreover, the background becomes conformally flat and hence maximally symmetric when the base space curvature is fine-tuned to $K = 4\lambda l^2$, in which case the full geometry reduces to a KS spacetime.\footnote{See \cite{Ort_Srini} for a more detailed analysis of the intersection of (A)dS-Taub-NUT spacetimes with the KS class in $n \geq 4$.} We thus find that the Weyl alignment properties of the full geometry and those of the background agree with the results of Sections \ref{Weyl_types}, \ref{subsection_remarks_alge_types} (see Table \ref{table_Weyl_types}). As for the Ricci types, the background and the full geometry, being Einstein spacetimes, are of type $D(O)$, consistent with Sections \ref{Ricci_types}, \ref{subsection_remarks_alge_types} (cf. propositions therein and Table \ref{table_Ricci_types}).

\subsubsection{Remarks on KS double copy}\label{double_copy}
 Before proceeding with the remarks on KS double copy for (A)dS-Taub-NUT spacetimes, we first make a general observation on Maxwell's equations in GKS spacetimes. It is easy to check that the metric determinant, $ g\equiv \det g_{ab}$, is invariant under a GKS transformation, i.e, $g=\Bar g$.
Moreover, for a vector potential, up to a gauge, of the form 
\begin{align}
  \mathbf{A}=\alpha \mathbf{k},\label{aligned_vector}  
\end{align} 
with $\mathbf{k}$ being the geodesic KS vector and $\alpha$ a scalar function, it can be shown that $g^{ac}g^{bd}F_{cd}=\Bar g^{ac}\Bar g^{bd}F_{cd}$. Therefore, the Maxwell equations for such a vector potential are invariant under a GKS transformation, i.e.,
\begin{align}
    (\sqrt{-g}g^{ac}g^{bd}F_{cd})_{,a}=0\iff (\sqrt{-\bar g}\bar g^{ac} \Bar g^{bd}F_{cd})_{,a}=0 , \label{Maxw_independ_H}
\end{align}
and hence the solution $\alpha$ is independent of $H$.\footnote{This property was already noted in \cite{Myers:1986un} in the context of KS spacetimes (cf. also \cite{Ort_Srini}).}

The invariance discussed above plays a crucial role in the Kerr-Schild double copy, as we now illustrate for (A)dS-Taub-NUT spacetimes. It can be easily checked that the (A)dS-Taub-NUT metric \eqref{GKS_NUT} admits a test-Maxwell solution of the form \eqref{aligned_vector}, with $\alpha = \frac{Q}{r^2+l^2}$ for an arbitrary constant $Q$, and $\mathbf{k}$ being the KS vector defined in \eqref{KS_vector_NUT}. Since $\mathbf{k}$ is geodesic, it follows from \eqref{Maxw_independ_H} that the test solution $\alpha$ is also a test solution in the background geometry \eqref{NUt_GKS_bg}. Therefore, upon choosing the test charge to be $Q = -M$, we have $\mathbf{A} = H\mathbf{k}$. As was already noted in \cite{Bahjat_double_copy_curved_bg_1}, this thus forms an example of the Kerr-Schild double copy in curved spacetime \cite{Bahjat_double_copy_curved_bg_1,double_copy-curved_3,Prabhu_double_copy_curved,doublecopy_curved_4}, where the test Maxwell solution in the background $\mathbf{\Bar{g}}$ is mapped to the GKS solution of the vacuum Einstein equations around the same background. In particular, \cite{Bahjat_double_copy_curved_bg_1} refer to it as the type B double copy.

Let us note that the notion of the ``type B'' double copy is also discussed in \cite{Srini_Kubiz_Hassaine} for the more general charged (A)dS-Kerr-NUT black holes in $n=4$, where, instead of the $M=0$ limit considered here, they use the extremal limit of the full metric as the background for the GKS form. The ``type B'' double copy for (A)dS-Kerr-NUT \cite{Bahjat_double_copy_curved_bg_1,Srini_Kubiz_Hassaine} makes use of the GKS structure of the metric, in contrast to the approach of \cite{Luna:2015paa}, which utilizes the double KS form \cite{PLEBANSKI197698}. The advantage of using the GKS form is that one does not have to switch to an unphysical signature, unlike in the double KS form. However, the downside is that the vector potential \eqref{aligned_vector} misses the (magnetic charge) contribution from the second ``null'' vector of the double KS form. With contributions from both null vectors, one can interpret the (A)dS-Kerr-NUT metric as being ``constructed'' via double copy around a  maximally symmetric background. In the GKS/``type B'' perspective of the double copy, the ``information'' associated with the missing magnetic part must be provided ad hoc in the form of the curved background (which, in our case, is the $M=0$ limit of the full geometry).

\section{Application of the results to the case of geodesic, expanding, twistfree, and shearfree $\mathbf{k}$} \label{RT_GKS_section}
When $\mathbf{k}$ of a GKS spacetime is geodesic, expanding, shearfree, and twistfree, the spacetime necessarily also belongs to the Robinson-Trautman class (cf. Appendix \ref{RTspacetimes_append} and references therein). However, given a GKS-Robinson-Trautman spacetime (i.e., a spacetime in the intersection of the GKS and Robinson-Trautman classes), it is a priori not obvious whether the KS vector coincides with a `privileged' Robinson-Trautman null direction which is geodesic, expanding, shearfree, and twistfree. To clarify this, we show in the following proposition that, in any GKS-Robinson-Trautman spacetime, the KS vector must necessarily be shearfree and twistfree. 
\begin{proposition}
\label{GKS_RT_prop}
Let $\mathbf{g}$ be a GKS-Robinson-Trautman spacetime with KS vector $\mathbf{k}$. Suppose that a null direction $\mathbf{l}$, distinct from the KS direction, defines the geodesic, expanding, shearfree, and twistfree Robinson-Trautman null congruence. Then, $\mathbf{k}$ must also be shearfree and twistfree.
\end{proposition}
\noindent \textbf{Proof.}
We follow exactly the same line of argument as in Propositions \ref{prop_l_AND} and \ref{prop_multiple_l_AND}. Because $\mathbf{l} \neq \mathbf{k}$, after appropriately normalizing $\mathbf{l}$, we choose it to be the second null vector, $\mathbf{n}$, of the null frame. Likewise, we choose $\Bar{\mathbf n}$ according to equation \eqref{n_bg}. The shear and twist of $\mathbf{n}$ are zero by assumption, and hence, upon using equation \eqref{Nij_GKS}, we have
\begin{align}
    &N_{[ij]}= \Bar{N}_{[ij]}- HA_{ij}=0, \\
    & (N_{(ij)}-\frac{\delta_{ij}}{n-2}N_{kk})= (\Bar N_{(ij)}-\frac{\delta_{ij}}{n-2}\Bar N_{kk})+ H \sigma_{ij}=0.
\end{align}
 As before, for the consistency of the above relations for all values of the parameters ``$\mu_\alpha$'' appearing in $H$, the $H$-dependent and independent terms must vanish separately. Therefore, $A_{ij} = 0 = \sigma_{ij}$, proving the claim of the proposition.\\

Thanks to the above proposition, under the assumption that $\mathbf{k}$ is geodesic,\footnote{Proposition \ref{GKS_RT_prop} does not make any conclusion about the expansion of $\mathbf{k}$ and hence, in general one has to assume that $\mathbf{k}$, in addition to being geodesic, is also expanding. However, we make a stronger (implicit) assumption that the spacetime is strictly Robinson-Trautman, and has no intersection with the Kundt class (cf. \cite{Griffiths_Podolský_2009,Podolsky:2008ec} for the definition of Kundt spacetimes). Therefore, the geodesicity assumption, along with Proposition \ref{GKS_RT_prop} suffices to guarantee that $\mathbf{k}$ is expanding.} one can conclude that it always forms a `privileged' null direction of GKS-Robinson-Trautman spacetimes. Therefore, GKS spacetimes with a geodesic, expanding, shearfree, and twistfree $\mathbf{k}$ are equivalent to GKS-Robinson-Trautman spacetimes with a geodesic $\mathbf{k}$. Their metric, in the Robinson-Trautman form \eqref{RT_general} adapted to $\mathbf{k}$, which naturally also coincides with the GKS form \eqref{GKS}, is given by\footnote{Hereafter, we will refer to these spacetimes simply as GKS-Robinson-Trautman, with the implicit assumption that $\mathbf{k}$ is geodesic. }
\begin{align} 
&\mathbf{g}= g_{ij}(r,u,x) dx^{i}dx^{j} + 2g_{ui}(r,u,x)dudx^{i}- 2dudr - 2f(r,u,x)du^2 - 2H(r,u,x) \mathbf{k}\otimes \mathbf{k}, \label{RT_GKS}\\
& \mathbf{k}=-du, \quad f(r,u,x)= g_{uu}(r,u,x)-H(r,u,x),\label{KS_RT_GKS}
\end{align}
where the coordinates $(u, r, x)$ adapted to $\mathbf{k}$ are as defined in Appendix \ref{RTspacetimes_append}. The background of the GKS form \eqref{RT_GKS}, clearly also forms a Robinson-Trautman geometry with $\mathbf{k}$ being a `privileged' Robinson-Trautman direction, which is consistent with the fact that the geodesicity and optical properties of $\mathbf{k}$ are invariant under a GKS transformation.

Let us note that for any Robinson-Trautman spacetime (not necessarily intersecting with GKS), a `privileged' Robinson-Trautman vector, by virtue of its shearfree and twistfree conditions, automatically satisfies the optical constraint \eqref{opt_constraint}. In addition, if it forms a Riemann (or equivalently, a Ricci) AND, then by choosing an affine parametrization $r$ for the corresponding geodesic, its expansion becomes $\theta=\frac{1}{r}$ (see Proposition \ref{prop_RT_AND}). For the case of GKS-Robinson-Trautman spacetimes, this is consistent with the canonical form of the optical matrix discussed in the previous section. Another noteworthy feature of GKS-Robinson-Trautman spacetimes is that $\mathbf{k}$ must necessarily also be a WAND (cf. Appendix \ref{RTspacetimes_append} and references therein). Hence, the results of Section \ref{Weyl_types}, \ref{subsection_remarks_alge_types} imply that $\mathbf{k}$ is also a WAND for the background geometry. This can also be independently verified by applying the general results of Robinson-Trautman spacetimes (cf. again Appendix \ref{RTspacetimes_append}) to the background Robinson-Trautman geometry.
\subsection{All vacuum GKS-Robinson-Trautman spacetimes in $n>4$}\label{vacu_RT}
We apply our results to obtain the complete family of GKS spacetimes in $n>4$ with a geodesic, expanding, twistfree, and shearfree $\mathbf{k}$ that satisfies the vacuum Einstein equations \eqref{vacuum_Einstein}. Since these solutions form a subset of the full family of vacuum Robinson-Trautman solutions obtained in \cite{Podolsky_Ort_RT} (cf. also \cite{Ortaggio_RT}), our discussion will largely rely on the results presented there.

We start with the GKS-Robinson-Trautman metric ansatz \eqref{RT_GKS} and impose the vacuum Einstein equations \eqref{vacuum_Einstein}. We assume that any dependence on the cosmological constant is fully absorbed into the background metric, so that the background also satisfies the vacuum Einstein equations with the same cosmological constant, i.e., $\Bar{R}_{ab}=\frac{2\Lambda}{n-2}\Bar{g}_{ab}$. We work in the null frame defined by \eqref{frame_RT}, with the metric function $g_{uu}$ given by \eqref{KS_RT_GKS} and the corresponding null frame for the background defined by \eqref{n_bg}, \eqref{frame_background}. Without loss of generality, we also assume that the spatial frame vectors $\mathbf{m}_{(i)}$ are parallel transported along $\mathbf{k}$, which simplifies the coefficients $m^{i}_{\hspace{1mm}j}(r,u,x)$ defined in \eqref{coframe_RT} to \eqref{gij_RT}. Using the Einstein equations for both the background and the full spacetime, we find that the $H$-dependent terms in each of the frame components of the Ricci tensor \eqref{R00_GKS}-\eqref{R11_GKS_1} must vanish on-shell. Before discussing the details of the equations for $H$, which will be our primary focus, we simplify the metric ansatz \eqref{RT_GKS} using the $H$-independent parts of \eqref{R00_GKS}-\eqref{R11_GKS_1}, i.e., the Einstein equations for the background.

To solve the $\Bar{R}_{00}=0$ equation, we make use of the results in Appendix \ref{RTspacetimes_append}, while for the remaining $H$-independent equations, we take a shortcut and directly use the results of \cite{Podolsky_Ort_RT} (cf. also \cite{Ortaggio_RT}). From Proposition \ref{prop_RT_AND}, we see that $\Bar R_{00}=0$ leads to $\theta=\frac{1}{r}$ and hence $g_{ij}=r^2 h_{ij}(u,x)$ (see equation \eqref{gij_simplifi_RT}). Additionally, Proposition \ref{prop_RT_AND} also implies that $\mathbf{k}$ is a Riemann AND of the background (and hence , from equation \eqref{R0i0j_GKS}, also of the full geometry). Upon using the remaining Einstein equations for the background, the metric ansatz \eqref{RT_GKS} simplifies to the following form \cite{Podolsky_Ort_RT}
\begin{align}
     \mathbf{g}= r^2 h_{ij}(u,x)dx^i dx^j - 2dudr - 2f(r,u,x) du^2 - 2H(r, u, x) \mathbf{k}\otimes \mathbf{k}.\label{RT_GKS_met_1}
    \end{align}
The transverse spatial metric $h_{ij}(u,x)$ is given by
\begin{align}
& h_{ij}(u,x)= P^{-2}(u,x) \gamma_{ij}(x), \quad \det \gamma_{ij}=1, \label{simpli_hij_RT}
\end{align}
with its Ricci tensor $\mathcal{R}_{ij}$ and the Ricci scalar $\mathcal{R}=h^{ij}\mathcal{R}_{ij}$ satisfying
\begin{align}
    \mathcal{R}_{ij}= \frac{\mathcal{R}}{n-2}h_{ij}, \label{trans_Einste_RT}
\end{align}
where $\mathcal{R}$ is a function of $u$ alone \cite{Podolsky_Ort_RT}. The metric function $f$ reads \cite{Podolsky_Ort_RT,Ortaggio_RT}
\begin{align}
    & 2f(r,u,x)=\left(\frac{\mathcal{R}(u)}{(n-2)(n-3)}-2r (\ln P(u,x))_{,u} - \frac{2\Lambda}{(n-2)(n-1)}r^2 \right), \label{f_for_vacu_RT}
\end{align}
where $\Lambda$ is the cosmological constant. Moreover, the spatial frame covectors simplify to $\mathbf{m}^{(i)}=\frac{r}{P}w^{i}_{\hspace{1mm}j}(x)dx^j$, or equivalently $m^{i}_{\hspace{1mm}j}= \frac{r}{P}w^{i}_{\hspace{1mm}j}(x)$, with the coefficients $w^{i}_{\hspace{1mm}j}$ satisfying $\delta_{ij} w^{i}_{\hspace{1mm}k} w^{j}_{\hspace{1mm}l}=\gamma_{kl}$. Furthermore, using the simplifications to the metric and the frame vectors, we obtain the following non-vanishing Ricci rotation coefficients from \eqref{Ricci_rot_Li1_RT}- \eqref{Ricci_rot_Mijk_RT}
\begin{align}  
&L_{ij}=S_{ij}=\theta \delta_{ij},  \quad N_{(ij)}= \frac{P_{,u}}{P}\delta_{ij} + (f+H)S_{ij},  \quad L_{11}= -(f_{,r}+H_{,r}),\label{Ricci_Rot_RT_vacu_1} \\
&N_{i1}= (f_{,j}+H_{,j})m_{i}^{\hspace{1mm}j},\quad  \overset{i}{M}_{ql}=m^{i}_{\hspace{1mm}j,k}m_{q}^{\hspace{1mm}j}m_{l}^{\hspace{1mm}k}  -\Gamma^{j}_{\hspace{1mm}kw}(r, u, x) m_{l}^{\hspace{1mm}k}m_{q}^{\hspace{1mm}w}m^{i}_{\hspace{1mm}j}, \label{Ricci_Rot_RT_vacu_2}
\end{align}
where $\Gamma^{j}_{\hspace{1mm}kw}(r, u, x)$ is as defined in \eqref{christoffel_RT_base_space}.

Let us now turn to the equations for $H$. The components \eqref{R00_GKS} are $H$-independent, while the $R_{ij}$ component \eqref{Rij_GKS} leads to equation \eqref{vacuum_Rij_eqn}, which fixes the $r$-dependence of $H$ as
\begin{align}
H= -\frac{M}{r^{n-3}}, \label{H_for_RT_GKS_1}
\end{align}
where $M=M(u,x)$ is an $r$-independent integration function. The solution \eqref{H_for_RT_GKS_1} can be obtained directly from \eqref{r-dpend_H_Einstein} by imposing the twistfree and shearfree conditions, i.e., $q=0, m=n-2$. Moreover, as already mentioned in Section \ref{GKS_in_GR_subsection}, the solution \eqref{H_for_RT_GKS_1} also automatically satisfies equation \eqref{D^2H_eqn_vacuum_GKS} (with $\omega=0$), resulting from the $R_{01}$ component \eqref{R01_GKS}. Therefore, it only remains to solve the equations for $H$ coming from the $R_{1i}$  and $R_{11}$ components \eqref{R1i_GKS}, \eqref{R11_GKS_1}. By using the Ricci rotation coefficients \eqref{Ricci_Rot_RT_vacu_1}, \eqref{Ricci_Rot_RT_vacu_2},  and the background Einstein equations, namely $\Bar{R}_{00}=0=\Bar{R}_{0i}$, we obtain 
\begin{align}
    &2H \Bar R_{010i} -\delta_i(DH)+ 2L_{ij}\delta_{j}H - L_{jj}\delta_{i}H=0, \label{H_eqn_1} \\
 & \delta_{i}\delta_{i}H +\overset{i}{M}_{kk}\delta_{i}H+\Bar N_{ii}DH-S_{ii}\Bar \Delta H  -2H\Big( \Bar \Delta S_{ii}+L_{ki} \Bar N_{ki}\Big)=0.   \label{H_eqn_2}
\end{align}
Using \eqref{Ricci_Rot_RT_vacu_1}, \eqref{Ricci_Rot_RT_vacu_2}, and the Ricci identity $(11b)$ of \cite{Ricci_in_HD}, it follows that $\Bar R_{010i}=0$. Therefore, upon substituting the expressions for $L_{ij}$ and $H$ from \eqref{Ricci_Rot_RT_vacu_1} and \eqref{H_for_RT_GKS_1} into equation \eqref{H_eqn_1}, we obtain
\begin{align}
    \delta_i M=0. \label{H_eqn3}
\end{align}
Finally, using $\Bar \Delta= f \nabla_r -\nabla_u$ (cf. \eqref{n_bg}, \eqref{frame_background}, \eqref{frame_RT}), the expression for $\Bar N_{ij}$ deduced from \eqref{Ricci_Rot_RT_vacu_2}, and the results \eqref{H_for_RT_GKS_1} and \eqref{H_eqn3}, one can simplify \eqref{H_eqn_2} as
\begin{align}
    (n-1)M (\ln P)_{,u}- M_{,u}=0.\label{H_eqn4}
\end{align}
As discussed in \cite{Podolsky_Ort_RT}, we now use the following reparametrization transformation, which leaves the metric \eqref{RT_GKS_met_1} unchanged
\begin{align}
&\Tilde u = \Tilde u(u),\hspace{1mm} \Tilde r= r/\Dot{\Tilde u}, \quad  P= \Tilde P \Dot {\Tilde u},\hspace{1mm}  \mathcal{R}= \Tilde{\mathcal{R}} \Dot {\Tilde{u}}^2,\hspace{1mm} M= \Tilde M \Tilde u^{n-1},   \label{repara_RT_coord}
\end{align}
where $\Dot{\Tilde u}=\frac{d \Tilde u}{du}$. Since $\Tilde{u}$ is an arbitrary function of $u$, we can set $\Tilde M$ to be a constant by absorbing any $u$-dependence it might have into the definition of the new coordinate $\Tilde u$. Therefore, equation \eqref{H_eqn4} in the new coordinates (after dropping all the tildes) reads
\begin{align}
    M \ln(P)_{,u}=0. \label{H_eqn_4_RT}
\end{align}
The definition of GKS spacetimes requires that $H \neq 0$, which for our choice of $f$ and $H$ (see \eqref{f_for_vacu_RT}, \eqref{H_for_RT_GKS_1}), implies that $M \neq 0$. However, we notice that for $\Lambda \neq 0$, we could absorb the $\Lambda$ term of $f$ into the definition of $H$, i.e., $f \rightarrow f + \frac{\Lambda}{(n-2)(n-1)}r^2$, $H \rightarrow H - \frac{\Lambda}{(n-2)(n-1)}r^2$. Such a redefinition leaves equation \eqref{H_eqn_4_RT} unchanged but allows us to have $M = 0$ when $\Lambda \neq 0$. Therefore, we have two possibilities for solving equation \eqref{H_eqn_4_RT}: (i) $M \neq 0$ with arbitrary $\Lambda$, and (ii) $M = 0 \neq \Lambda$, leading to two branches of solutions, which we elaborate below.

\subsubsection*{Case (i): $M\neq 0$, $\Lambda$ arbitrary}
In this case, equation \eqref{H_eqn_4_RT} implies that $P = P(x)$, i.e., the transverse spatial metric $h_{ij}$ is independent of $u$, and hence $\mathcal{R}$ is a constant. Further, when $\mathcal{R}\neq0$, one can again use the reparametrization freedom \eqref{repara_RT_coord}, to rescale it as $\mathcal{R}=\pm (n-2)(n-3)$ \cite{Podolsky_Ort_RT}. Therefore, the solution in this case is given by \eqref{RT_GKS_met_1}, with $h_{ij}=h_{ij}(x)$ forming a Riemannian Einstein space (see \eqref{trans_Einste_RT}) and the metric function $g_{uu}=(f+H)$ given by \cite{Podolsky_Ort_RT,Ortaggio_RT}
\begin{align}
    & 2g_{uu}=2(f+ H)=K   -\frac{2M}{r^{n-3}} - \frac{2\Lambda}{(n-2)(n-1)}r^2, \label{guu_RT_M_nonzero}
\end{align}
where $K=0,\pm 1$ and $M$ is a constant. For $\Lambda \neq 0$, one can split $f$ and $H$ in three possible ways: (a) absorb $\Lambda$ term into $f$ and keep $M$ term in $H$, (b) absorb $M$ term into $f$ and keep $\Lambda$ term in $H$, and (c) keep both $\Lambda$ and $M$ terms in $H$. However, since in our case $\Lambda$ is arbitrary and can also be vanishing, it is natural to absorb $\Lambda$ term into the background and define the functions $f$ and $H$ as
\begin{align}
    2f=K- \frac{2\Lambda}{(n-2)(n-1)}r^2, \quad   H= -\frac{M}{r^{n-3}} . \label{H,f_for_RT_Mnonzero}
\end{align}
Let us analyze the algebraic properties of the solution. For the above choice of $f$ and $H$, the background and the full geometry are of Ricci type $D(O)$, with the geometries becoming Ricci flat for $\Lambda=0$. Thus, the combinations of Ricci types are consistent with Table \ref{table_Ricci_types}. Moreover, since we are dealing with Einstein spacetimes, any null direction is a double (quadruple) Ricci AND. Using the Ricci rotation coefficients \eqref{Ricci_Rot_RT_vacu_1}, \eqref{Ricci_Rot_RT_vacu_2}, and the Ricci identities \cite{Ricci_in_HD}, the non-vanishing frame components of the Weyl tensor can be found to be \cite{Podolsky_Ort_RT,Ortaggio_RT}
\begin{align}
 & C_{0101}=  -M\frac{(n-3)(n-2)}{r^{n-1}}, \quad C_{0i1j}= -M \frac{(n-3)}{r^{n-1}}\delta_{ij}, \label{Weyl_1_RT}\\
 &C_{ijkl}=\frac{1}{r^2}\left(\mathcal{R}_{\hat{i}\hat{j}\hat{k}\hat{l}} - 2\big(K -\frac{2M}{r^{n-3}}\big)\delta_{i[k}\delta_{l]j} \right),\label{Weyl_2_RT}
\end{align}
 where $\mathcal{R}_{\hat{i}\hat{j}\hat{k}\hat{l}}$ denote the frame components of the Riemann tensor of the transverse space taken along its orthonormal frame vectors $\mathbf{\hat m}_{(i)}= \hat m_{i}^{\hspace{1mm}j}\partial_{j}$ defined by $\hat m_{i}^{\hspace{1mm}j}=r  m_{i}^{\hspace{1mm}j}$.\footnote{Note that the coefficients $\hat m_{i}^{\hspace{1mm}j}$ satisfy $\hat m_{i}^{\hspace{1mm}k}\hat m_{j}^{\hspace{1mm}l}h_{kl}=\delta _{ij}$ or equivalently the inverse coefficients, $\hat m^{i}_{\hspace{1mm}l}=\frac{1}{r}m^{i}_{\hspace{1mm}l}$, satisfy $\hat m^{i}_{\hspace{1mm}k}\hat m^{j}_{\hspace{1mm}l}\delta_{kl}=h _{ij}$.} Therefore, the full geometry is of Weyl type $D$ with $\mathbf{k}$ and $\mathbf{n}= dr+(f+H)du$ being the two double WANDs \cite{Podolsky_Ort_RT,Ortaggio_RT}. The background on the other hand, has only the following non-zero Weyl component 
 \begin{align}
     \Bar C_{ijkl}=\frac{1}{r^2}\left(\mathcal{R}_{\hat{i}\hat{j}\hat{k}\hat{l}} - 2K \delta_{i[k}\delta_{l]j} \right), \label{Weyl_bg_RT_1}
 \end{align}
which can be obtained by setting $M=0$ in \eqref{Weyl_1_RT}, \eqref{Weyl_2_RT}. From \eqref{Weyl_bg_RT_1} we see that the background geometry is conformally flat iff the transverse metric $h_{ij}$ defines a space of constant curvature $K$ \cite{Podolsky_Ort_RT,Ortaggio_RT}. Therefore, the background is of Weyl type $D$ or $O$, with $\mathbf{k}$ and $\mathbf{\bar n}= dr + f du $ being the two double WANDs in the former case. In the case of a conformally flat background, we obtain the (A)dS-Schwarzschild–Kottler–Tangherlini black holes \cite{Tangherlini:1963bw}, which belong to the KS class \cite{HD_KS,Malek:2010mh}. All these aspects related to the Weyl types are consistent with Table \ref{table_Weyl_types}.\\

It is interesting to note that the full geometry, and hence, due to the geodesicity of $\mathbf{k}$, also the background (see \eqref{Maxw_independ_H}), admit a test Maxwell field solution given (up to a gauge) by\footnote{It is easy to see that \eqref{test_Max_RT} forms a test solution by setting the magnetic part to zero in the Maxwell field solution of the higher-dimensional charged Robinson-Trautman spacetimes \cite{Ort_Zof_Pod_2_from_RT}, and subsequently taking the weak-field limit of the matter coupling, i.e., $\kappa = 0$ (see \eqref{Einstein_eqns}). These test solutions, as well as the resultant KS double copy, were already noted in \cite{Ort_Srini} for the special case of Robinson-Trautman spacetimes intersecting with the KS class.}  
\begin{align}
    \mathbf{A}= \frac{Q}{r^{n-3}}\mathbf{k}. \label{test_Max_RT}
\end{align}
Hence, if we choose the test charge to be $Q = -M$, we get $\mathbf{A} = H\mathbf{k}$. Therefore, similar to the (A)dS-Taub-NUT example, this also fits into the notion of a ``type B'' double copy in a curved background \cite{Bahjat_double_copy_curved_bg_1}.

\subsubsection*{Case (ii): $M= 0 \neq \Lambda$}
In this case, equation \eqref{H_eqn_4_RT} identically vanishes, and hence one cannot conclude that $P$ is independent of $u$. However, as in the previous case, one could use the reparameterization freedom \eqref{repara_RT_coord} to rescale $\mathcal{R}$ as $\mathcal{R}=\pm (n-2)(n-3)$. Therefore, the solution in this case is given by \eqref{RT_GKS_met_1} with $h_{ij}=h_{ij}(u,x)$, defined by \eqref{simpli_hij_RT}, satisfying \eqref{trans_Einste_RT}. The metric function $g_{uu}=(f+H)$ is given by \cite{Podolsky_Ort_RT,Ortaggio_RT,Pravda_type_D} 
\begin{align}
& 2g_{uu}=2(f+H) = \left(K - 2r (\ln P(u,x)){,u}\right) - \frac{2\Lambda}{(n-2)(n-1)}r^2. \label{guu_fn_for_RT_case_2}
\end{align} 
Unlike the previous case, the GKS splitting here relies solely on $\Lambda$. Therefore, the only possible way to define $f$ and $H$ is
\begin{align}
&2f = \left(K - 2r (\ln P(u,x)){,u}\right),\quad H= - \frac{2\Lambda}{(n-2)(n-1)}r^2.  \label{H_fn_for_RT_case_2}
\end{align}
The background, defined by the limit $\Lambda=0$, is Ricci flat. The full geometry, on the other hand, being an Einstein spacetime with a non-zero $\Lambda$, is of Ricci type $D$. The only non-zero frame component of the Weyl tensor is \cite{Podolsky_Ort_RT,Ortaggio_RT} 
\begin{align} 
C_{ijkl} = \frac{1}{r^2}\left(\mathcal{R}_{\hat{i}\hat{j}\hat{k}\hat{l}} - 2K \delta_{i[k}\delta_{l]j} \right), \label{Weyl_case_2_RT} \end{align} 
where $\mathcal{R}_{\hat{i}\hat{j}\hat{k}\hat{l}}$ is as defined in the $M \neq 0$ case. Since \eqref{Weyl_case_2_RT} is independent of $\Lambda$, the background Weyl tensor is the same as that of the full geometry. Therefore, the background and the full geometry are both of Weyl type $D(O)$, with the geometries reducing to conformally flat when the transverse metric defines a space of constant curvature $K$. The multiple WANDs of the full geometry are given by $\mathbf{k}$ and $\mathbf{n}= dr + (f + H)du$ \cite{Podolsky_Ort_RT,Ortaggio_RT}, while those of the background are given by $\mathbf{k}$ and $\mathbf{n} = dr + fdu$. As in the previous case, the algebraic types are consistent with the results of Section \ref{GKS_with_geo_k}.\\

To summarize this subsection, we have identified the subset of the higher-dimensional vacuum Robinson-Trautman solutions obtained in \cite{Podolsky_Ort_RT} that forms the complete family of vacuum GKS spacetimes in $n>4$ with a geodesic, expanding, shearfree, and twistfree $\mathbf{k}$. The solutions consist of the two branches above, and the algebraic types of the spacetimes are consistent with Section \ref{GKS_with_geo_k}. We also note that, for these solutions, $\mathbf{k}$ is a Riemann AND, and $R_{ij} - \Bar{R}_{ij} \propto S_{ij}$.\footnote{For the $M\neq0$ branch, since $\Lambda$ is absorbed into the background we have $R_{ij} - \Bar{R}_{ij}=0$, and hence Proposition \ref{prop_opt_contr_2} can equivalently be applied. On the other  hand, the $M=0$ branch satisfies $\Bar{R}_{ij}=0$, $R_{ij}=\frac{2\Lambda}{n-2}\delta_{ij}$.} By Proposition \ref{prop_opt_contr}, this is consistent with the fact that the optical matrix, $L_{ij} = S_{ij} = \theta \delta_{ij}$, satisfies the optical constraint \eqref{opt_constraint}.\footnote{It is worth noting that the full family of higher-dimensional vacuum Robinson-Trautman spacetimes consists of the two branches of solutions in the GKS-Robinson-Trautman class, along with the $\Lambda =0$ limit of the $M=0$ branch \cite{Podolsky_Ort_RT,Ortaggio_RT,Pravda_type_D}.} 
\section{Discussion}
\label{discussion}
In this paper, we have studied GKS spacetimes in a theory-independent setting. Restricting to the case of geodesic $\mathbf{k}$, we derived several geometric and algebraic properties of GKS spacetimes. Regarding the algebraic features, we specifically obtained all kinematically allowed Ricci and Weyl types of GKS spacetimes. Furthermore, we showed that, unlike KS spacetimes, GKS spacetimes need not be algebraically special; rather, their algebraic types are constrained by those of the background geometry. Moreover, in contrast to KS spacetimes \cite{HD_KS,Malek:2010mh}, the KS vector of GKS spacetimes need not form a WAND. An interesting example of these features is provided by the Schwarzschild-Melvin black holes (cf. Sections \ref{matter-matter}, \ref{subsection_remarks_alge_types}), which are of Weyl type $I$, with $\mathbf{k}$ not being a WAND.

For the assumption of expanding $\mathbf{k}$, we also obtained the conditions under which it satisfies the optical constraint. Noting that the optical constraint in the context of KS spacetimes \cite{HD_KS,Malek:2010mh} helped classify all higher-dimensional electrovacuum KS spacetimes (with $\mathbf{A}\propto \mathbf{k}$ and expanding $\mathbf{k})$ \cite{Ort_Srini} , it would be interesting to see if the same can be repeated for GKS spacetimes satisfying the optical constraint. We also briefly discussed the GKS structure of the five-dimensional charged rotating black hole solutions of CCLP \cite{CCLP}. Recently, these solutions were generalized to all odd dimensions \cite{Deshpande:2024vbn}. Therefore, it would also be interesting to see if these generalized solutions also possess a GKS structure and, in the affirmative case, if it could provide new insights.

Our study has omitted the analysis of the possible constraints on the optical matrix for the case of non-expanding $\mathbf{k}$. The results of \cite{Ortaggio_Jose_Jakub,HD_KS,Malek:2010mh} show that Einstein-KS spacetimes with a non-expanding $\mathbf{k}$ are equivalent to type $N$ Kundt spacetimes. In contrast, for the case of Einstein-GKS spacetimes, it can be checked that $\mathbf{k}$ being non-expanding does not automatically lead to Kundt. Moreover, even the specific case of Einstein-Kundt in the GKS class can be less special than type $N$ \cite{Kuchynka}. Therefore, it would be worth exploring the non-expanding case in detail and investigating these aspects further. In this context, it would also be interesting to see if the double copy results of \cite{Ort_Pra_Pra_KS_double}, derived for vacuum KS spacetimes with a non-expanding $\mathbf{k}$, can be extended to the more general Einstein-GKS class.

While analyzing the GKS structure of all the examples, we have defined the GKS splitting in terms of a family of finite number of parameters. Particularly, in most examples the GKS splitting is done with respect to the mass parameter $M$, with the background being the $M=0$ limit of the full metric. However, it is interesting to note that when the metric involves multiple parameters, one could define an infinite family of GKS splitting, which we briefly illustrate below for the simple example of Kerr spacetimes.

The Kerr metric admits a standard KS form given by
\begin{align}
\mathbf{g}= \bet + \frac{2Mr}{r^2+a^2\cos^2\theta} \mathbf{k}\otimes \mathbf{k},
\end{align}
with the flat background $\bet$ and the KS vector given respectively by equations \eqref{flatbg} and \eqref{KS_vector_kerr}. The metric has two parameters, $M$ and $a$, and can be rewritten as
\begin{align}
\mathbf{g}&= \Big(\bet +\frac{2f(a)r}{r^2+a^2\cos^2\theta} \mathbf{k}\otimes \mathbf{k}\Big) + \frac{2(M-f(a))r}{r^2+a^2\cos^2\theta} \mathbf{k}\otimes \mathbf{k} \nonumber = \mathbf{\Bar{g}}_{\tiny{ f}} + \frac{2(M-f(a))r}{r^2+a^2\cos^2\theta} \mathbf{k}\otimes \mathbf{k},
\end{align}
where $f(a)$ is an arbitrary function of $a$. The background, $\mathbf{\Bar{g}}_{\tiny{ f}}$, in the above `modified' form, is the $M=f(a)$ limit of the full metric, and the GKS splitting `parameter' is now given by $\mu\equiv M-f(a)$. Since there are infinitely many possible ways to choose the function $f$, one could think of this as an infinite family of GKS splittings. Although, we illustrated the idea for the Kerr case, as already mentioned it can be applied to any GKS metric having multiple parameters.

An interesting special choice for the function $f$ in the context of black hole solutions admitting a GKS form is $f=M_{\tiny{extremal}}$. For instance, in the case of Kerr, the extremal mass is given by $M_{\tiny{extremal}}=a$. This choice was explored in \cite{Srini_Kubiz_Hassaine} for various black hole metrics (including the CCLP example \cite{CCLP}) where the corresponding GKS form was referred to as the extremal Kerr-Schild form.

\section*{Acknowledgments}
I am grateful to Marcello Ortaggio for suggesting this project and for many valuable discussions, as well as for reading several versions of the draft and providing helpful comments for improvement. I would also like to thank David Kubiz\v n\'ak for useful discussions. I thank Tomáš Málek for discussions on extended Kerr-Schild spacetimes and for sharing his Mathematica notebook related to the topic. I acknowledge the use of xAct Mathematica packages \cite{xAct} for performing some of the computations. This work is supported by the GAČR 25-15544S grant from the Czech Science Foundation, the Charles University Research Center Grant No. UNCE24/SCI/016, and the research plan RVO 67985840 of the Institute of Mathematics, Czech Academy of Sciences.

\renewcommand{\thesection}{\Alph{section}}
\setcounter{section}{0}

\renewcommand{\theequation}{{\thesection}\arabic{equation}}

\section{Further examples of GKS spacetimes with expanding $\mathbf{k}$ }
\setcounter{equation}{0}
\label{Examples_apeendix}
In this appendix, we discuss some more examples of GKS spacetimes with a geodesic and expanding $\mathbf{k}$ and use them to further illustrate our general results (cf. Sections \ref{gen_props}, \ref{GKS_with_geo_k}, \ref{expanding ks}). We present one example for each of the following four categories of GKS transformation 
\begin{enumerate}
    \item \textbf{Vacuum-Vacuum}: The background and the full geometry are both solutions to the vacuum Einstein equations. 
    \item \textbf{Vacuum-Matter}: The background is vacuum, whereas the full geometry satisfies the Einstein equations with matter. \item \textbf{Matter-Matter}: Both the background and the full geometry are solutions to the Einstein equations with matter. \item  \textbf{GR-Beyond GR}: The background is a solution to the Einstein equations while the full geometry forms a solution in a theory beyond GR.
\end{enumerate}
 Let us note that among the examples presented in the main text, the four-dimensional (A)dS-Taub-NUT spacetimes (cf. Section \ref{NUT_4d}) and the vacuum GKS-Robinson-Trautman (cf. Section \ref{vacu_RT}) belong to the Vacuum-Vacuum category, whereas the CCLP solution (cf. Sections \ref{subsection_remarks_alge_types}, \ref{GKS_in_GR_subsection}) belongs to the Matter-Matter category of the GKS transformation.
\subsection{Kerr-NUT-(A)dS spacetimes in $n\geq 4$ (Vacuum-Vacuum)} \label{Kerr-NUT_section}
 The Kerr-NUT-(A)dS solutions of \cite{Chen_Pope_Kerr_NUT_multi} provide the NUT generalization of rotating black holes \cite{Myers:1986un,Hawking:1998kw,Klemm:1997ea,Gibbons_rot_HD_cosmol} in any dimensions. The Kerr-NUT-(A)dS spacetimes, in the Euclidean signature, admits the following multi-KS form \cite{Kerr_NUT_multi_KS} (cf. also \cite{Frolov:2017kze}) 
\begin{align}
   & \mathbf{g}= \mathbf{g}'- 2\sum_{\nu=1}^{k} \frac{b_\nu x_{\nu}^{1-\varepsilon}}{U_\nu}\mathbf{v}^{\nu}\mathbf{v}^{\nu}, \label{multi-KS_NUT}
\end{align}
with 
\begin{align}
  &   \mathbf{g}'=\sum_{\rho=1}^{k} \frac{X_{\rho}}{U_{\rho}} \mathbf{v} ^{\rho}\mathbf{v}^{\rho}-2i \sum_{\rho=1}^{k} \mathbf{v}^{\rho}dx_{\rho} + \varepsilon  \frac{c}{A^{(k)}} (\sum_{w=0}^{k} A^{(w)}d\psi_{w})^2, \\
    &U_{\nu}= \prod_{\substack{\rho=1\\\rho\neq \nu}}^{k}(x_{\rho}^2-x_{\nu}^2), \quad \mathbf{v} ^{\nu}= \sum_{w=0}^{k-1} A_{\nu}^{(w)}d\psi_{w}+ \frac{i U_{\nu}}{X_{\nu}}dx_{\nu},
\end{align}
 where $n=2k+ \varepsilon$ with $\varepsilon=0$ for even and $\varepsilon=1$ for odd $n$, and  
 $(x_1,x_2,\dots, x_k)$, $(\psi_1,\psi_2,\dots, \psi_{k+\varepsilon})$ are the $2k+\varepsilon$ coordinates of the metric. The parameters $b_\nu$'s are related to the $(k-1)$ NUT charges and the mass. The parameter $c_k$ is related to the cosmological constant as $c_k = \frac{2(-1)^k \Lambda}{(n-1)(n-2)}$. The parameters $(c_1,\dots,c_{k-1})$ correspond to the $(k-1)$ black hole rotation parameters in even dimensions $n=2k$, and $(c_1,\dots,c_{k-1},c)$ correspond to those in odd dimensions $n=2k+1$. All the parameters in the metric are real. The definitions of the other functions appearing in the above multi-KS form are as given in \cite{Frolov:2017kze} and are independent of the mass and NUT parameters.\footnote{We are using the multi-KS form given by equations $(4.75)$ and $(4.76)$ of \cite{Frolov:2017kze} (with minor typo corrections applied to the version available while writing this article). The coordinates $(\psi_1,\psi_2,\dots, \psi_{k+\varepsilon})$, the functions $X_\rho$, the $1$-forms $\mathbf{v}^\nu$, and the metric $\mathbf{g}'$ correspond to $(\mathring{\psi}_1, \mathring{\psi}_2,\dots, \mathring{\psi}_{k+\varepsilon})$, $\mathring{X}_\rho$, $\bmu ^\nu$, and $\mathring{\mathbf{g}}$ of \cite{Frolov:2017kze}, respectively.}

It can be seen that $\mathbf{v}^\rho$ are null, but complex $1$-forms. Likewise, the background, $\mathbf{g}'$, of the multi-KS form is also complex. However, their combination as given by the multi-KS form results in the full metric being real \cite{Frolov:2017kze}. The Euclidean metric \eqref{multi-KS_NUT} can be brought to the Lorentzian signature by a Wick rotation of the coordinate $x_k$ and the parameter $b_k$ as $ x_k=ir$ and $b_k=(i)^{1+\varepsilon} M$, with $M$ identified as the mass parameter. Due to the Wick rotation, the null $1$-form $\mathbf{v}^k$ becomes real, which we identify with the KS vector $\mathbf{k}$. Therefore, the (A)dS-Kerr-NUT metric in the Lorentzian signature can be cast into the GKS form \eqref{GKS} with
\begin{align}
    &\mathbf{\Bar g}= \mathbf{g}'- 2\sum_{\nu=1}^{k-1} \frac{b_\nu x_{\nu}^{1-\varepsilon}}{U_\nu}\mathbf{v}^{\nu}\mathbf{v}^{\nu}, \quad \mathbf{k}= \sum_{w=0}^{k-1} A_{k}^{(w)}d\psi_{w}- \frac{ U_{k}}{X_{k}}dr, \quad H=-\frac{M r^{1-\varepsilon}}{U_k},  \label{Ks_vector_Kerr-NUT}
\end{align}
where $U_k= \prod_{\rho=1}^{k-1}(x_{\rho}^2+r^2)$.

Since the cosmological constant is absorbed into the background and the function $H$ depends only on the mass parameter, we have $R_{ab} - \Bar{R}_{ab} = 0$. Hence, not only is $\mathbf{k}$ geodesic (cf. proposition \ref{prop_geod}), but the example also aligns with Section \ref{GKS_in_GR_subsection}. Moreover, it was shown in \cite{Pravda_type_D} that $\mathbf{k}$ satisfies the optical constraint. Therefore, from Proposition \ref{prop_opt_contr_2} (see also Remark \ref{remark_optic_contr_2}), $\mathbf{k}$ must be a WAND. It indeed turns out that $\mathbf{k}$ is a double WAND \cite{Hamamoto:2006zf}. We also see that $H$ matches the form given in \eqref{r-dpend_H_Einstein} with $m=2q+\varepsilon$ and $q=k-1$.

It was shown in \cite{Hamamoto:2006zf} that the spacetime is of Weyl type $D$ (cf. also \cite{Pravda_type_D}), with the other double WAND being $\mathbf{n}=\Big(\frac{X_k}{2U_k}\left(\sum_{w=0}^{k-1} A_{k}^{(w)}d\psi_{w}+ \frac{U_{k}}{X_{k}}dr\right)+ \frac{Mr^{1-\varepsilon}}{U_k}\mathbf{k}\Big)$. Since the background is the $M=0$ limit of the full geometry, it must therefore be of Weyl type $D$ or $O$,\footnote{The four-dimensional (A)dS-Taub-NUT spacetimes intersect with the Kerr-NUT-(A)dS class, under a certain limit of the latter. As shown in Section \ref{NUT_4d}, fine-tuning the base curvature can make the background of the (A)dS-Taub-NUT metrics conformally flat. Similar but more sophisticated conditions exist for higher-dimensional (A)dS-Taub-NUT spacetimes \cite{Ort_Srini}, which also have a non-trivial intersection with the Kerr-NUT-(A)dS class \cite{6D_MP_uniqueness}. Therefore, the subset of Kerr-NUT-(A)dS spacetimes with a conformally flat GKS background is non-empty.} with the double WANDs being $\mathbf{k}$ and $\mathbf{\Bar{n}}=\mathbf{n}|_{M=0}$, which is consistent with Sections \ref{Weyl_types}, \ref{subsection_remarks_alge_types}.

\subsection{Generalized Vaidya spacetimes in $n\geq4$ (Vacuum-Matter)}\label{vacu_matter_section}

Generalization of Vaidya metrics to $n\geq4$ belong to the GKS-Robinson-Trautman class (cf. Section \ref{RT_GKS_section}) \cite{Podolsky_Ort_RT} and are given by equation \eqref{RT_GKS_met_1}, with\footnote{Analogous to this example, one can also identify subsets of Robinson-Trautman spacetimes with a 2-form \cite{Ort_Zof_Pod_2_from_RT} or, more generally, a $p$-form electromagnetic field \cite{Ortaggio:RT_pform}. These would also form examples of a Vacuum-Matter GKS transformation as well as the GKS-Robinson-Trautman spacetimes.
}
\begin{align}
    2(f(r)+H)=\left(K - \frac{2\Lambda}{(n-2)(n-1)}r^2 -\frac{2M(u)}{r^{n-3}} \right), \label{f_for_Vaidya}
\end{align}
where $M(u)$ is the time dependent mass function and the other quantities appearing in \eqref{f_for_Vaidya} and the Vaidya metric remain unchanged from that of the vacuum GKS-Robinson-Trautman spacetimes (cf. Section \ref{vacu_RT}).
The generalized Vaidya metrics solve the Einstein equations \eqref{Einstein_eqns} with an energy-momentum tensor given by
   $ \mathbf{T}=\frac{n^2(u)}{r^{2-n}}du\otimes du$,
so that upon setting $\kappa=8\pi$ in the Einstein equations \eqref{Einstein_eqns}, one gets the relation $M_{,u}=\frac{8\pi n^2(u)}{n-2}$.

The metric can be cast into the GKS form with a background and the KS vector identical to those of the $M\neq0$ branch of the vacuum GKS-Robinson-Trautman, so that the function $H$ becomes
\begin{align}
    & H=- \frac{M(u) }{r^{n-3}}. \label{H_fn_Vaidya}
\end{align}
The full geometry is of Weyl type $D$, with the mWANDs given by $\mathbf{k}$ and $\mathbf{n} = \Big(dr + (f(r) + H)du\Big)$ \cite{Podolsky_Ort_RT}. The Weyl and Ricci types of the background, as well as the arguments related to consistency with Proposition \ref{prop_opt_contr} concerning the optical constraint, remain unchanged from Section \ref{vacu_RT}. The frame components of the Ricci tensor are given by
\begin{align}
    R_{01}=\frac{2\Lambda}{n-2}, \quad R_{ij}=\frac{2\Lambda}{n-2}\delta_{ij}, \quad R_{11}=T_{11}=\frac{n^2(u)}{r^{2-n}}.
\end{align}
Therefore, the full geometry is of Ricci type $II$, with $\mathbf{k}$ being the double Ricci AND, consistent with Table \ref{table_Ricci_types}.
\subsection{Schwarzschild-Melvin black hole solutions in $n=4$ (Matter-Matter)} \label{matter-matter}

The following solutions of the Einstein-Maxwell theory\footnote{The energy momentum-tensor of the Maxwell field is given by $T_{ab}=(F_{ac}F_{b}^{\hspace{1mm}c}-\frac{1}{4}g_{ab}F_{cd}F^{cd})$. In this example, the coupling constant $\kappa$ (in the Einstein equations \eqref{Einstein_eqns}) is set equal to $2$.} obtained by Ernst \cite{Ernst:1976mzr}, known as the Schwarzschild-Melvin black holes, represent a black hole immersed in a Bonnor-Melvin magnetic universe \cite{WBBonnor_1954,Melvin:1963qx,Melvin}\footnote{The metric \eqref{Melvin_BH_metric_2} is related to the standard form (cf. \cite{Griffiths_Podolský_2009}) by a coordinate transformation given by $du=dt - \mathcal{H}^{-1} dr$.}
\begin{align}
   & \mathbf{g}=P^{2}\Big[-\mathcal{H} du^2 -2dudr + r^2 d\theta^2 \Big] + P^{-2}r^2 \sin^2 \theta d\phi^2,\label{Melvin_BH_metric_2}\\
   & \mathbf A=A_{\phi}d\phi= \frac{1}{2}P^{-1}Br^2 \sin^2\theta d\phi, \quad \mathbf F= P^{-2}Br\sin \theta (\sin \theta dr + r\cos \theta d\theta) \wedge d \phi, \\\label{FandA_for_Melvin_BH} 
   & \mathcal{H}=1- \frac{2M}{r}, \quad
    P= 1+ \frac{1}{4} B^2 r^2 \sin^2 \theta, 
\end{align}
where $M$ and $B$ are respectively the constant mass and magnetic field strength parameters. The metric can be cast into a GKS form,\footnote{In \cite{Adolfo_Melvin_BH}, a generalization of the Melvin spacetimes, as well as their black hole extensions, were constructed in the Einstein-Modmax theory \cite{Modmax_1,Modmax_2,Modmax_3}. Further, it was shown there that the generalized Schwarzschild-Melvin black holes admit a GKS form.} with the background, KS vector and the function $H$ given by\footnote{The solutions obtained in \cite{Ort_Magnetic_BH}, which form a higher-dimensional generalization of the Schwarzschild-Melvin black holes in the Einstein-Maxwell theory, also readily admit a GKS form, exactly as in the four-dimensional case presented here. The implications of its GKS structure will be studied elsewhere.}
\begin{align}
   & \mathbf{\Bar g}=P^{2}\Big[-du^2 -2dudr + r^2 d\theta^2 \Big] + P^{-2}r^2 \sin^2 \theta d\phi^2, \label{Melvin_metric}\\
    & 2H= -P^{2} \frac{2M}{r}, \quad \mathbf{k}= du. \label{KS_vec_Melvin_BH}
\end{align}

The background, which forms the $M=0$ limit of the full solution, represents the Bonnor-Melvin magnetic universe and has exactly the same on-shell energy-momentum tensor as the full geometry. Hence, from Proposition \ref{prop_geod}, $\mathbf{k}$ must be geodesic, which can indeed be confirmed explicitly. The full geometry is of Weyl type $I_i$ with four distinct WANDs \cite{Bose_Ernst_algebra_type} (cf. also \cite{Ort_Melvin}). On taking the massless limit $M=0$, the four single WANDs degenerate to two double WANDs, and hence the Melvin spacetimes are of Weyl type $D$ \cite{Wild_Melvin_type_D, Griffiths_Podolský_2009}. Therefore, from Proposition \ref{prop_double_WAND_k}, we conclude that the KS vector \eqref{KS_vec_Melvin_BH} cannot define a WAND of these spacetimes, which is consistent with the result of \cite{Bose_Ernst_algebra_type}.

As for the Ricci types, the background is of type $D$ with its two double Ricci ANDs coinciding with its two double WANDs \cite{Griffiths_Podolský_2009}, and hence $\mathbf{k}$ cannot be a Ricci AND. Moreover, as $R_{ab}=\Bar{R}_{ab}$, the full geometry is also of Ricci type $D$. Since $\mathbf{k}$ is not a Ricci AND (and thus not a Riemann AND), Proposition \ref{prop_opt_contr} cannot be applied, and Proposition \ref{prop_opt_contr_2} does not provide any new insights regarding the optical constraint or the optical matrix. However, it can be explicitly checked that $\mathbf{k}$ forms a shearing (expanding-twistfree) null congruence with a non-degenerate optical matrix and, therefore, does not satisfy the optical constraint (cf. Section \ref{Optical constraint} for comments linking the shear and the optical constraint in $n=4$).

\subsection{Static black hole in the Einstein-Gauss-Bonnet theory (GR-Beyond GR) }\label{beyond_GR}
For the case of GKS spacetimes belonging to the GR-Beyond GR category, we consider the example of a static vacuum black hole solution in the Einstein-Gauss-Bonnet (EGB) theory in arbitrary higher dimensions \cite{Boulware:1985wk,Wheeler:1985qd,Dotti:2005rc,Hervik:2019gly} (cf. also \cite{ Anabalon:2009kq}),\footnote{One could straightforwardly extend the current discussion to some of the other beyond GR examples studied in \cite{Hervik:2019gly}.} and show that it admits a GKS representation around a GR black hole background. The metric ansatz for static black holes, studied in \cite{Hervik:2019gly}, is given by the following simplified form of the GKS-Robinson-Trautman metrics (cf. Section \ref{RT_GKS_section} and also equation \eqref{RT_GKS_met_1})
\begin{align}
  \mathbf{ g}=  r^2 h_{ij}(x)dx^i dx^j -2dudr -2g_{uu}(r)du^2, \label{Modified_grav_soln_ansatz}
\end{align}
 where the $(n-2)$-dimensional base manifold $h_{ij}(x)$ is restricted to be a universal space,\footnote{See Appendix A of \cite{Hervik:2019gly} and the references therein for the definition of a universal space.} which, in particular, implies that it is an Einstein space with constant (curvature) scalar invariants.

The vacuum equations of the EGB theory are given by
\begin{align}
    \frac{1}{\kappa}\left( R_{ab} -\frac{1}{2}Rg_{ab}+\Lambda g_{ab}\right) + 2\gamma \left(RR_{ab}-2R_{acbd}R^{cd}+R_{acde}R_{b}^{\hspace{1mm}cde}-2 R_{ac}R_{b}^{\hspace{1mm}c}-\frac{1}{4}I_{\mbox{\tiny GB}}g_{ab} \right)=0, \label{GB_EOM}
\end{align}
where $I_{\mbox{\tiny GB}}=R_{abcd}R^{abcd}-4R_{ab}R^{ab}+ R^2$, and $\kappa$, $\gamma$ are constants. Upon using the metric ansatz \eqref{Modified_grav_soln_ansatz}, the vacuum EGB equations lead to the following solution for the metric function $g_{uu}$ \cite{Hervik:2019gly,Dotti_GB,Maeda_GB,Dotti_GB_2,Bogdanos_GB} 
\begin{align}
    2g_{uu}=K + \frac{r^2}{2\kappa \hat{\gamma}}\Bigg[ 1\pm \sqrt{1+ 4\kappa \hat{\gamma}\left( \frac{2\Lambda}{(n-2)(n-1)}+\frac{2 M}{r^{n-1}}\right)-\frac{4\kappa^2 \hat{\gamma}^2\Tilde{I}^2_W}{r^4}}\Bigg], \label{GB_soln}
\end{align}
where $K=0,\pm1$ is related to the base space Ricci scalar as in the vacuum GKS-Robinson-Trautman example (cf. Section \ref{vacu_RT}), and $M$ is the constant mass parameter. The other quantities in \eqref{GB_soln} are defined as
\begin{align}
    \hat{\gamma}=(n-3)(n-4)\gamma, \quad (n-2)(n-3)(n-4)(n-5)\Tilde{I}_W^2= \Tilde{C}_{\beta_1 \dots \beta_4}\Tilde{C}^{\beta_1 \dots \beta_4},
\end{align}
with $\Tilde{C}_{\beta_1 \dots \beta_4}$ being the Weyl tensor of the base manifold $h_{ij}$. The branch of the solution \eqref{GB_soln} with the minus sign admits a GR black hole limit, which is obtained by taking $\hat{\gamma} \to 0$ and is given by
\begin{align} 
2f_{GR}\equiv 2g_{uu}|_{\tiny \hat{\gamma} \to 0}=K-\frac{2\Lambda}{(n-2)(n-1)} -\frac{2M}{r^{n-3}}. \end{align}
Therefore, we can perform a GKS split of this specific branch of the EGB solution around the Einstein black hole (EBH) as\footnote{One can also define a GKS splitting using the background $\mathbf{\Bar g}=r^2 h_{ij}(x)dx^i dx^j -2dudr -Kdu^2$ and the scalar function $2H= (2g_{uu}-K)$. This splitting works for both branches of the solution \eqref{GB_soln}; however, the background does not represent a black hole.}
\begin{align} \mathbf{\Bar g}{\mbox{\tiny {EBH}}}=r^2 h_{ij}(x)dx^i dx^j -2dudr -2f_{GR}(r)du^2, \quad H= (g_{uu}-f_{GR}), \label{GB_GKS_bg_1} \end{align}
with the KS vector identified as $\mathbf{k}=-du$.

We see that the background represents a vacuum black hole solution in the GKS-Robinson-Trautman class and, therefore, must form a subclass of the $M \neq 0$ branch of Section \ref{vacu_RT},\footnote{It forms a subclass because the assumption of the base, $h_{ij}$, being a universal space is more restrictive than it being an Einstein space.} whose algebraic properties are already analyzed there. As for the full geometry, it was shown in \cite{Hervik:2019gly} that it is of Weyl type $D$, with the two mWANDs being $\mathbf{k}$ and $\mathbf{n} = dr + g_{uu}du$. Moreover, since $\mathbf{k}$ is geodesic and satisfies the optical constraint in the background, it must also satisfy these properties in the full EGB black hole geometry, owing to their invariance under a GKS transformation (cf. equations \eqref{Li0_GKS}). Consistent with this, it turns out that $R_{00}=0=\Bar{R}_{00}$ (cf. $(B.7)$ of \cite{Hervik:2019gly}), and $R_{ij}\propto \Bar{R}_{ij}\propto S_{ij}=\frac{1}{r}\delta_{ij}$ (cf. $(B.6)$ and the subsequent parts of \cite{Hervik:2019gly}). Finally, we note that the full geometry is also of Ricci type $D$, with the Ricci double ANDs coinciding with the mWANDs (cf. again $(B.6),(B.7)$ of \cite{Hervik:2019gly}).

\section{Curvature components (geodesic $\mathbf{k}$) }
\setcounter{equation}{0}
\label{Curvature}
In this appendix, we provide the frame components of the Riemann, Ricci, and Weyl tensors for GKS spacetimes when $\mathbf{k}$ is geodesic and affinely parametrized. 
\subsection*{Riemann tensor components} \label{Riemann}
\begin{align}
      R_{0i0j}=&  \Bar R_{0i0j}, \quad R_{010i}= \Bar R_{010i}, \quad R_{0ijk}   =  \Bar R_{0ijk}, \label{R0i0j_GKS}\\
     R_{0101}=&  \Bar R_{0101}+ D^2 H, \quad    R_{ijkl} = \Bar R_{ijkl} + 4H(A_{ij}A_{kl}+ A_{k[j}A_{i]l} + S_{l[i}S_{j]k}),\label{R0101_GKS}\\
       R_{0i1j} =& \Bar R_{0i1j} - H \Bar R_{0i0j}- L_{ij}DH- 2HL_{kj}A_{ki}, \quad R_{01ij} =\Bar R_{01ij} + 2 A_{ji} DH+ 4HS_{k[j}A_{i]k},\label{R0i1j_GKS}\\
    R_{011i}  =&  \Bar R_{011i} + H \Bar R_{010i} -\delta_i (DH)+ 2L_{[i1]} D H + L_{ji}\delta_jH + 2H(2L_{ji}L_{[1j]}+ L_{j1}A_{ji}),\label{R011i_GKS}\\
    R_{1ijk}=& \Bar R_{1ijk} + H \Bar R_{0ijk} + 2L_{[j|i|}\delta_{k]}H+ 2A_{jk}\delta_i H
     -  2H \Big( \delta_{i}A_{kj}+ L_{1j}L_{ki} -L_{1k}L_{ji}  \nonumber\\
     &  -L_{j1}A_{ki}+L_{k1}A_{ji}+ 2L_{[1i]}A_{kj}+ A_{lj}\overset{l}{M}_{ki}-A_{lk}\overset{l}{M}_{ji}\Big),\label{R1ijk_GKS_1}\\
  R_{1i1j}= & \Bar R_{1i1j}+H^{2}\Bar R_{0i0j} +2H^{2}(A_{ik}S_{kj}-S_{ik}A_{kj}) \nonumber \\
 & +\delta_{(i}\delta_{j)}H+\overset{k}{M}_{(ij)}\delta_{k}H +(2L_{1j}-L_{j1})\delta_{i}H+(2L_{1i}-L_{i1}\big)\delta_{j}H+\Bar N_{(ij)}DH-S_{ij}\Bar \Delta H \nonumber \\
 & +2H\Big(\delta_{(i|}L_{1|j)}- \Bar \Delta S_{ij}-2L_{1(i}L_{j)1}+2L_{1i}L_{1j}-L_{k(i|} \Bar N_{k|j)}+L_{1k}\overset{k}{M}_{(ij)}-L_{k(i} \overset{k}{\Bar M}_{j)1}-L_{(i|k}\overset{k}{\Bar M}_{|j)1}\Big).\label{R1i1j_GKS_1}
\end{align}
Let us note in \eqref{R1i1j_GKS_1} that a sufficient condition for the term $2H^{2}(A_{ik}S_{kj}-S_{ik}A_{kj})$ to vanish is the optical constraint \eqref{opt_constraint}, while the equivalent condition is that $S_{ij}$ and $A_{kl}$ commute, i.e., $[S,A]=0$.
\subsection*{Ricci tensor components} \label{Ricci}
\begin{align}
      R_{00}=&\Bar R_{00} , \quad  R_{0i}=  \Bar R_{0i}\label{R00_GKS}\\
       R_{ij} =&\Bar R_{ij} - 2H \Bar R_{0i0j} + 2HL_{ik}L_{jk} -2S_{ij}\left[DH+ (n-2)\theta H\right],\label{Rij_GKS}\\
     R_{01}  =& \Bar R_{01}- H \Bar R_{00} - (D^2H+ (n-2)\theta DH+2H\omega^2),\label{R01_GKS}\\
     R_{1i}  =& \Bar R_{1i} + H \Bar R_{0i} + 2H \Bar R_{010i} -\delta_i (DH)+ 2L_{[i1]} D H + 2L_{ij}\delta_{j}H - L_{jj}\delta_{i}H\nonumber\\
    &+ 2H(\delta_{j}A_{ij}+A_{ij}\overset{j}{M}_{kk}-A_{kj}\overset{i}{M}_{kj}- L_{jj}L_{1i} + 3L_{ij}L_{[1j]} + L_{ji}L_{(1j)}),\label{R1i_GKS}\\
    R_{11}= & \Bar R_{11}+H^{2}\Bar R_{00}   +\delta_{i}\delta_{i}H +(4L_{1i}-2L_{i1}+\overset{i}{M}_{kk})\delta_{i}H+\Bar N_{ii}DH-S_{ii}\Bar \Delta H \nonumber \\
 & +2H\Big(\delta_{i}L_{1i}- \Bar \Delta S_{ii}+4L_{1i}L_{[1i]}-L_{ki} \Bar N_{ki}+L_{1k}\overset{k}{M}_{ii}-2S_{ik}\overset{k}{\Bar M}_{i1}\Big).\label{R11_GKS_1}
\end{align}
It is evident from \eqref{R11_GKS_1} that the lower Ricci tensor $R_{ab}$ is no longer linear in $H$ for GKS spacetimes with a general background (even with geodesic $\mathbf{k}$) \cite{martin1986petrov, bilge}, unlike the KS class \cite{HD_KS, Malek:2010mh, Ort_Srini}. However, it was noted in \cite{Grses1975LorentzCT, Taub:1981evj, Dereli:1986cm, Stephani:2003tm} that the mixed tensor $R^a_{\hspace{1mm}b}$ remains linear in $H$. 

The Ricci scalar is given by
\begin{align}
    R=  \Bar{R}  - 4H \Bar R_{00} -2 \left[D^2H+ 2(n-2)\theta D H + H(n-2)(n-3)\theta^2 + H (\omega^2- \sigma^2)\right].
\end{align}
\subsection*{Weyl tensor components}
The frame components of the Weyl tensor can be straightforwardly obtained from the standard definition, using the expressions for the Riemann and Ricci tensors given in the previous subsections. Since the precise expressions are not particularly illuminating, we present below only the essential structure important for the analysis in Section \ref{GKS_with_geo_k}. The Weyl components of the full geometry with $+2$ and $+1$ boost weights are identically equal to the respective background Weyl components. In particular, we have
\begin{align}
    & C_{0i0j}= \Bar{C}_{0i0j}, \quad   C_{010i}= \Bar C_{010i}, \quad C_{0ijk}= \Bar C_{0ijk}. \label{C0i0j}
    \end{align}
    As for the components with boost weights ranging from $0$ to $-1$, they are equal to the corresponding background Weyl components upto terms linear in H
    \begin{align}
  & C_{01ij}= \Bar{C}_{01ij}   + \text{terms linear in } H, \dots,\hspace{1mm} C_{1ijk}= \Bar C_{1ijk}+  \text{terms linear in } H .\label{C0101}
     \end{align}
For the boost weight $-2$ component, similar to expressions for $R_{1i1j}$ and $R_{11}$, there is also additionally a piece quadratic in $H$
     \begin{align}
    C_{1i1j} =& \Bar C_{1i1j}+H^{2} \Big(\Bar C_{0i0j} +2(A_{ik}S_{kj}-S_{ik}A_{kj}) \Big) + \text{terms linear in } H . \label{C1i1j}
    \end{align}
The terms linear in $H$, whose exact structures are not elaborated above, also involve derivatives of $H$. For example, in the case of $C_{01ij}$, the linear terms are given by $(2 A_{ji} DH+ 4HS_{k[j}A_{i]k})$.
\section{A review of selected results on Robinson-Trautman spacetimes}
\setcounter{equation}{0}
\label{RTspacetimes_append}
Robinson-Trautman geometries are characterized by the presence of a geodesic, expanding, twistfree, shearfree null congruence generated by a null vector field $\mathbf{l}$ \cite{Griffiths_Podolský_2009,Stephani:2003tm}. In coordinates adapted to $\mathbf{l}$, the Robinson-Trautman metric reads \cite{Podolsky_Ort_RT,Podolsky_Svarc_RT}
\begin{align} ds^2= g_{ij}(r,u,x) dx^{i}dx^{j} + 2g_{ui}(r,u,x)dudx^{i}- 2dudr - 2g_{uu}(r,u,x)du^2, \label{RT_general} \end{align}
where $u = \text{constant}$ defines null hypersurfaces normal to $\mathbf{l}$, $r$ is an affine parameter of the geodesic $\mathbf{l}$ (i.e., $l^a\partial_a = \partial_r$ and hence $l_a dx^a = -du$ ), and $x^i$ (denoted collectively as $x$) are $(n-2)$ coordinates on the spatial hypersurface defined by fixing $u$ and $r$.

Let us introduce the following null coframe adapted to the metric \eqref{RT_general}
\begin{align} 
 \mathbf{m}^{(0)}=n_a dx^{a}\equiv dr+g_{uu}du-g_{ui}dx^{i}, \quad \mathbf{m}^{(1)}=l_a dx^{a}= -du, \quad \mathbf{m}^{(i)} = m^{i}_{\hspace{1mm}j}(r,u,x)dx^{j}, \label{coframe_RT}
\end{align} 
where the coefficients $m^{i}_{\hspace{1mm}j}$ are defined by $\delta_{ij}m^{i}_{\hspace{1mm}k}m^{j}_{\hspace{1mm}l}=g_{kl}$, so that $\mathbf{m}^{(i)}$ form an orthonormal coframe adapted to the $(n-2)$-dimensional spatial metric $ g_{ij}$. The dual frame is then given by
\begin{align}
    \mathbf{m}_{(0)}=\mathbf{l}=\partial_r, \quad  \mathbf{m}_{(1)}=\mathbf{n}= g_{uu}\partial_r-\partial_u, \quad \mathbf{m}_{(i)}= m_{i}^{\hspace{1mm}j}\partial_{j}+g_{uj}m_{i}^{\hspace{1mm}j} \partial_r ,\label{frame_RT}
\end{align}
where $m_{i}^{\hspace{1mm}j}=(m^{-1})^{j}_{\hspace{1mm}i}$. The non-vanishing Ricci rotation coefficients corresponding to the null-frame read\footnote{The Ricci rotation coefficients defined by the covariant derivative of $\mathbf{l}$ are denoted by $L_{ab}$ here. This is in anticipation of the identification of $\mathbf{l}$ with the KS vector of GKS-Robinson-Trautman spacetimes in Section \ref{RT_GKS_section}.}
\begin{align}
    &L_{i1}= L_{1i}= N_{i0}= - \frac{1}{2}g_{uj,r}m_{i}^{\hspace{1mm}j},\quad \overset{i}{M}_{j0}=\frac{1}{2}(m^{i}_{\hspace{1mm}l,r}m_{j}^{\hspace{1mm}l}-m^{j}_{\hspace{1mm}l,r}m_{i}^{\hspace{1mm}l}), \quad
S_{ij}=\frac{1}{2}(m^{i}_{\hspace{1mm}l,r}m_{j}^{\hspace{1mm}l}+m^{j}_{\hspace{1mm}l,r}m_{i}^{\hspace{1mm}l}), \label{Ricci_rot_Li1_RT} \\
& N_{(ij)}= -\frac{1}{2}(m^{i}_{\hspace{1mm}q,u}m_{j}^{\hspace{1mm}q}+m^{j}_{\hspace{1mm}q,u}m_{i}^{\hspace{1mm}q}) + g_{uu}S_{ij}, \quad N_{[ij]}= (g_{ul,k} - g_{uk,r}g_{ul}) m_{[i}^{\hspace{1mm}k}m_{j]}^{\hspace{1mm}l},  \quad L_{11}= -g_{uu,r}, \\
    & \overset{i}{M}_{j1}=-\frac{1}{2}(m^{i}_{\hspace{1mm}q,u}m_{j}^{\hspace{1mm}q}-m^{j}_{\hspace{1mm}q,u}m_{i}^{\hspace{1mm}q})+g_{uu} \overset{i}{M}_{j0}-N_{[ij]}, \quad N_{i1}= (g_{uu,r}g_{uj}+g_{uu,j}+g_{uj,u}-g_{uj,r}g_{uu})m_{i}^{\hspace{1mm}j},\\
   &  \overset{i}{M}_{ql}= (m^{i}_{\hspace{1mm}j,r}g_{uk}+m^{i}_{\hspace{1mm}j,k})m_{q}^{\hspace{1mm}j}m_{l}^{\hspace{1mm}k}  -\Gamma^{j}_{\hspace{1mm}kw}(r, u, x) m_{l}^{\hspace{1mm}k}m_{q}^{\hspace{1mm}w}m^{i}_{\hspace{1mm}j}  -\frac{1}{2}(g_{wj,r}g_{uk} +g_{kj,r}g_{uw}-g_{kw,r}g_{uj})m_{l}^{\hspace{1mm}k}m_{q}^{\hspace{1mm}w}m_{i}^{\hspace{1mm}j}, \label{Ricci_rot_Mijk_RT}\\
 & \Gamma^{l}_{\hspace{2mm}kw}(r, u, x)= \frac{1}{2} g^{jl}  (g_{wj,k}+g_{kj,w}-g_{kw,j}). \label{christoffel_RT_base_space}
\end{align}
Assuming the spatial frame vectors $\mathbf{m}_{(i)}$ to be parallel transported along $\mathbf{l}$, which can always be achieved by means of a local $SO(n-2)$ rotation of $\mathbf{m}_{(i)}$ (cf. equation $7$ of \cite{Ricci_in_HD}), we obtain the following simplifications for $\overset{i}{M}_{j0}$ and $S_{ij}$.
\begin{align}
    &\overset{i}{M}_{j0}=0, \quad S_{ij}=m^{i}_{\hspace{1mm}l,r}m_{j}^{\hspace{1mm}l}=m^{j}_{\hspace{1mm}l,r}m_{i}^{\hspace{1mm}l}.\label{Sij_RT}
\end{align}
Further, since $\mathbf{l}$ is shearfree, we have $S_{ij} = \theta(r, u, x) \delta_{ij}$. Therefore, on integrating equation \eqref{Sij_RT}, one finds \cite{Podolsky_Ort_RT,Podolsky_Svarc_RT}
\begin{align}
    m^{i}_{\hspace{1mm}j}= \exp \Big(\int dr \theta(r,u,x)\Big)  \hat {m}^{i}_{\hspace{1mm}j}(u,x) \iff g_{ij}= \exp \Big(2 \int dr \theta(r,u,x)\Big) h_{ij}(u,x), \label{gij_RT}
\end{align}
where  $\delta_{ij}\hat m^{i}_{\hspace{1mm}k}\hat m^{j}_{\hspace{1mm}l}=h_{kl}$. Although in the above relation we have used parallelly transported spatial frame vectors $\mathbf{m}_{(i)}$, the relation $\delta_{ij} m^{i}_{\hspace{1mm}k} m^{j}_{\hspace{1mm}l} = g_{kl}$ remains invariant under $SO(n-2)$ rotations of the spatial frame vectors. Therefore, the expression for $g_{ij}$ given in \eqref{gij_RT} holds true regardless of whether the spatial frame vectors are parallel transported along $\mathbf{l}$ or not.

One can calculate the frame components of the Riemann curvature using the expressions for the Ricci rotation coefficients \eqref{Ricci_rot_Li1_RT}-\eqref{christoffel_RT_base_space}, and Ricci identities given in \cite{Ricci_in_HD}.\footnote{See \cite{Podolsky_Svarc_RT} for a full list of coordinate components.} To keep things simple, we give only the $R_{0i0j}$ component and a consequent result. Using Ricci identity (11g) of \cite{Ricci_in_HD}, and the relevant expressions for Ricci rotation coefficients, one obtains 
\begin{align}
    R_{0i0j}=-(\theta_{,r} +  \theta^2)\delta_{ij}.\label{R0i0j_RT}
\end{align}
From the above equation, one also deduces that
\begin{align}
    &R_{00}=-(n-2)(\theta_{,r} +  \theta^2),\label{R00_RT}\\
    & C_{0i0j}=0.\label{C0i0j_RT}
\end{align}
The expressions \eqref{R0i0j_RT}-\eqref{C0i0j_RT} were already obtained in \cite{Podolsky_Svarc_RT}, where it was concluded that Robinson-Trautman spacetimes are necessarily of Weyl type $I$ or more special with $\mathbf{l}$ being a WAND. From equations \eqref{R0i0j_RT} and \eqref{R00_RT}, it is straghtforward to make the following observation (cf. Theorem $1$ of \cite{Ortaggio:RT_pform} and references therein)
\begin{proposition}\label{prop_RT_AND}
    Let $\mathbf{l}$ generate an expanding, shearfree, twistfree, geodesic null congruence of a Robinson-Trautman spacetime, and let $r$ be an affine parameter along the geodesic. Then, the following are equivalent
    \begin{enumerate}
        \item $\mathbf{l}$ defines a Riemann AND of the spacetime.
        \item  $\mathbf{l}$ defines a Ricci AND of the spacetime.
        \item The expansion of $\mathbf{l}$ is given, up to gauge, by $\theta= \frac{1}{r}$. 
    \end{enumerate}
\end{proposition}
As a consequence of the above proposition, when $\mathbf{l}$ is a Riemann AND, the expression for $g_{ij}$ given by \eqref{gij_RT} simplifies to 
\begin{align}
   g_{ij} = r^2 h_{ij}(u,x). \label{gij_simplifi_RT}  
\end{align}
\section{A result on symmetric rank-$2$ tensors of type $D$ }
\setcounter{equation}{0}
\label{gen_result_type_D_tensors}
\begin{proposition}\label{Prop_rank_2_type_D}
  Let  $\mathbf{T}_{ab}$ be a symmetric rank-$2$ tensor of type $D$. Let $\{\mathbf{m}_{(0)} = \mathbf{l}, \mathbf{m}_{(1)} = \mathbf{n}, \mathbf{m}_{(i)}\} $ be a null frame adapted to the repeated ANDs $\mathbf{l}$ and $\mathbf{n}$ of $\mathbf{T}_{ab}$. Let $\{E_1,E_2,\dots, E_{n-2}\}$ be the eigenvalues of $T_{ij}$. Then, $\mathbf{T}_{ab}$ admits a single AND iff there exists at least a pair of eigenvalues $E_{i_1}, E_{i_2}$ such that $E_{i_1}>T_{01}>E_{i_2}$.
\end{proposition}
\noindent \textbf{Proof.}  In a null frame adapted to a type $D$ tensor $\mathbf{T}_{ab}$, by definition \cite{HD_alg_review}, all the frame components with non-zero b.w. must vanish, and at least one of the zero b.w. components $T_{01}$, $T_{ij}$ must be non-zero. In addition,  as $T_{ij}$ is symmetric, one can diagonalize it by performing a spatial rotation. Therefore, without loss of generality, let us assume that the adapted frame $\{\mathbf{m}_{(0)} = \mathbf{l}, \mathbf{m}_{(1)} = \mathbf{n}, \mathbf{m}_{(i)}\} $ already diagonalizes $T_{ij}$. In order to investigate the conditions for an additional single AND of $\mathbf{T}_{ab}$, let us perform the following null rotations about $\mathbf{n}$ \cite{Ricci_in_HD}

\begin{align}
     \mathbf{\hat{n}} & = \mathbf{n}, \quad
   \mathbf{\hat{l}}  = \mathbf{l} + z_{i}\mathbf{m}^{(i)} - \frac{1}{2} z^2 \mathbf{n}, \quad
  \mathbf{\hat{m}^{(i)}}  = \mathbf{m}^{(i)} - z_i \mathbf{n},\label{null_rot_n}
\end{align}
where $z_i$ are $(n-2)$ real functions. The b.w. $+2$ and $+1$ components in the new frame are given respectively by
\begin{align}
 &\Hat T_{00}= z_{i} z_{j} T_{ij} - z^2 T_{01}= \sum_{i=1}^{n-2} z_{i}^2 (E_i- T_{01}), \label{T_00_hat}\\
   & \Hat T _{0i}= z_{j}T_{ij}- z_{i}T_{01}= z_{i} (E_i -T_{01}) \quad (\text{no summation over } i).\label{T_0i_hat}
\end{align}
Let $j_1$, $j_2$, and $j_3 = (n-2) - (j_1 + j_2)$ be the number of positive, negative, and zero entries, respectively, of the matrix $\text{diag}(E_1 - T_{01}, \dots, E_{n-2} - T_{01})$. By reordering the spatial vectors $\mathbf{m}^{(i)}$ of the old frame, one can arrange
\begin{align}
   & E_{i}-T_{01} >0\quad \text{for } i=1,\dots, j_1, \nonumber \\
   & E_{i}-T_{01} <0\quad \text{for } i=j_1+1,\dots, j_1+j_2, \nonumber \\
   & E_{i}-T_{01} =0\quad \text{for } i=j_1+j_2+1,\dots, n-2. 
\end{align}
One can therefore rewrite \eqref{T_00_hat} as
\begin{align}
 &\Hat T_{00}= \sum_{i=1}^{j_1} z_{i}^2 (E_i- T_{01})+\sum_{i=j_1+1}^{j_1+j_2} z_{i}^2 (E_i- T_{01})\label{T_00_hat_2}.
\end{align}
Note that each term in the first sum is positive and each term in the second sum is negative.

Let us now proceed to prove the claim of the proposition. First, assume that $\mathbf{\hat{l}}$ is a single AND of $\mathbf{T}_{ab}$. By definition \cite{HD_alg_review}, this implies that $\hat{T}_{00} = 0$ and not all of the $\hat{T}_{0i}$ components vanish. From \eqref{T_0i_hat}, we see that if $j_1 = 0$ and $j_2 = 0$, then $\hat{T}_{0i}$ would identically vanish. Therefore, we require that at least one of $j_1$ or $j_2$ is non-zero. Additionally, for $\hat{T}_{00} = 0$, it is necessary that \eqref{T_00_hat_2} contains both positive and negative terms, thereby implying that $j_1 \neq 0\neq j_2$, i.e., there exists at least one pair of eigenvalues $E_{1}$ and $E_{j_1+1}$ such that $E_{1} > T_{01} > E_{j_1+1}$.

Conversely, suppose that $j_1 \neq 0 \neq j_2$. One can then set $\hat{T}_{00} = 0$ by choosing

\begin{align}
& z_{i} = \sqrt{\frac{1}{j_1 (E_i - T_{01})}} \quad \text{for }i = 1, \dots, j_1, \label{z_i_cond_1} \\
& z_{i} = \sqrt{\frac{1}{j_2 (T_{01} - E_i)}}\quad \text{for } i = j_1 + 1, \dots, j_1 + j_2, \label{z_i_cond_2}\\
& z_{i}= \text{arbitrary}\quad \text{for } i=j_1+j_2,\dots, n-2.\label{z_i_cond_3}
\end{align}
Moreover, the non-vanishing b.w. $+1$ components are given by

\begin{align}
    \Hat T _{0i} =& \sqrt{\frac{(E_i -T_{01})}{j_1} } \quad \text{for }i = 1, \dots, j_1, \\
    \Hat T _{0i} =& -\sqrt{\frac{( T_{01}-E_i)}{j_2} }\quad \text{for } i = j_1 + 1, \dots, j_1 + j_2.
\end{align}
Therefore, $\mathbf{\hat{l}}$ defined by \eqref{z_i_cond_1}-\eqref{z_i_cond_3} form a $j_3$-dimensional surface of single ANDs for $\mathbf{T}_{ab}$.
\begin{remark}\label{remark_single_RAND}
For $n=3$, $T_{ij}$ is just one component. Therefore, one can see from \eqref{T_00_hat} and \eqref{T_0i_hat} that $\Hat T_{00}=0 \iff \Hat T_{0i}=0 $. Thus, in $n=3$, a single AND is forbidden for a symmetric rank-$2$ tensor of type $D$.
    \end{remark}
    \begin{remark}
    \label{rem_Eins}
 It can be seen that for Einstein spaces, defined by \eqref{vacuum_Einstein}, the only non-zero components of $R_{ab}$ in any null frame are the b.w. zero components $R_{01}$ and $R_{ij}$. Therefore, Einstein spacetimes are of Ricci type $D$. However, they do not admit a single Ricci AND, as every null direction is doubly degenerate. This also agrees with Proposition \ref{Prop_rank_2_type_D}, as in this case, all the eigenvalues of $R_{ij}$ are equal.
        
    \end{remark}

    \begin{remark}\label{remark_single_WAND_typeD_4d}
        Similar to rank-$2$ symmetric tensors, type $D$ Weyl tensors may also admit single WANDs under special circumstances\cite{Ortaggio:RT_pform}.
  However, for $n = 4$, it is well known that the Weyl tensor has exactly four WANDs, counting also the multiplicities (cf. \cite{Stephani:2003tm} and the references therein). Therefore, $n>4$ is a necessary condition for type $D$ Weyl tensors to possess a single WAND.
\end{remark}



%
%
%
%
%

\end{document}